\documentclass[english,aps, usenatbib,times]{mn2e}
\usepackage[T1]{fontenc}
\usepackage[latin9]{inputenc}
\usepackage{hyperref}
\usepackage{amsmath, amssymb}
\usepackage{esint}
\usepackage{graphicx}
\usepackage{subfigure}
\usepackage[english,english]{babel}
\usepackage{amssymb,amsfonts,textcomp}
\usepackage{supertabular}
\usepackage{hhline}
\usepackage{hyperref}
\usepackage[usenames]{color}
\usepackage{amsmath}
\def\genus{Euler characteristic}
\def\horizon{{{\sc horizon} $4\pi$\,}}

\hypersetup{dvips,colorlinks=true,breaklinks=true,bookmarks=true,hypertexnames=false,bookmarksnumbered=true,backref=page,pagebackref=true,citecolor= black,linkcolor= black,bookmarksopen=false,bookmarks=true,bookmarksnumbered=true}

 \global\long\def\vk{{\bf k}}
 
\def\gtrsim{\lower.5ex\hbox{$\; \buildrel > \over \sim \;$}}
\def\lwsim{\lower.5ex\hbox{$\; \buildrel < \over \sim \;$}}

\newcommand{\dd}{\mathrm{d}}

\begin{document}

\onecolumn

\title[Non-Gaussian  Minkowski functionals  \&
extrema counts in redshift space]{Non-Gaussian  Minkowski functionals  \&
extrema counts \\ in redshift space}
\author[S. Codis,  C. Pichon, D. Pogosyan, F. Bernardeau, T. Matsubara  ]{S. Codis$^{1,2}$,  C. Pichon$^{1,2}$,  D. Pogosyan$^{3}$, F. Bernardeau$^{2,1}$, T. Matsubara$^{4}$. \\
 $^{1}$Institut d'Astrophysique
de Paris \& UPMC (UMR 7095), 98, bis boulevard Arago , 75 014, Paris,
France.\\
$^{2}$ Institut de Physique Th\'eorique, Orme des Merisiers, b\^{a}timent 774, CEA/Saclay
F-91191 Gif-sur-Yvette, France.\\
$^{3}$  Department of Physics, University of Alberta, 11322-89 Avenue, Edmonton, Alberta, T6G 2G7, Canada.\\
$^{4}$  Kobayashi-Maskawa Institute for the Origin of Particles and the Universe, Nagoya University, Chikusa-ku, Nagoya, 464-8602, Japan\\
}
\maketitle

\begin{abstract}

In the context of upcoming large-scale structure surveys such as Euclid, 
it is of prime importance to quantify the effect of peculiar velocities on geometric probes.
Hence the formalism to compute {\sl  in redshift space} the geometrical and topological one-point statistics 
of mildly non-Gaussian 2D and 3D cosmic fields is developed. 
Leveraging the partial isotropy of the target statistics, the Gram-Charlier expansion of the joint probability distribution of the field and its derivatives is reformulated
 in terms of the corresponding anisotropic variables. 
In particular, the cosmic non-linear evolution of the Minkowski functionals,
 together with the statistics of extrema  are investigated in turn for 3D catalogues and 2D slabs.
The amplitude of the non-Gaussian redshift distortion correction is estimated for these geometric probes. 
In 3D, gravitational perturbation theory is implemented in redshift space to predict the cosmic evolution of all  relevant Gram-Charlier coefficients.
Applications to the estimation of the cosmic parameters $\sigma(z)$ and $\beta=f/b_{1}$  from upcoming surveys is discussed. 
Such  statistics are of interest for anisotropic fields beyond cosmology.
\end{abstract}
\setcounter{tocdepth}{3}
\section{Introduction}
In modern cosmology,  random fields (3D or 2D) are fundamental ingredients in
the description of the 
large-scale matter density  and the Cosmic Microwave Background (CMB). The large-scale structure of  the matter distribution
in the Universe is believed to be the result of the gravitational  growth of primordial
nearly-Gaussian small perturbations originating from quantum fluctuations. 
Deviations from Gaussianity inevitably arise due to the non-linear dynamics of the growing structures, but may also be
present at small, but potentially detectable levels in the initial seed inhomogeneities. Thus, studying non-Gaussian signatures
in the random fields of cosmological data provides methods to learn both the details of early Universe's physics
and mechanisms for structure's growth, addressing issues such as the matter content of the Universe \citep{Zunckel11}, the role of bias between 
galaxies and dark matter distributions \citep{Desjacques}, and whether it is  dark energy or a modification to Einstein's gravity \citep{Wang12} 
that is responsible for the acceleration of the Universe's expansion.

With upcoming high-precision surveys, it has become necessary to revisit alternative tools  to investigate
the statistics of random cosmological fields so as to handle observables with different sensitivity.
 Minkowski functionals  \citep{Mecke94} have been being actively used
\citep{Gott87,Weinberg87,Melott88,Gott89, Hikage02, Hikage03,Park,Gott2007,Planck} as an alternative
to the usual direct measurements of higher-order moments and N-point correlation
functions \citep[amongst many other studies]{Scoccimarro98,Percival07,Gaztanaga09,Nishimichi10}
as they will present different biases and might be more robust, e.g. with regards to rare events.
These functionals are topological and geometrical estimators involving the critical sets of the field. As such they are invariant under a monotonic transformation of the field (and in particular bias-independent if the bias is monotonic!). If their expression for Gaussian isotropic fields has long been known  \citep{Doro70,Adler81,BBKS,Hamilton86,Bond87,Ryden88,Ryden89}, their theoretical prediction have also been computed more recently for mildly non-Gaussian fields  (\cite{Matsubara0}, \cite{PGP2009} and \cite{Matsubara10} for the first and second non-Gaussian corrections using a multivariate Edgeworth expansion, and more recently by \cite{Gay12} to 
all orders in non-Gaussianity). However, one major assumption in these results
has been the isotropy of the underlying field.

Indeed, the assumptions of homogeneity and isotropy of our observable Universe
are central to our understanding of the universe. This focuses our
primary attention on statistically homogeneous and isotropic random fields as the description of cosmological 2D and 3D data.
These statistical symmetries provide essential guidance for the theoretical  
description of the geometry of the cosmological random fields. Recent
papers \citep{PGP2009,PPG2011,Gay12} leverage the isotropy of the target
statistics to
develop non-Gaussian moment expansion to all orders for popular and novel statistics such as the above-mentioned
Minkowski functionals, but also extrema counts and the skeleton of the filamentary structures.
This formalism, generalizing the earlier works \citep{BBKS, Matsubara0, Matsubara, pogoskel},
 allows for controlled comparison of the non-Gaussian deviations
of geometrical measures in theoretical models, simulations, and the
observations, assuming homogeneity and isotropy. Nevertheless, 3D surveys are
conveyed in redshift space where the hypothesis of isotropy breaks down.
Indeed in astrophysical observations, the three dimensional positions of  structures are frequently not accessible directly.
While angular positions on the sky can be obtained precisely, the radial
line-of-sight (hereafter LOS) position of objects is determined by proxy,
e.g., via measurements of the LOS velocity component.  This implies that galaxy distribution data are presented
in redshift space  in cosmological studies.

Figures~{\ref{fig:field} and~\ref{fig:4pifield} illustrates the amplitude of such redshift distortion as  a function of scale.
On small, non-linear  scales, the well-known finger-of-God effect
\citep{Jackson72,Peebles1980}  stretches the collapsing clusters
along the line-of-sight  whereas on larger scales (see Figure~\ref{fig:4pifield})
the redshift space distortion flattens
the voids along the LOS \citep{Sargent77}, in accordance to the linear 
result of \cite{Kaiser87}.
\begin{figure}
 \begin{center}
 \includegraphics[width=0.4\textwidth]{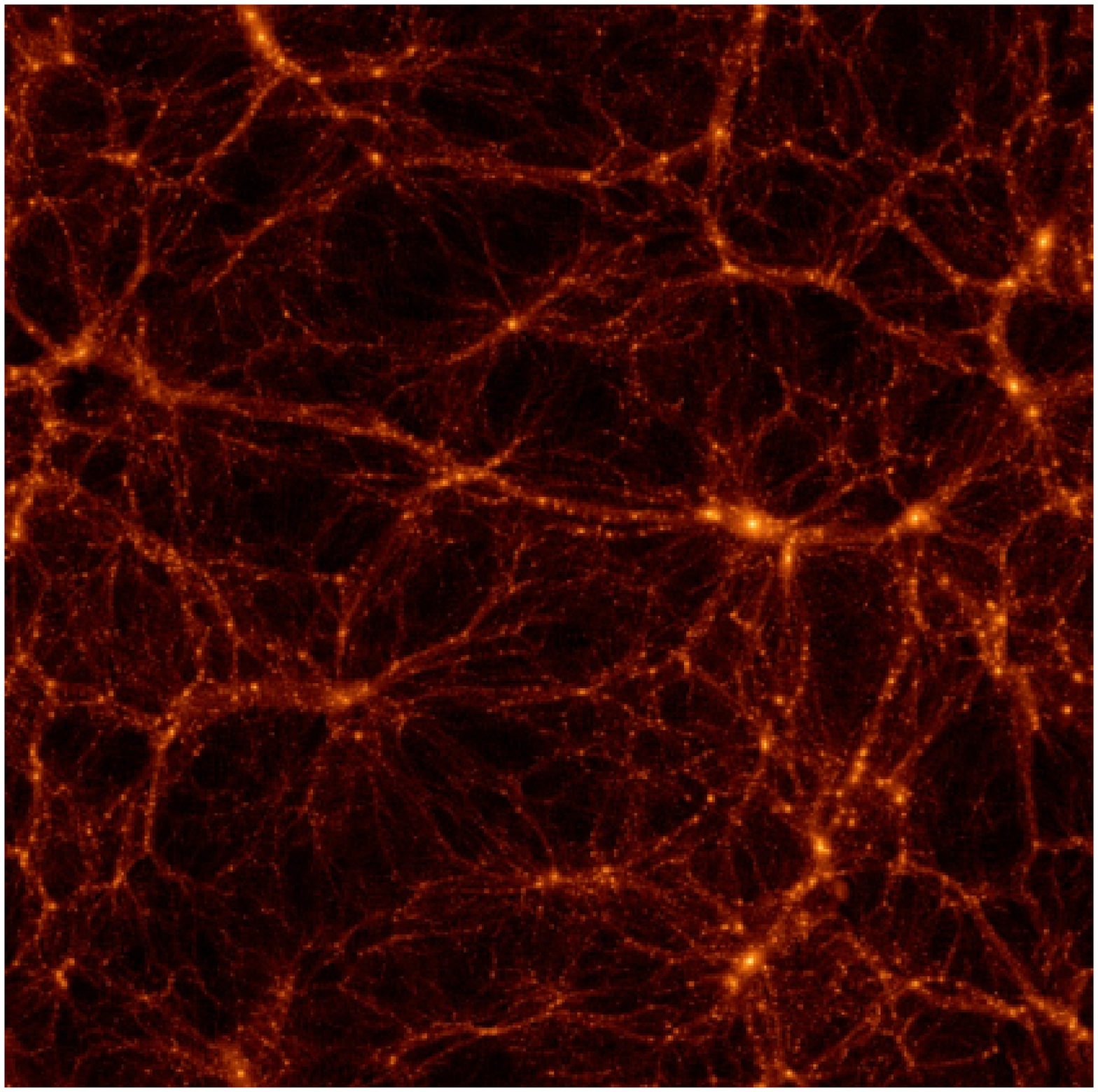}
 \includegraphics[width=0.4\textwidth]{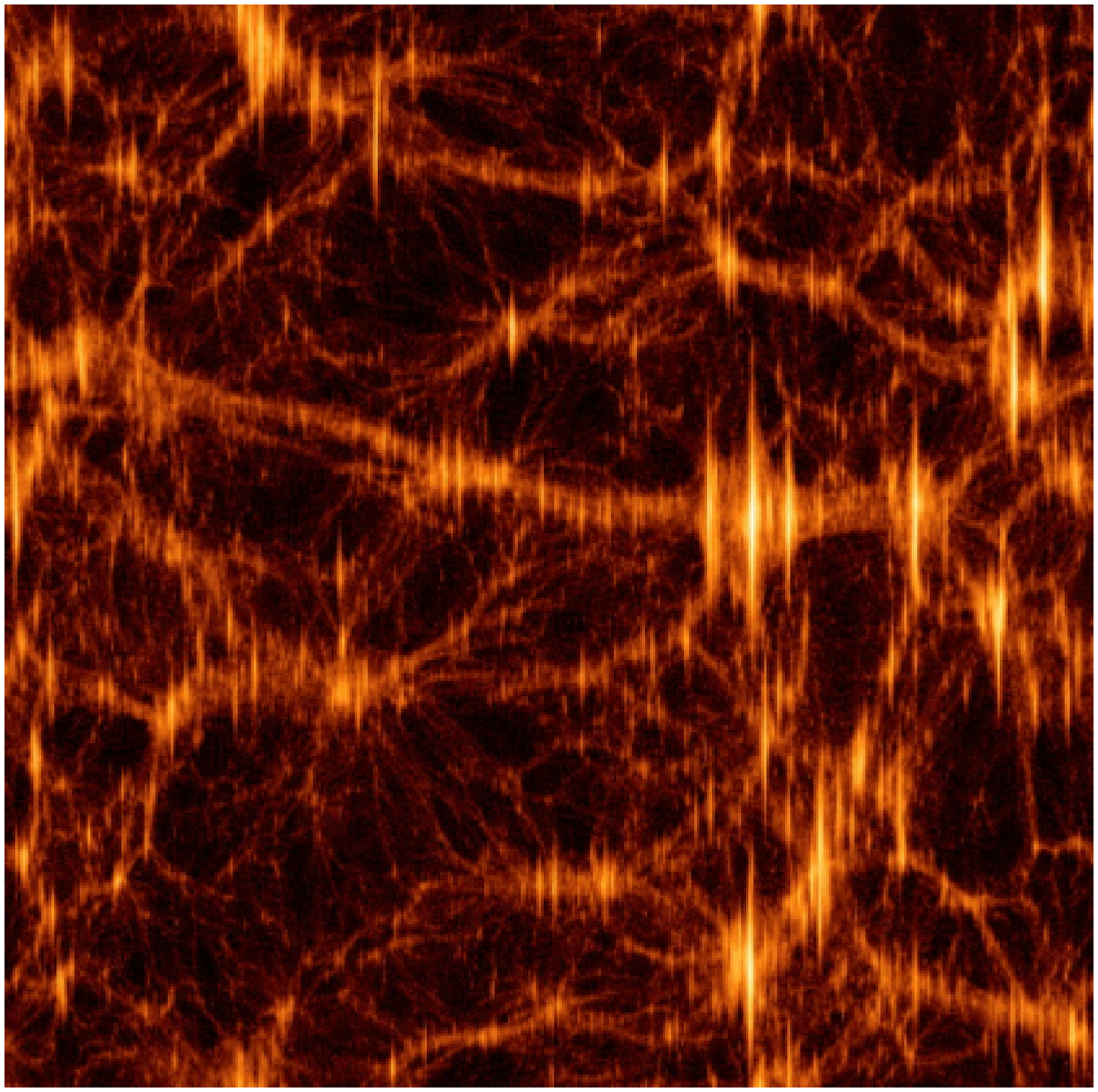}
 \caption{{\sl Left panel}: an example of a slice through a $512^3$ dark matter particles $\Lambda$CDM simulations  at redshift 0 ($\Omega_m=0.3$, $\Omega_\Lambda=0.7$, $\sigma_8=0.92$) in real space.  The boxsize is 100 Mpc$/h$, the slice thickness 10 Mpc$/h$.  {\sl   Right panel}  shows the same field when  redshift distortion  has been applied along the ordinate. Fingers of God are quite visible on that slice (see also Figure~\ref{fig:4pifield}).
 \label{fig:field} 
}
 \end{center}
\end{figure}
But anisotropy in data does not affect only cosmological surveys: this issue is ubiquitous in physics: e.g neutral hydrogen distribution is mapped in Position-Position-Velocity
cubes in studies of the interstellar or intergalactic medium, and turbulent distribution of magnetic field are mapped via Faraday rotation measure of the
synchrotron emission \citep{heyvaerts2013}.  Thus, as a rule, the underlying isotropy of  structures in real space is broken in the space where
data is available. Hence developing techniques to recover properties of
underlying fields from distorted datasets, anisotropic in the
LOS direction, is a fundamental problem in astrophysics.  
 
In this paper we thus present a theory of the non-Gaussian Minkowski functionals and
extrema counts for $z$-anisotropic cosmological fields,
targeted for application to  data sets in redshift space in the so-called
plane-parallel approximation.  Hence this theory
is a generalization of the formalism for mildly non-Gaussian but isotropic
fields of \cite{PGP2009,Gay12} on one hand, and the theory of
anisotropic redshift space
 effects on statistics in Gaussian limit developed in 
 \cite{Matsubara96} on the other hand.
The effects of anisotropy and non-Gaussianity are simultaneously important
 for a precise description of the large scale structures of the Universe (LSS hereafter) as mentioned in \cite{Matsubara1995}, where N-body simulations suggest that redshift space distortion has noticeable impact on the shape of the genus curve in the weakly non-linear regime.
At the same time the theory presented in this paper is general and applicable to  mildly
non-Gaussian homogeneous and statistically axisymmetric random fields of any
origin, for example for extending Velocity Channel Analysis of HI maps \citep{LazPog00} to  account for non-Gaussian
density compressibility, or describing cosmological perturbations in anisotropic Bianchi models of the Universe.

As an anticipation, Figure~\ref{fig:CGvsRDS} illustrates the importance of modelling appropriately the anisotropy
 on the particular example of the 3D Euler characteristics of a mildly non-Gaussian scale-invariant cosmological field.
 Indeed, the theoretical prediction assuming isotropy given in equation~(\ref{eq:chibar})  below (in purple) significantly
 fails to describe the measurement unlike the prediction proposed in equation~(\ref{eq:3DNGgenus}) (in red).
 This suggests that in redshift space it is of great importance to properly account for  anisotropy. This will be the topic of this paper.
\begin{figure}
 \begin{center}
 \includegraphics[width=0.45\textwidth]{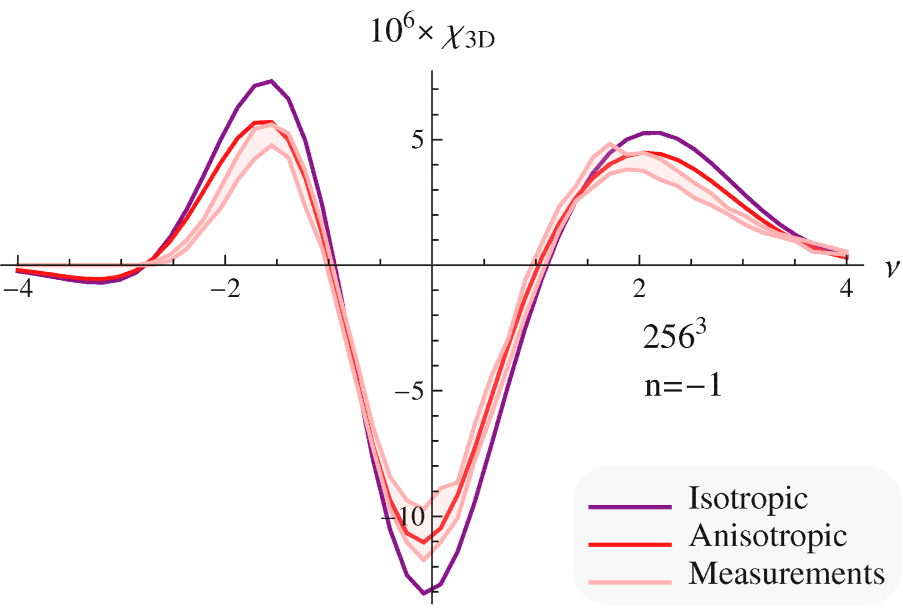}
 \includegraphics[width=0.45\textwidth]{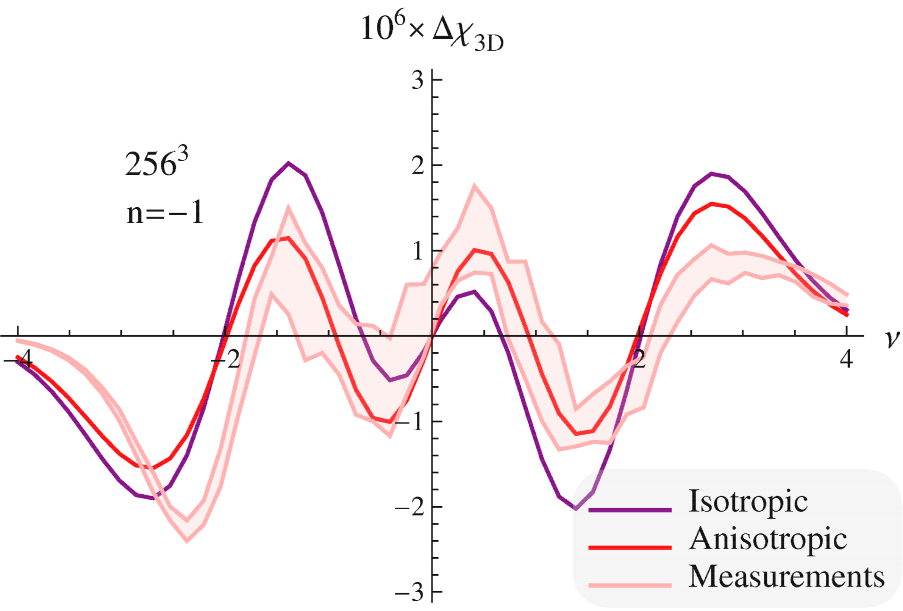}
\caption{{\sl  Left panel:} 
 3D Euler characteristic of a mildly non-Gaussian scale invariant cosmological field ($n=-1, f=1$, $\sigma=0.18$).  The prediction to first order in non-Gaussianity given by equation~(\ref{eq:3DNGgenus}) corresponds to the red solid line and the prediction assuming isotropy (see equation~(\ref{eq:chibar})) to the purple line. The one sigma shaded area is the measurement. {\sl  Right panel:} 
 same as left panel for the first order correction to the Gaussian distribution only.
\label{fig:CGvsRDS}
}
 \end{center}
\end{figure}
Section \ref{sec:minkowski} defines the statistics used in this paper, namely
Minkowski functionals and critical points counts. 
Section \ref{sec:formalism} introduces the formalism to deal with the joint probability density function of the density field and its derivatives up to second order in redshift space. In particular, it shows how rotational invariance is used to introduce a relevant set of variables which diagonalizes the Gaussian JPDF;  a Gram-Charlier expansion of the non-Gaussian JPDF is written there. 
 Section \ref{sec:Min23} presents Minkowski functionals and extrema counts in 2 and 3D.
 Section \ref{sec:genus} presents the full non-Gaussian expression for 3D and 2D Euler characteristic in redshift space, while
section \ref{sec:otherminkowski} is devoted to the last three Minkowski functionals (Area of isosurfaces, length of isocontours and contour crossing) for which we present expressions in redshift space up to first order in non-Gaussianity. Section \ref{sec:ext} sketches the derivation for extrema counts in two and three dimensions. 
Section~\ref{sec:nuf} re-expresses these functionals as a function of the filling factor threshold, while  Section~\ref{sec:invariance} investigates the implications of 
the topological invariance for the corresponding set of cumulants of the field.
Section \ref{sec:cosmo} analyses implication for the estimation of cosmological parameters. 
It shows how to compute the relevant three point cumulants in redshift space (e.g. skewness and its generalization to the derivatives of the density field) at tree order within perturbation theory (PT). In particular, it provides a way to compute analytically the angular part of the integrals under consideration. 
It then discusses which features of the bispectrum are  robustly measured using these 
non-Gaussian critical sets  and sketches two main applications: measuring $\sigma$ (hence dark energy) of the possibly masked underlying field, and measuring $\beta\equiv f/b_{1}\simeq\Omega_m^{0.55}/b_1$.
 Appendix~\ref{sec:appendixA}  lists properties of the relevant cumulants.
Appendix~\ref{sec:toys} presents briefly a   set of  ``$f_\mathrm{{nl}}$''  anisotropic non-Gaussian field  toy models which we use to validate our theory.
Finally Appendix~\ref{sec:chi3Dappendix} derives the 3D Euler characteristic at all orders in non-Gaussianity.
%
\section{Geometrical statistics in 2 and 3D }
\label{sec:minkowski}
Our prime focus is the geometrical and topological properties of cosmological fields.
One way to probe these  is to look at isocontours of the field at different thresholds and use Minkowski functionals to 
describe them  \citep{Mecke94}.
These Minkowski functionals are known to be the only morphological descriptors in Integral Geometry
 that respect motion-invariance, conditional continuity and additivity \citep{Hadwiger57}.
 As a result, they form a robust and meaningful set of observables. In $d$ dimensions,
 there are $d+1$ such functionals (4 in 3D and 3 in 2D), namely in 3D: the
encompassed volume, $f_V$, the surface area, 
${\cal N}_3$,
 the integral mean curvature and the integral Gaussian curvature (closely related to the Euler-Poincar\'e characteristic, $\chi$).
For random fields these functionals are understood as densities, i.e quantities per unit volume of space.

Especially when studying anisotropic field, complimentary information can be
obtained by using the geometrical statistics
and Minkowski functionals for the field obtained on lower dimensional sections of the 3D field. For example, in addition to 3D isocontour area statistics,
one can introduce the length of 2D isocontours on a planar sections ${\cal N}_2$, and contour crossings by
a line through 3D space, ${\cal N}_1$.  These statistics for cosmology were first introduced by \cite{Ryden88,Ryden89}.
 In the isotropic limit, they are trivially related: $2{\cal N}_{1}=4{\cal N}_{2}/\pi={\cal N}_{3}$;
  this relation does not hold anymore for an anisotropic field as it will be shown in sections~\ref{sec:N3}, \ref{sec:N2}
 and \ref{sec:N1}.  Similarly, in addition to the full Euler characteristic $\chi_\mathrm{ 3D}$ of 3D excursion sets,
we shall consider the 2D Euler characteristic, $\chi_\mathrm{ 2D}$, on  planar sections through the field.

 All geometrical measures can be expressed as averages over the joint
probability density function (JPDF)
 of the field and its derivatives. In the following, let us call $x$ the field under consideration and,
without loss of generality, assume that it has zero mean. In cosmological applications, 
this field, for instance,  can be the 3D density contrast.
Collecting the well-known results from an extensive literature \citep[e.g.][]{Rice44,Rice45,Ryden88,Matsubara96} in a compact form, 
we have for the first two Minkowski functionals
\begin{eqnarray}
\label{eq:filling_factor}
 f_{V}(\nu) &=& \left\langle \Theta(x-\nu) \right\rangle ~~,\\
{\cal N}_{3}(\nu)\!\!&=& 
\left\langle|\nabla x| \delta_\mathrm{ D}(x-\nu)\right\rangle\,,\quad
{\cal N}_{2}(\nu)= 
\left\langle|\nabla _{\!{\cal S}}\, x| \delta_\mathrm{ D}(x-\nu)\right\rangle\,, \quad
{\cal N}_{1}(\nu)
=\left\langle|\nabla  _{\!{\cal L}}\,x| \delta_\mathrm{ D}(x-\nu)\right\rangle\label{eq:N1}\,, 
\end{eqnarray}
where the $\delta_{\rm D}$-function in the statistical averaging signifies evaluation at the given threshold $x=\nu$,
while the step function reflects the cumulative averaging over the values above the threshold $x \ge \nu$.
We see that the family of threshold-crossing statistics is given by the average gradient of the field for ${\cal N}_3$
or its restriction to the plane ${\cal S}$ or line ${\cal L}$ for ${\cal N}_2$ and ${\cal N}_1$, respectively.

The average of the Gaussian curvature on the isosurface is, via the Gauss-Bonnet theorem, its topological Euler characteristic
 $\chi$, which thus can be expressed directly as 
 $\chi(\nu)=\left\langle \delta_\mathrm{ D}(x-\nu)\delta_\mathrm{ D}(\nabla_{1}x)\delta_\mathrm{ D}(\nabla_{2}x)|\nabla_{3}x|(\nabla_{1}\nabla_{1}x \nabla_{2}\nabla_{2}x -(\nabla_{1}\nabla_{2}x )^{2})\right\rangle$
\citep{Hamilton86,Matsubara96}. In this paper we use the Euler characteristic, $\chi_\mathrm{ 3D}$, of the excursion set
encompassed by the isosurface (which in 3D is just one half of the Euler charasteristic of the isosurface itself,
and is equal to minus the genus for the definitions used in cosmology, 
see detailed discussions for such conventions  in \cite{Gay12}).  Being the alternating sum of Betti numbers, $\chi_\mathrm{ 3D}$
is related via  Morse theory \citep[e.g.,][]{Jost08} to the alternating sum of the number of critical points
in the excursion volume. For a random field \citep{Doro70,Adler81,BBKS} \footnote{Note that this expression comes from
 $\delta_\mathrm{ D}(\nabla x)=\sum_{x_{0}|\nabla x_{0}=0}\delta_\mathrm{ D}(x-x_{0})/|\det\! \left({\nabla_{i}\nabla_{j}  x}\right)|$
 and the absolute value of the Hessian can be dropped because we are interested in the {\sl alternating} sum of critical points.
}
\begin{equation}
\label{eq:3DEuler}
\chi_{3\mathrm{ D}}(\nu)\!\!=
- \left\langle \det\! \left({\nabla_{i}\nabla_{j}  x}\right)\delta_\mathrm{ D}(\nabla x) \Theta(x-\nu) \right\rangle 
\;, \quad  i,j\in\{1,2,3\}\,. 
\end{equation}
The Euler characteristic of the excursion sets of a 2D field (in particular,  2D slices of a 3D random field) is given
by a similar expression \citep{Adler81,Bond87,Coles88,Melott89,Gott90}
\begin{equation}
\label{euler2D}
\chi_{2\mathrm{ D}}(\nu)\!\!=
\left\langle \det\! \left({\nabla_{i}\nabla_{j}  x}\right)\delta_\mathrm{ D}(\nabla x) \Theta(x-\nu) \right\rangle 
\;, \quad  i,j\in\{1,2\}\,.
\end{equation}
For an anisotropic 3D field along the third direction, $\chi_\mathrm{ 2D}$ depends on the angle $\theta_{\cal S}$ between this
direction and the plane (${\cal S}$) under consideration \citep{Matsubara96}.
equation~(\ref{euler2D}) is then understood as
\begin{equation}
\label{eq:2DEuler}
\chi_{2\mathrm{ D}}(\nu,\theta_{\cal S})=\left\langle
\det\! \left({\tilde\nabla_{i}\tilde\nabla_{j}  x}\right)\delta_\mathrm{ D}(\tilde\nabla x)\Theta(x-\nu)\right\rangle \;, \quad i,j\in\{1,2\},
\end{equation}
where $\tilde\nabla$ is the gradient on the plane i.e. $\tilde\nabla_{1}=\nabla_{1}$ and $\tilde\nabla_{2}=\sin \theta_{\cal S}\nabla_{2}+ \cos \theta_{\cal S}\nabla_{3}$.
The freedom associated with the choice of plane will be further discussed in Section~\ref{sec:cosmo}.

With the same formalism, it is easy to compute the critical points counts \citep{Adler81,BBKS}.
Equation~(\ref{eq:3DEuler}) leads to a cumulative counting above a given threshold
for maxima, two type (filamentary and wall-like) saddle points and minima
\begin{eqnarray}
n_\mathrm{ max, 3\mathrm{ D}}(\nu)&=&-\left\langle \det\! \left({\nabla_{i}\nabla_{j}  x}\right) \delta_\mathrm{ D}(\nabla x) \Theta(-\lambda_1) 
\Theta(x-\nu) \right\rangle, \label{eq:max}
 \\
n_\mathrm{ sadf, 3\mathrm{ D}}(\nu)&=&+\left\langle \det\! \left({\nabla_{i}\nabla_{j}  x}\right) \delta_\mathrm{ D}(\nabla x) \Theta(\lambda_1) \Theta(-\lambda_2)\Theta(x-\nu)\right\rangle, \\
n_\mathrm{ sadw, 3\mathrm{ D}}(\nu)&=&-\left\langle\det\! \left({\nabla_{i}\nabla_{j}  x}\right) \delta_\mathrm{ D}(\nabla x) \Theta(\lambda_2) \Theta(-\lambda_3)\Theta(x-\nu)\right\rangle, \\
n_\mathrm{ min, 3\mathrm{ D}}(\nu)&=&+\left\langle  \det\! \left({\nabla_{i}\nabla_{j}  x}\right) \delta_\mathrm{ D}(\nabla x) \Theta(\lambda_3)\Theta(x-\nu)\right\rangle\,,
\label{eq:min}
\end{eqnarray}
where averaging conditions are set by the signs of sorted eigenvalues $\lambda_1 \ge \lambda_2 \ge \lambda_3$ of the Hessian matrix of the field. 
Taking alternating sum eliminates the constraints on  signs of eigenvalue,  leading to the  $\chi_\mathrm{ 3D}$ statistics. Similar expressions as equations~(\ref{eq:max})-(\ref{eq:min}) apply for 
2D  extrema.

Extrema counts provide us with information on peaks (dense regions), minima (under-dense regions), and saddle points.
In some applications there is symmetry between extrema (e.g. in CMB studies  minima and maxima of the temperature field
are equivalent); in others, they describe very different structures, e.g. in LSS  dense peaks correspond to 
graviationally collapsing objects like galactic or cluster haloes while minima seed the regions devoid of structures.
Saddle-type extrema are also interesting in their own right,  being related to the underlying filamentary structures
(bridges connecting peaks through saddles), which in turn can also be characterized by the skeleton
\citep{Novikov2006,Gay12} of the Cosmic Web.
A particular advantage of the described geometrical statistical estimates is that they are invariant under monotonic
transformation of the underlying field $x \to f(x)$, provided one maps the threshold correspondingly $\nu \to f(\nu)$.
For cosmological data this means that these statistics are formally invariant with respect to any monotonic local bias between
the galaxy and matter distributions. We demonstrate this formally, order by order, in section~\ref{sec:invariance}.
\begin{figure}
 \begin{center}
 \includegraphics[width=0.4\textwidth]{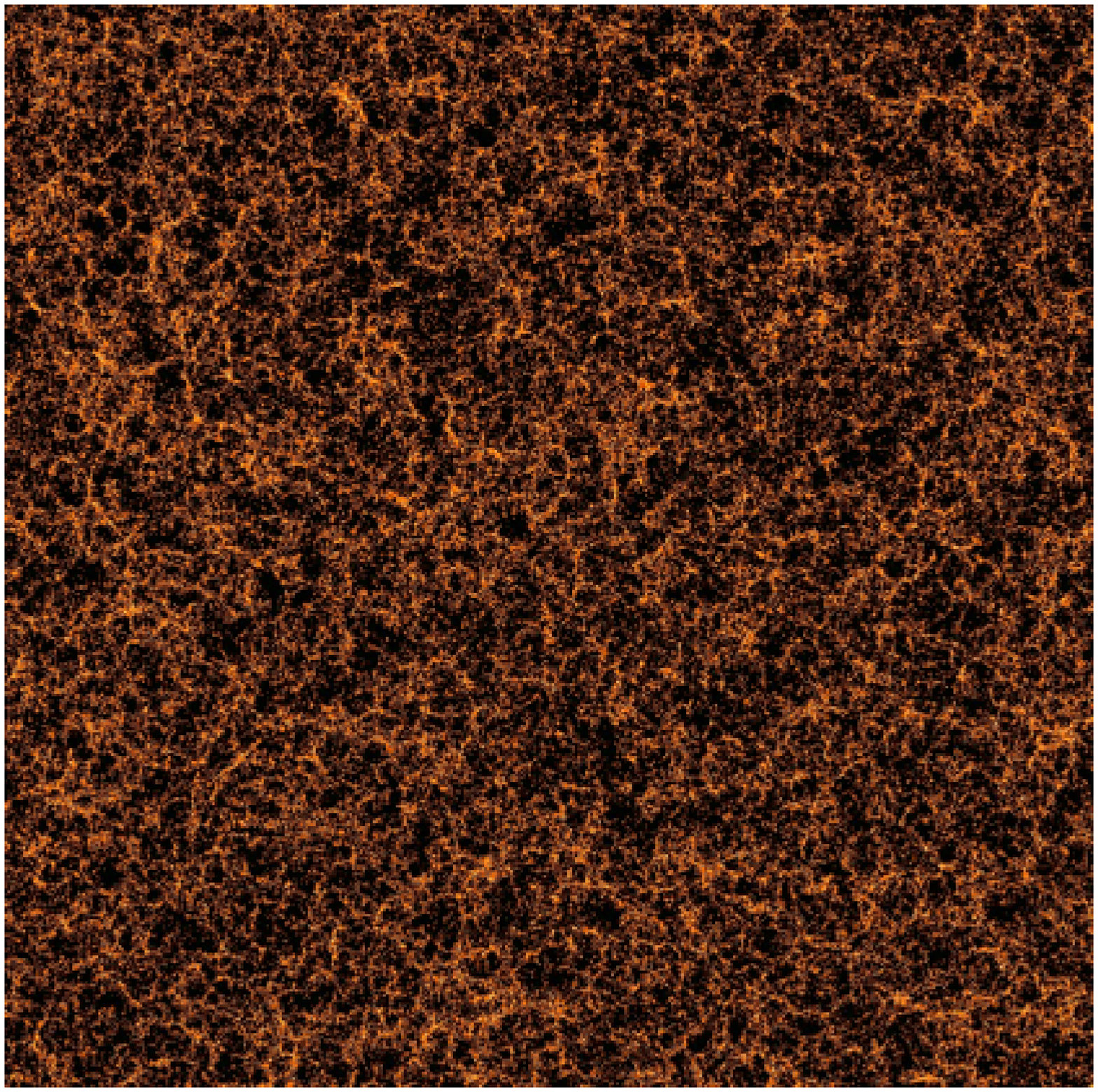}
 \includegraphics[width=0.4\textwidth]{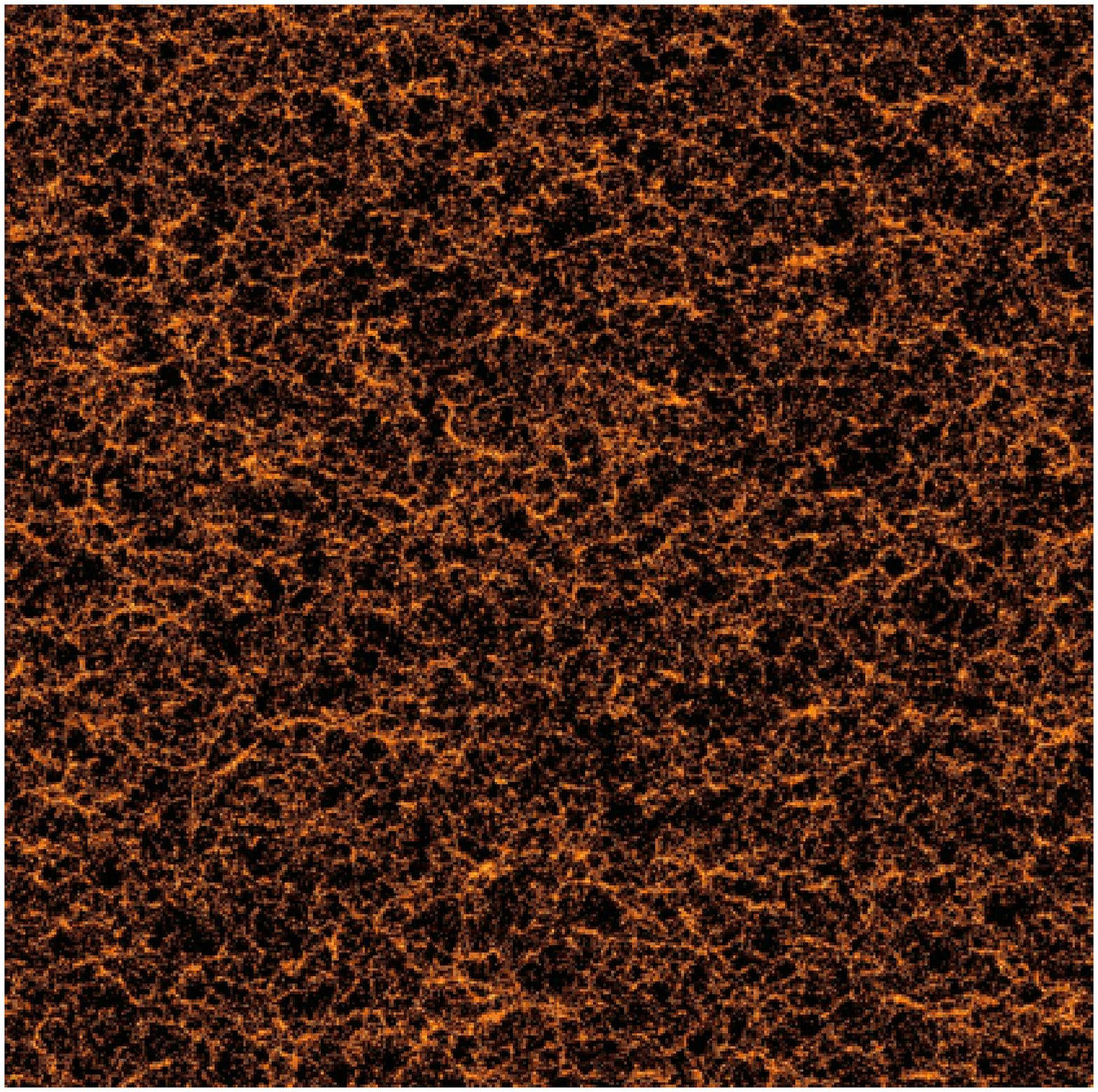}
 \caption{A slice through the  {\sc horizon} $4\pi$ halo catalog  at redshift zero without ({\sl left panel}) and with ({\sl right panel}) redshift distortion (along the ordinate).  The boxsize is 2~Gpc$/h$, the slice thickness 40 Mpc$/h$.  The area is 20$\times$20 larger than the slice presented in Figure~\ref{fig:field}. Each dot represents a halo colour-coded by its mass.
 Note the clear preferred horizontal elongation of structures in redshift space.
\label{fig:4pifield} 
}
 \end{center}
\end{figure}
%
\section{The joint PDF: rotational invariance and Gram-Charlier expansion}
\label{sec:JPDF}
\label{sec:formalism}
Evaluating  the expectations in equations~(\ref{eq:filling_factor}-\ref{eq:min}) requires a model for the joint probability distribution function (JPDF)
 of the field and its derivative up to  second  order, $P(x, \nabla_i x, \nabla_i \nabla_j x)$.
Let us now proceed to developing this JPDF  for a mildly non-Gaussian and anisotropic field such
as the cosmological density field in redshift space, starting with a formal definition of redshift space.
\subsection{Statistically anisotropic density field in redshift space}
\label{sec:redshift_space}
In an astrophysical context,  we focus on the statistics of isodensity contours of matter  in the redshifted Universe. The
estimation of position via  redshift  assigns to a given object   
the ``redshift'' coordinate, ${\bf s}$,
\begin{equation}
 {\bf s} = {\bf r} + H^{-1} {\bf v} \cdot {\bf \hat r}
\label{eq:redshiftmap}
\end{equation}
shifted from the true position ${\bf r}$ by the projection of the peculiar velocity 
${\bf v}$ along the  line-of-sight direction   ${\bf \hat r}$. 

On large scales, in the linear regime of density evolution, the mapping to redshift coordinates induces
an anisotropic change in mass density contrast \citep{Kaiser87}, best given in Fourier space 
\begin{equation}
\hat \delta^{(s)}(\mathbf{k})=(1+f\mu^{2}) \hat \delta^{(r)}(\mathbf{k}), 
\label{eq:Kaiser_mass}
\end{equation}
that has dependency on the angle $\mu={\bf k} \cdot {\bf \hat r}/k$
between the direction of the wave ${\bf k}$ and the line of sight, and the amplitude, $f$, tracing
the growth history of linear inhomogeneities $D(a)$, 
$f= \dd \log D/\dd \log a \approx \Omega_{m}^{0.55}$, \citep{Peebles1980}.  The main qualitative effect of this distortion
is  the enhancement of clustering via the squeezing overdense regions and the stretching underdense voids along the line of sight.
If  matter is traced by biased halos (e.g. galaxies), the redshift space distortion 
$\delta_g$ can be modeled  by a linear bias $b_1$ factor \citep{Kaiser84}
\begin{equation}
\label{eq:Kaiser}
\hat \delta_{g}^{(s)}(\mathbf{k})=(1+\beta \mu^{2})\hat\delta_{g}^{( r )}(\mathbf{k})\,, \quad \beta \equiv f/b_1\,.
\end{equation}
Depending on  context, we shall use either $f$ or $\beta$ to parametrize the linearized redshift distortions, and either equation~(\ref{eq:Kaiser_mass}) or (\ref{eq:Kaiser}) 
to generate maps.

In the mildly non-linear regime,  the focus of this investigation,  redshift distortions 
interplay with  non-Gaussian corrections that develop with the growth of non-linearities. 
 \cite{Scoccimarro99,Bernardeau} established the framework for a perturbative approach to this regime,
which we built upon in Section~\ref{sec:PT}.
In redshift space, the density field
is statistically anisotropic, with  line-of-sight expectations
differing  from  expectations in the perpendicular  directions (in the plane of the sky).
We shall consider the Minkowski functionals and extrema statistics
in the  plane-parallel approximation, where the LOS direction is identified with
the Cartesian third, ``$z$'', coordinate.
As the first step, we extend the non-Gaussian formalism introduced in \cite{PGP2009,Gay12} to partly
anisotropic, axisymmetric fields, establishing several formal results for all orders in non-Gaussianity.
\subsection{3D formalism}
Following \cite{PGP2009}, 
 we choose a non-Gaussian Gram-Charlier expansion of the JPDF 
\citep{Cramer46,Kendall58,Chambers,Jusz95,Amendola96,Blinnikov}
(for  a first application to CMB see, e.g., \cite{Scara91}) using polynomial variables that are invariant with respect to the statistical
symmetries of the field.
In the presence of anisotropy in the direction along the LOS, all statistical measures should be independent  with respect to  sky rotations in the plane
 perpendicular to the LOS. 

Let us denote the field variable as $x$ for the density contrast and $x_{i}$, $x_{ij}$ for its first and second derivatives.
These field variables  are separated into the following groups based on their behaviour under sky rotation:
$x$ and $x_{33}$ are scalars, $x_3$ is a pseudoscalar, $(x_1,x_2)$ and 
$(x_{13}, x_{23})$ are two vectors and $(x_{11},x_{22},x_{12})$ is a symmetric
$2 \times 2$ tensor.  We can construct eight  2D sky rotation invariant
polynomial quantities: four linear ones
$x$, $x_{33}$, $x_3 $  and $J_{1\perp} \equiv x_{11} +
x_{22}$; three quadratic: $q^2_{\perp} \equiv x_1^2 + x_2^2$, 
$Q^2 \equiv x_{13}^2 + x_{23}^2$  and $J_{2\perp} \equiv 
(x_{11} - x_{22})^2 + 4 x_{12}^2 $; and one cubic,
$\Upsilon \equiv \left( x_{13}^2 - x_{23}^2 \right) \left(x_{11} - x_{22} \right) + 4 x_{12}x_{13} x_{23}$, which is directly
related to  the 2D polar angle $\psi$ between $(x_{13}, x_{23})$
vector and the eigendirection of the $(x_{11},x_{22},x_{12})$ matrix
\footnote{There is no quadratic
combination that would represent this angle via a scalar product of two vectors. The reason is that
any ``vector'' build linearly from  $x_{ij}$, such as the ``Q,U'' one
$(x_{11}-x_{22},2 x_{12})$, rotates with 
twice the rotation angle when the real vectors, e.g., $(x_{13},x_{23})$ rotate 
 normally. The combination $\Upsilon$ can be seen to be a scalar product of
``vectors'' $(x_{11}-x_{22},2 x_{12})$ and $(x_{13}^2-x_{23}^2,2 x_{13}x_{23})$.
}
, $\Upsilon = Q^2 \sqrt{J_{2\perp}} \cos 2 \psi$.

To build the Gram-Charlier expansions, we start with the Gaussian limit to the one-point
JPDF of all the invariant variables, which then
serves as a kernel for defining orthogonal polynomials  into which the deviations
from Gaussianity are expanded (see \cite{Gay12} for further
details).  The Gaussian limit is determined as the
limit in which distribution of field variables $x,x_{i},x_{ij}$ is approximated
by Gaussian JPDF.  The field variables are defined to have zero mean. Their
covariance matrix in anisotropic case contains the variances, defined as
\begin{eqnarray}
&&\left\langle x^2 \right\rangle \equiv \sigma^2, ~~
\left\langle x_1^2 \right\rangle = \left\langle x_2^2 \right\rangle \equiv
\frac{1}{2} \sigma_{1\perp}^2, ~~
\left\langle x_3^2 \right\rangle \equiv \sigma_{1 \parallel}^2 ,
  \nonumber \\&& 
\left\langle x_{11}^2 \right\rangle = \left\langle x_{22}^2 \right\rangle
\equiv \frac{3}{8} \sigma_{2\perp}^2 , ~~
\left\langle x_{12}^2 \right\rangle \equiv \frac{1}{8} \sigma_{2 \perp}^2 , ~~
\left\langle x_{33}^2 \right\rangle \equiv \sigma_{2 \parallel}^2, ~~
 \left\langle x_{13}^2 \right \rangle = \left\langle x_{23}^2 \right \rangle
\equiv
\frac{1}{2} \sigma_Q^2 ,
\end{eqnarray}
and the  cross-correlations, amongst which the non-zero are
\begin{equation}
\langle x x_{11} \rangle = \langle x x_{22} \rangle \equiv
-\frac{1}{2} \sigma_{1\perp}^{2}, ~~
\langle x x_{33} \rangle \equiv -\sigma_{1\parallel}^{2},~~
\langle x_{11} x_{22} \rangle\equiv \frac 1 8  \sigma_{2\perp}^2, ~~
\langle x_{11} x_{33} \rangle =  \langle x_{22} x_{33} \rangle =
\frac{1}{2} \sigma_Q^2.
\end{equation}
These properties translate into the
following lowest moments for 2D rotation invariant quantities 
\begin{equation}
 \langle x \rangle  = \langle J_{1\perp} \rangle = \langle x_{33} \rangle =
\langle \Upsilon \rangle = 0, ~~\left\langle x^2 \right\rangle = \sigma^2, ~~
\langle J_{1\perp}^2 \rangle = \sigma_{2\perp}^2, ~~ \left\langle x_{33}^2
\right\rangle = \sigma_{2 \parallel}^2, ~~
\langle q^2_{\perp} \rangle =
\sigma_{1\perp}^2, ~~ \left\langle Q^2 \right\rangle = \sigma_Q^2, ~~ \left
\langle J_{2\perp} \right\rangle = \sigma_{2\perp}^2.
\end{equation}
The cross-correlations of invariants are limited to the  $x,J_{1\perp},x_{33}$
subset,
$\left\langle x J_{1\perp} \right\rangle = -\sigma_{1\perp}^{2}$, $\left\langle
x x_{33} \right\rangle =-\sigma_{1\parallel}^{2}$, $\left\langle J_{1\perp}
x_{33} \right\rangle = \sigma_Q^2 $ and the coupling of $\Upsilon$ with
$Q^2$ and $J_{2\perp}$. The scalar part of the gradient $x_3$ remains
uncorrelated with the rest of the variables due to its pseudo-scalar nature
(changing sign when $z$ flips). 
From now on, we shall consider field variables to be normalized by
the corresponding  $\sigma$, $\sigma_{1\perp}$, $\sigma_{1\parallel}$,
$\sigma_{2\perp}$, $\sigma_{2\parallel}$ and $\sigma_Q$. The
linear invariant combinations
are normalized by their standard deviations and the quadratic ones by
their mean values. Correlations are then described by the dimensionless
coefficients
$\gamma_\perp \equiv \sigma_{1\perp}^2 /(\sigma \sigma_{2\perp})$,
$\gamma_\parallel \equiv \sigma_{1\parallel}^2 /(\sigma \sigma_{2\parallel})$
and $\gamma_2 \equiv \sigma_{Q}^2 /(\sigma_{{2}\perp} \sigma_{{ 2}\parallel})$,
 corresponding to generalized shape parameters \citep{Bernardeau}.
Note that in the isotropic limit, 
\begin{equation}
\label{isosigmas}
 \sigma^2_{1\perp} = 2/3 \sigma^2_1 , ~~~
                   \sigma^2_{1\parallel}      = 1/3 \sigma^2_1, ~~~
                    \sigma^2_{2\perp} =8/15 \sigma^2_2, ~~~
                    \sigma^2_{2\parallel}       = 1/5 \sigma^2_2, ~~~
                    \sigma^2_Q        = 2/15 \sigma^2_2, ~~~
 \gamma_\perp    =  \sqrt{5/6} \gamma, ~~~
                    \gamma_\parallel   = \sqrt{5/9} \gamma, ~~~
                    \gamma_2      = 1/\sqrt{6},
\end{equation}
where $\sigma_1^2 \equiv \left\langle x_1^2 + x_2^2 + x_3^2 \right\rangle =
\langle \nabla x \cdot \nabla x \rangle$, $\sigma_2^2 = \left\langle (x_{11} +
x_{22}+x_{33})^2 \right\rangle = \langle ( \Delta x )^2 \rangle $ and $\gamma
\equiv \sigma_1^2/(\sigma \sigma_2)$.

Let us now also introduce a  decorrelated set of
invariant
variables $(x, \xi, \zeta)$  with the following combinations of  $x,J_{1\perp},x_{33}$:
\begin{equation}
x_{33} = \xi \sqrt{\frac{1 - \gamma_\perp^2 - \gamma_{\parallel}^2 - \gamma_2^2
+ 2 \gamma_\perp \gamma_{\parallel}\gamma_2}
{ 1 - \gamma_\perp^2}}
+ \frac{\gamma_2 - \gamma_{\parallel} \gamma_{\perp}}{\sqrt{1-\gamma_{\perp}^2}}
\zeta - \gamma_{\parallel} x\;, \quad
J_{1\perp}= \zeta  \sqrt{1-\gamma_{\perp}^2} -\gamma_{\perp} x.
\end{equation}
The resulting Gaussian distribution $G=G(x,q^2_{\perp},x_3,\zeta, J_{2\perp},
\xi, Q^2, \Upsilon)$ then simply reads in terms of these variables
\begin{equation}
G\,\dd x\, \dd q_{\perp}^2 \dd x_3\,
 \dd \zeta\, \dd J_{2\perp} \dd \xi\, \dd Q^2 \dd \Upsilon  =\frac{1}{4\pi^3} e^{-\frac{1}{2} x^2  - q_{\perp}^2  - \frac{1}{2} x_3^2 - \frac{1}{2} \zeta^2 
- J_{2\perp} - \frac{1}{2} \xi^2 - Q^2 } \dd x\, \dd q_{\perp}^2 \dd x_3\, \dd \zeta\, \dd J_{2\perp} \dd \xi\,
 \dd Q^2  \frac{d\Upsilon}{\sqrt{Q^4 J_{2\perp} - \Upsilon^2}}\,,
\label{eq:3DPDF_Gauss}
\end{equation}
where $x,x_3,\xi,\zeta$ vary in the range $\left]-\infty,\infty\right[$ ,  $q_\perp^2,Q^2,J_{2\perp}$ span positive values
$[0,\infty[$ ,  and $\Upsilon$ is limited to $\left[-Q^2\sqrt{J_{2\perp}},Q^2\sqrt{J_{2\perp}}\right]$.

The Gram-Charlier polynomial expansion for a non-Gaussian JPDF is then obtained by using polynomials that are orthogonal with respect to the 
kernel provided by equation~(\ref{eq:3DPDF_Gauss}). The remaining coupling between   $\Upsilon$ and the $(Q^2, J_{2\perp})$ variables
in equation~(\ref{eq:3DPDF_Gauss}) introduces a 
technical complexity in building an explicit set of such polynomials. Fortunately, for most of the geometrical statistics
considered in this paper, the $\Upsilon$-dependence is trivial and we can limit ourselves to PDF's marginalized
over $\Upsilon$. After $\Upsilon$ marginalization all the remaining variables
are uncorrelated in the Gaussian limit, and 
the non-Gaussian JPDF, $P(x,q^2_{\perp},x_3,\zeta, J_{2\perp}, \xi, Q^2)
$, can be expanded in a series of direct products of the familiar Hermite (for which we use the
`probabilists' convention)
and Laguerre polynomials
\begin{equation}
P= 
G\! \left[  1 +\!   \sum_{n=3}^\infty \sum_{\sigma_{n}} 
\!\frac{(-1)^{j+l+r}}{i!\;j!\; k!\; l!\;m!\;r!\;p!} 
\!\left\langle x^i {q^{2j}_{\perp}} {\zeta}^k J^{l}_{2\perp} x_3^m Q^{2r} \xi^p \right\rangle_{\!\mathrm{GC}}\!\!\!
H_i\!\left(x\right)\! L_j\!\left(q_{\perp}^2\right)\!
H_k\!\left(\zeta\right) \!L_l\!\left(J_{2\perp}\right)\!
H_m\!\left(x_3\right)\! L_r\!\left(Q^2\right)\! H_p\!\left(\xi\right)
\right], 
\label{eq:defP}
\end{equation}
where $\sum_{\sigma_n}$ is the sum over all combinations of indices 
$\sigma_{n}=\{(i,j,k,l,m,p,r)\in \mathbb{N}^{7}|i+2 j+k+2 l + m + 2r + p =n\}$ such that powers of the field add up to $n$,
and $G$ is given by equation~(\ref{eq:3DPDF_Gauss}) after integration over $\Upsilon$. The  terms within the 
expansion~({\ref{eq:defP}}) are sorted in the order of the power in the field variable $n$.
The Gram-Charlier coefficients are defined by
\begin{equation}
\label{eq:GC}
\left\langle x^i {q^{2j}_{\perp}} {\zeta}^k J^{l}_{2\perp} x_3^m Q^{2r} \xi^p \right\rangle_{\!\mathrm{GC}}\equiv(-1)^{j+l+r}j!\,l!\,r! \left\langle H_{i}(x) L_{j}({q^{2}_{\perp}})H_{k}( {\zeta})L_{l}( J_{2\perp})H_{m}( x_3)L_{r}( Q^{2}) H_{p}(\xi) \right\rangle\,,
\end{equation}
normalized so that $\left\langle x^i {q^{2j}_{\perp}} {\zeta}^k J^{l}_{2\perp} x_3^m Q^{2r} \xi^p \right\rangle_{\!\mathrm{GC}}=\left\langle x^i {q^{2j}_{\perp}} {\zeta}^k J^{l}_{2\perp} x_3^m Q^{2r} \xi^p \right\rangle+
\textrm{products of lower order moments}$.
The advantage of using strictly polynomial variables is that all the moments that appear in the Gram-Charlier coefficients
can be readily related to the moments of the underlying field, and can be obtained if the theory of the latter
(for example perturbation theory of gravitational instability) is known.  As shown in \cite{Gay12}, at the lowest
non-Gaussian ($n=3$) order 
the Gram-Charlier coefficients are just equal to the moments of the corresponding variables,
while in the next two orders they coincide with their cumulants, defined as ``field cumulants''\footnote{``field cumulant''
means the cumulant computed after expressing the non-linear variable through the field quantities, e.g.,  
$\langle q_\perp^4 \rangle_{{\rm field}\; c} \equiv \langle (x_1^2+x_2^2)^2 \rangle_c$. We drop the prefix ``field'', always assuming
``field'' cumulants.}
if the variable is non-linear \citep[for details, see][]{Gay12}.

Expression~(\ref{eq:defP}) is somewhat simplified under the condition of  zero gradient, arising, e.g. when investigating the Euler characteristic and extrema densities,
\begin{equation}
P_{\textrm{ext}}(x,\zeta, J_{2\perp}, \xi, Q^2) \!
=\!  G_\mathrm{ ext}\! \!\left[ 1 +  
 \sum_{n=3}^\infty \sum_{\sigma_{n}}
\!\frac{(-1)^{j+l+r+m}2^{-m}}{i!\;j!\; k!\; l!\;m!\;r!\;p!} 
\!\left\langle x^i q^{2j}_{\perp} {\zeta}^k J^{l}_{2\perp} x_3^{2m} Q^{2r} \xi^p \right\rangle_{\!\mathrm{GC}}\!\!\!
H_i\!\left(x\right) \!
H_k\!\left(\zeta\right)\! L_l\!\left(J_{2\perp}\right)\!
 L_r\!\left(Q^2\right) \!H_p\!\left(\xi\right)\!
\right], \label{eq:PDF3Dext}
\end{equation}
since $L_{j}(0)=1$, $H_{2m}(0)=(-1)^{m}(2m)!/(2^{m} m!)$ and $H_{2m+1}(0)=0$ ;
here $G_\mathrm{ ext}=\frac{1}{\pi} G(x,0,0,\zeta,J_{2\perp},\xi,Q^{2})$, where
the $1/\pi$ comes from the use of polar $(q^{2}_{\perp})$ vs cartesian coordinates $(x_{1},x_{2})$. Note that
$m$ is then replaced by $2m$ in the power count in $\sigma_n$.
\subsection{Theory on 2D planes}
One of our purposes  is to study Minkowski functionals on 2D planar sections of 3D fields. 
Let us therefore  introduce the anisotropic 2D (on the plane) formalism. In a 2D planar slice through anisotropic 
space no residual symmetries are left, so we may use the field variables
directly. The emphasis then is on relating
the field properties on the plane to the three dimensional ones.
Denoting the basis vectors of 3D space as $\mathbf{u}_{1}$, $\mathbf{u}_{2}$,
and $\mathbf{u}_{3}$, with $\mathbf{u}_{3}$ directed along the LOS, let us
introduce the coordinate system on the 2D plane using the pair of basis
vectors such that $\mathbf{s}_{1}=\mathbf{u}_{1}$ is perpendicular to the LOS
and $\mathbf{s}_{2}=\cos \theta_{\cal S}\mathbf{u}_{3}+\sin \theta_{\cal
S}\mathbf{u}_{2}$, where $\theta_{\cal S}$ is the angle between the LOS and that
plane. 
We label the field variables on the plane with tilde, using the
set $({\tilde x},{\tilde x_{1}},\tilde x_{2},{\tilde x_{11}},{\tilde
x_{22}},{\tilde x_{12}}$),
where the relation to 3D spatial derivatives is established by
$\partial_{s_{1}}=\partial_{1}$ and $\partial_{s_{2}}=\cos \theta_{\cal
S}\partial_{3}+\sin \theta_{\cal S}\partial_{2}$.
Hereafter, the first direction corresponds to $s_{1}$, the second direction to
$s_{2}$. Note that ${\tilde x}$, ${\tilde x_{1}}$ and ${\tilde x_{11}}$ coincide
with their 3D counterparts.
We use variables rescaled by their respective variance
denoted, using the same notation as in the 3D case, as $({\tilde \sigma},{\tilde
\sigma_{1\perp}},{\tilde \sigma_{1\parallel}},{\tilde \sigma_{2\perp}},{\tilde
\sigma_{2\parallel}},{\tilde \sigma_{Q}})$. These variances involve the plane
orientation in the following way
${\tilde \sigma}=\sigma$,
${\tilde \sigma_{1\perp}}^{2}=\left\langle  x_{1}^{2}\right\rangle  =\frac 1 2 \sigma_{1\perp}^{2}$,
${\tilde \sigma_{1\parallel}}^{2}=\left\langle  {\tilde x_{2}}^{2}\right\rangle  =\cos^{2} \!\theta_{\cal S}\sigma_{1\parallel}^{2}+\frac 1 2\sin^{2} \!\theta_{\cal S}\sigma_{1\perp}^{2}$,
${\tilde \sigma_{2\perp}}^{2}=\left\langle  x_{11}^{2}\right\rangle  =\frac 3 8 \sigma_{2\perp}^{2}$,
${\tilde \sigma_{2\parallel}}^{2}=\left\langle  {\tilde x_{22}}^{2}\right\rangle   =
\cos^{4}\!\theta_{\cal S}\sigma_{2\parallel}^{2}+\frac 3 8\sin^{4}\!\theta_{\cal S}\sigma_{2\perp}^{2}+3\cos^{2}\!\theta_{\cal S}\sin^{2}\!\theta_{\cal S}\sigma_{Q}^{2}$ and 
${\tilde \sigma_{Q}}^{2}=\left\langle  {\tilde x_{12}}^{2}\right\rangle  =\frac 1 2 \cos^{2}\!\theta_{\cal S}\sigma_{Q}^{2}+\frac 1 8 \sin^{2}\!\theta_{\cal S}\sigma_{2\perp}^{2}$\,.
As previously, in order to diagonalize the JPDF we introduce $\tilde \xi$ and
$\tilde\zeta$ such that
\begin{equation}
{\tilde x_{22}}=\sqrt{\frac{1-{\tilde \gamma_{\perp}}^{2}-{\tilde \gamma_{\parallel}}^{2}-{\tilde \gamma_{2}}^{2}+2{\tilde \gamma_{\perp}}{\tilde \gamma_{\parallel}}{\tilde \gamma_{2}}}{1-{\tilde \gamma_{\perp}}^{2}}}{\tilde \xi}
-\frac{{\tilde \gamma_{\parallel}}{\tilde \gamma_{\perp}}-{\tilde \gamma_{2}}}{\sqrt{1-{\tilde \gamma_{\perp}}^{2}}}{\tilde \zeta}
-{\tilde \gamma_{\parallel}}x \;,\quad
{\tilde x_{11}}=\sqrt{1-{\tilde \gamma_{\perp}}^{2}}{\tilde \zeta} -{\tilde \gamma_{\perp}}x\,,
\end{equation}
 where ${\tilde \gamma_{\perp}}={\tilde \sigma_{1\perp}}^{2}/({\tilde \sigma}{\tilde \sigma_{2\perp}})$,
 ${\tilde \gamma_{\parallel}}={\tilde \sigma_{1\parallel}}^{2}/({\tilde \sigma}{\tilde \sigma_{2\parallel}})$ 
 and ${\tilde \gamma_{2}}={\tilde \sigma_{Q}}^{2}/({\tilde \sigma_{2\parallel}}{\tilde \sigma_{2\perp}})$.
In terms of these variables, the resulting Gaussian distribution is simply
\begin{equation}
 G_{2\textrm{D}}(\tilde x,\tilde x_{1},\tilde x_{2},\tilde\zeta, \tilde\xi,\tilde x_{12}) = \frac{1}{8\pi^{3}} \exp \left[{-\frac{1}{2} \tilde x^2  -  \frac 12 \tilde x_{1}^2-\frac 1 2 \tilde x_{2}^{2}- \frac{1}{2} \tilde \zeta^2 
- \frac 1 2\tilde \xi^{2}-\frac 1 2 \tilde x_{12}^{2} } \right]\,,
\end{equation}
 and  the fully Non-Gaussian JPDF can be written using a Gram-Charlier expansion in  Hermite polynomials only
\begin{equation}
P_{2\textrm{D}}(\tilde x,\tilde x_{1},\tilde x_{2},\tilde\zeta,\tilde \xi,\tilde x_{12})\! = \!G_{2\textrm{D}}\! \left[ 1\!+\!\sum_{n=3}^\infty \sum_{\sigma_{n}}
\frac{1}{i!\; j!\;k!\; l!\;m!\;p!} 
\left\langle \tilde x^i \tilde x_{1}^{j} \tilde x_{2}^{k}{\tilde \zeta}^l \tilde\xi^{m}\tilde x_{12}^{p} \right\rangle_{\!\mathrm{GC}}\!\!\!
H_i\left(\tilde x\right) H_{j}( \tilde x_{1})H_{k}(\tilde x_{2})
H_l(\tilde\zeta) H_m(\tilde\xi)H_{p}(\tilde x_{12})
\right]\!, \label{eq:PDF2D}
\end{equation}
where $\sigma_{n}=\{(i,j,k,l,m,p)\in \mathbb{N}^{6}|i+j+k+l+m+p=n\}$ and the Gram-Charlier coefficients are given by $\left\langle \tilde x^i \tilde x_{1}^{j} \tilde x_{2}^{k}{\tilde \zeta}^l \tilde\xi^{m}\tilde x_{12}^{p} \right\rangle_{\!\mathrm{GC}}=\left\langle H_{i}( \tilde x)H_{j}( \tilde x_{1}) H_{k}(\tilde x_{2})H_{l}({\tilde \zeta}) H_{m}(\tilde\xi)H_{p}(\tilde x_{12}) \right\rangle$.

Equations (\ref{eq:PDF3Dext}) and~(\ref{eq:PDF2D}) fully characterize the one-point statistics of possibly
anisotropic weakly non-Gaussian field in 2 and 3D.  These expansions apply whatever the origin of the anisotropy, and  in particular for redshift-induced anisotropy in a 
cosmic environment.  

\section{Prediction for Minkowski functionals and extrema counts} \label{sec:Min23}
Topological and geometrical measures in  redshift space 
that we are investigating are obtained
by integrating  the suitable quantities over the distributions (\ref{eq:PDF3Dext}) and~(\ref{eq:PDF2D}), in accordance to
equations~(\ref{eq:filling_factor}-\ref{eq:min}). We first collect the results for Minkowski
functionals, for which such integrals can be carried analytically, and then discuss extrema counts, where 
one has to resort to numerical integration.  The Gram-Charlier expansion leads to series representation, e.g.
$\chi_\mathrm{ 3D} = \chi_\mathrm{ 3D}^{(0)} + \chi_\mathrm{ 3D}^{(1)} +  \chi_\mathrm{ 3D}^{(2)} + \ldots$ for which we give here
the zero (Gaussian) and the first (non-Gaussian) order terms.  $ \chi_\mathrm{ 3D}^{(1)}$ correspond to first order terms in 
the variance $\sigma$ in the cosmological perturbation series. The Gaussian terms, e.g. $\chi_\mathrm{ 3D}^{(0)}$, in redshift space have been first investigated in \cite{Matsubara96}, while the first non-Gaussian corrections
are  novel results of this paper. Higher order terms can also be readily obtained within our formalism  (see  Appendix~\ref{sec:chi3Dappendix}).
Most of our results are general for arbitrary weakly non-Gaussian fields with axisymmetric statistical properties.

The presented statistics fall into two families.   One group is the statistics of the 3D field as a whole, namely the
3D Euler characteristic, $\chi_\mathrm{ 3D}(\nu)$,  the area of the isosurfaces in 3D space, ${\cal N}_3(\nu)$,
the volume above a threshold, $f_{V}$, and the  differential count of extrema in 3D.  
The other group consists of measures on lower dimensional cuts through 3D volumes. These correspond to  measures on 
planar 2D cut of the 2D Euler characteristic,
$\chi_\mathrm{ 2D}(\nu)$, and  the length of isocontours on 2D plane, ${\cal N}_2(\nu)$, and statistics of zero crossings,
${\cal N}_1(\nu)$, along pencil-beam 1D lines through the volume, as well as the corresponding differential extrema counts.
The lower-dimensional statistics in an anisotropic space give additional leverage to study anisotropy properties, 
e.g., in the cosmological context, the magnitude of the redshift distortion, through their dependence on the direction
in which the section of the volume is taken. 
Indeed we show below  (equations~(\ref{eq:chi2D}), (\ref{eq:N2_ani_n3})...) that these statistics
follow the following generic form (again using $\chi_\mathrm{ 2D}$ as an example and omitting numerical constant factors) 
\begin{equation}
 \chi_\mathrm{ 2D}(\nu,\theta_{\cal S}) \propto  P(\nu) A(\beta_\sigma,\theta_{\cal S}) 
\left( H_i(\nu) + \sigma_z \sum_{m=\pm1,\pm3} 
\left(\tilde S^{(m),\perp}_{\chi_{\textrm{2D}},z} +\tilde S^{(m),\parallel}_{\chi_{\textrm{2D}},z}
\right) H_{i+m}(\nu) + \ldots \right)\,,
\label{eq:2Dstatform}
\end{equation}
where the overall amplitude ${\cal A}$ depends in an angle-sensitive way on the 
anisotropy parameter 
\begin{equation}
\beta_\sigma \equiv 1 - {\sigma_{1\perp}^2}/{2 \sigma_{1\parallel}^2}\,,\label{eq:defbeta}
\end{equation}
 that measures the difference between the rms values of
the line-of-sight and perpendicular components of the gradient, and with $\theta_{\cal S}$ the angle 
 between the line of sight and the 2D slice under consideration.
 In case of isotropy,  $\beta_\sigma=0$.
In anisotropic situations, $\beta_\sigma$ is positive, spanning the range $0 <  \beta_\sigma \le 1 $,
when the field changes faster in the z-direction 
($\sigma_{1\parallel} > \frac{1}{2} \sigma_{1\perp}$)
as is the case,  for instance, 
in the linear regime of redshift corrections where
 $\beta_\sigma = \frac{4}{5} \beta ({1+3\beta/7})/({1+6\beta/5+3\beta^2/7})$,  with $\beta=f/b_{1}$.
When the line-of-sight variations are smoother than the  perpendicular one ($\sigma_{1\parallel} < \frac{1}{2} \sigma_{1\perp}$),
$\beta_\sigma$ is negative, $ -\infty \le \beta_\sigma < 0 $, as is the case in the 
non-linear ``finger of God'' regime.  

The Gaussian and subsequent contributions have distinct signatures in this Hermite decomposition.
While the Gaussian term is described by an appropriate single Hermite mode $ H_i(\nu)$,
the first order corrections excite the modes that have parity opposite to that of  the leading Gaussian order.
The amplitude of non-Gaussianity is proportional to the variance $\sigma$ of the field and 
combinations of third order cumulants
$\tilde S^{(m)}_{\chi_{\textrm{2D}},z}$ that can be different if measured in the directions parallel, $\tilde S^{(m),\parallel}_{\chi_{\textrm{2D}},z}$ or
perpendicular, $\tilde S^{(m),\perp}_{\chi_{\textrm{2D}},z}$, to the line-of-sight (e.g. $\langle x q_\perp^2 \rangle$ versus $\langle x x_3^2\rangle$).
This suggests the following strategy: a)  determine  the $\beta$ parameter from the amplitude of the 2D or 1D statistics
taken at different angles to the line-of-sight; b)   determine the amplitude of non-Gaussianity, and
correspondingly $\sigma_z$ in redshift space, by comparing the Gaussian and non-Gaussian contributions through
fitting distinct Hermite functions to the measurements; and c) 
 determine the real-space $\sigma$  by combining the results of the two previous steps. This strategy is implemented on a fiducial experiment 
 in Section~\ref{sec:measuringf}.
The 3D statistics have a  representation similar to that of  equation~(\ref{eq:2Dstatform}), but without any control over the angle  parameter.
Therefore they are less sensitive to the effects of anisotropy, but should provide more robust measurements of the non-Gaussian
corrections than any given lower-dimensional subset of data.

\subsection{PDF of the field  and the  filling factor, $f_V$}\label{sec:filling}
The simplest statistic and Minkowski functional  is the filling factor, $f_V$, of the excursion set, i.e
the volume fraction occupied by the region above the threshold $\nu$.
Derived from the PDF of the field alone, \citep{Jusz95,Bernardeau95}
\begin{equation}
P(\nu)=\frac 1 {\sqrt{2\pi}}e^{-\nu^{2}/2}\left[1+\sum_{i=3}^{\infty}\frac{1}{i!}\left\langle x^{i}\right\rangle_{\textrm{GC}}H_{i}(\nu)\right]
=\frac 1 {\sqrt{2\pi}}e^{-\nu^{2}/2}\left[1+
\sigma \frac{S_{3}}{6} H_{3}(\nu)+\sigma^{2}\left(\frac{S_{4}}{24}H_{4}(\nu)
+\frac{S_{3}^{2}}{72}H_{6}(\nu)\right)\cdots\right],
\end{equation}
the functional $f_V(\nu) = \int_\nu^\infty P(x) d x$ is given by
\begin{eqnarray}
f_V(\nu) &=& \frac{1}{2} \mathrm{Erfc}\left(\frac{\nu}{\sqrt{2}}\right) + 
\frac 1 {\sqrt{2\pi}}e^{-\nu^{2}/2}\sum_{i=3}^{\infty}\frac{1}{i!}\left\langle x^{i}\right\rangle_{\textrm{GC}}H_{i-1}(\nu)
\nonumber \\
&=&
\frac{1}{2} \mathrm{Erfc}\left(\frac{\nu}{\sqrt{2}}\right) + 
\frac 1 {\sqrt{2\pi}}e^{-\nu^{2}/2}\left[\sigma \frac{S_{3}}{6}H_{2}(\nu)+\sigma^{2}\left(\frac{S_{4}}{24}H_{3}(\nu)
+\frac{S_{3}^{2}}{72}H_{5}(\nu)\right)\cdots\right],
\end{eqnarray}
where  the terms from the Gram-Charlier expansion are rearranged in 
powers of
$\sigma$ 
(forming the Edgeworth expansion)
using the usual skeweness $S_{3}=\left\langle x^3      \right\rangle/\sigma$, 
kurtosis $S_4 = (\left\langle x^4 \right\rangle_c-3)/ \sigma^2$ and subsequent scaled cumulants of the field $x$.

There are several advantages to use the value of the filling factor $f_V$ instead of $\nu$ as a variable in which
to express all other statistics.  Indeed, the fraction of volume occupied by a data set is readily available
from the data, whereas specifying $\nu$ requires prior knowledge of the variance $\sigma$, a quantity which is typically  the main unknown
for such investigations. 
Following  \cite{Gott87,Gott88,Gott89}, \cite{Seto00} and \cite{Matsubara},  
let us therefore introduce the threshold $\nu_{f}=\sqrt 2\,\textrm{erfc} ^{-1}(2 f_V)$ to be used 
as an observable alternative to $\nu$.
The mapping between $\nu$ and $\nu_{f}$ is monotonic and is implicitly given by the identity
\begin{equation} \label{eq:degauss}
\sqrt{2\pi}\,f_V = \int_{\nu_{f}}^{\infty}\dd y \,e^{-y^{2}/2}=
\int_{\nu}^{\infty} \dd x \,e^{-x^{2}/2}+\sum_{i=3}^{\infty}\frac{1}{i!}\left\langle x^{i}\right\rangle_{\textrm{GC}}H_{i-1}(\nu)e^{-\nu^{2}/2}\,,
\end{equation}
which when inverted implies
\begin{equation}
\nu=\nu_{f}+\sigma \frac {S_{3}}{ 6}H_{2}(\nu_{f})+{\cal O}(\sigma^{2}).
\label{eq:nufromnuf}
\end{equation}
If truncated to the first non-Gaussian order, this relation remains monotonic  for $\nu_f > -3/\left\langle x^3 \right\rangle$. All Minkowski functionals derived below will be
re-expressed in terms of $\nu_f$ in Section~\ref{sec:nuf}.

\subsection{ Euler characteristic in two and three anisotropic dimensions}
\label{sec:genus}
\subsubsection{3D Euler characteristic : $\chi_\mathrm{ 3D}$}
\label{sec:genus3D}
From equation~(\ref{eq:3DEuler}), the 3D Euler characteristic reads
\begin{equation}
\chi_\mathrm{ 3D} =- \frac 1 {\sigma_{1\perp}^{2}\sigma_{1\parallel}} 
\int \dd x\, \dd \zeta\, \dd J_{2\perp} \dd \xi\, \dd Q^2\dd \Upsilon P_{\textrm{ext}}
(x,\zeta,J_{2\perp}\xi,Q^{2},\Upsilon) I_{3}\,,
\label{eq:defchi3D}
\end{equation}
 where $I_3 = \det x_{ij} $.
 In terms of the  variables defined in section~\ref{sec:JPDF}, the Hessian is
\begin{equation}
I_3=\frac{1}{4} x_{33} ( J^{2}_{1\perp} - J_{2\perp} )-\frac 1 2 \gamma_2 \left( Q^2 J_{1\perp} - \Upsilon \right)~.
\label{eq:I3}
\end{equation}
The integration is easily performed using
 the orthogonality properties of Hermite and Laguerre polynomials to give the expression for the 
3D Euler characteristic \textit{to all orders in non-Gaussianity} that can be found in Appendix \ref{sec:chi3Dappendix}.
The result in equation~(\ref{eq:chi3d}) has the form
 \begin{equation}
\chi_\mathrm{ 3D}(\nu)=\frac 1 2\textrm{Erfc}\left(\frac{\nu}{\sqrt{2}}\right)\chi_\mathrm{ 3D}(-\infty)+\chi_\mathrm{ 3D}^{(0)}(\nu)+\chi_\mathrm{ 3D}^{(1)}(\nu)+\cdots\,,
\label{eq:Chi3D}
\end{equation}
where the asymptotic limit $\chi_{\mathrm{3D}}(-\infty)$ is zero in the infinite simply-connected 3D space, but
can reflect the Euler characteristic of the mask if data is available only in sub-regions with a complex mask
\footnote{Here by ``masking'' we understand the procedure that excludes some regions of space from observations
 without modifying the underlying statistical properties of the field.}.

At Gaussian order, the 3D Euler characteristic reads, in accordance with \cite{Matsubara96} 
\begin{equation}
 \chi_\mathrm{ 3D} ^{(0)}(\nu)=\frac {\sigma_{1\parallel}\sigma_{1\perp}^{2}}
{\sigma^{3}}\frac {H_{2}(\nu)}
 {8\pi^{2}} e^{-\nu^{2}/2} \; .
\end{equation}
At first non-Gaussian order, the 3D Euler characteristic is  (using relations between cumulants
 listed in Appendix \ref{sec:momentsrelation})
\begin{equation}
\label{eq:3DNGgenus}
\chi_\mathrm{ 3D}^{(1)}(\nu)
=\frac{\sigma_{1\parallel}\sigma_{1\perp}^{2}}{\sigma^{3}}\frac {e^{-\nu^{2}/2}}{8 \pi^{2}}\! \!\left[  
 \frac{ H_{5}(\nu)}{3!}\left\langle x^3      \right\rangle 
+ H_{3}(\nu)\left\langle x({q^2_{\perp}}+x_{3}^{2}/2)  \right\rangle 
-\frac { H_{1}(\nu)} {\gamma_{\perp}}\left\langle J_{1\perp}\!({q^2_{\perp}}+x_{3}^{2})     \right\rangle 
\right],
 \end{equation}
 where, when we expand our cumulants in terms of the field variables
 \begin{eqnarray}
  \left\langle x({q^2_{\perp}}+x_{3}^{2}/2) \right\rangle &=&  \frac{\sigma^{2}}{\sigma_{1\perp}^{2}} 
\left\langle  x  \left( \vphantom{\nabla_{\parallel}} \nabla_{\!\perp}x\cdot \nabla_{\!\perp}x \right) \right\rangle + 
\frac{\sigma^{2}}{2\sigma_{1\parallel}^{2}} \left\langle x 
\left( \nabla_{\parallel}x\cdot \nabla_{\parallel}x \right) \right\rangle, 
 \\
 \left\langle J_{1\perp}({q^2_{\perp}}+x_{3}^{2}) \right\rangle &=&
 \frac{\sigma^3}{\sigma_{2\perp} \sigma_{1\perp}^{2} }\!\left\langle  \Delta_{\perp} x \;
\left( \vphantom{\nabla_{\parallel}} \nabla_{\!\perp}x\cdot  \nabla_{\!\perp}x \right) \right\rangle +
 \frac{\sigma^3}{\sigma_{2\perp} \sigma_{1\parallel}^{2} }\! 
\left\langle  \Delta_{\perp} x \; \left( \nabla_{\parallel}x\cdot 
\nabla_{\parallel}x \right) \right\rangle
\, .
 \end{eqnarray}
Note that the $n=3$ Gram-Charlier coefficients are
actually equal to the cumulants of field variables,
therefore the brackets without the label $\mathrm{GC}$ are used, meaning
these are standard cumulants.
This correspondence is preserved at the next order as well, but will eventually
be broken for the higher
order Gram-Charlier coefficients. We refer again to \cite{Gay12} for a detailed discussion.

The isotropic limit can be put in a concise form using equation~(\ref{isosigmas}) and  the relationships $\left\langle  \Delta_{\perp} x \; \left( \nabla_{\parallel}x\cdot \nabla_{\parallel}x \right) \right\rangle=\left\langle  \Delta_{\perp} x \; \left( \nabla_{\perp}x\cdot \nabla_{\perp}x \right) \right\rangle= \left\langle \left( \nabla x\cdot \nabla x\right)\Delta x\right\rangle/3$:
\begin{equation}
 \chi_\mathrm{ 3D}^{\textrm{iso}}(\nu)
=\frac{e^{-\nu^{2}/2} }{(2\pi)^{2}}
\frac{\sigma_1^3}{3\sqrt{3}\sigma^3}
\left[H_{2}(\nu)+\frac{1}{3!}H_{5}(\nu)\left\langle  x^3\right\rangle
+\frac{3}{2}\frac{\sigma^{2}}{\sigma_{1}^{2}}H_{3}(\nu)\left\langle x\left( \nabla x\cdot \nabla x\right)\right\rangle
-\frac{9}{4}\frac{\sigma^{4}}{\sigma_{1}^{4}} H_{1}(\nu)  \left\langle 
\left( \nabla x\cdot \nabla x\right)\Delta x\right\rangle
+{\cal
O}(\sigma^{2})\right]\,,
\label{eq:chibar}
\end{equation} 
which is in exact agreement with \cite{Gay12} 
 (see Appendix \ref{sec:momentsrelation}).

These predictions for the 3D Euler characteristics  are now first validated on toy models  for which simulations are straightforward and cumulants simply analytic
 (see Appendix~{\ref{sec:toys}} for details about this $f_\mathrm{{nl}}$ toy model). In this non-dynamical model the redshift correction
is simulated by transforming the density according to the linear equation~(\ref{eq:Kaiser_mass}) with an $f$ factor chosen freely.
Figure~\ref{fig:genusGfnl} and \ref{fig:genusNGfnl} presents a good match between the theoretical prediction of equations~(\ref{eq:3DNGgenus}) and~(\ref{eq:chibar})  for the  3D non-Gaussian anisotropic Euler characteristic to 
simulated fields. It also shows 
 the evolution  as a function of $f_\mathrm{{nl}}$ (i.e as a function of the non-Gaussianity)
 for two values of the anisotropy parameter, $f=0$ (real space) and $f=1$ (redshift space). 

Figure~\ref{fig:minkowski-gravity} summarizes comparison of many results of this section with measurement
on the density fields in dark matter only, $256^3$,
scale-invariant LSS simulations with power index $n=-1$ and $\Omega_m=1$. To reproduce the redshift effects, the
positions of the particles are shifted  along $z$  according to equation~(\ref{eq:redshiftmap}).
For dark matter we have $f = \beta = 1$. 
The lowest right panel in Figure~\ref{fig:minkowski-gravity} compares the prediction 
of equation~(\ref{eq:3DNGgenus}) to the  ${\chi}_\mathrm{ 3D}$ statistics.
The measurements are done at  epochs for which $\sigma=0.18$ in real space.
For these fields, non-Gaussianity comes from gravitational clustering
but also from the mapping into redshift space which is intrinsically non-linear. The theoretical formula uses
the cumulants measured from simulation itself, so we do not test how accurately the cumulants  can be predicted,
e.g., using perturbation theory.
Figure~\ref{fig:minkowski-gravity}  shows that the theoretical prediction at next-to-leading order
(NLO, i.e. at first non-Gaussian order) mimics very well the measurement for intermediate contrasts as expected
(note that we plot the odd part of the signal to suppress the Gaussian term and focus on deviation to Gaussianity only).
To better fit the data at the tails of the distribution, one has to take into account higher-order corrections,
which, as previous real-space studies show, are non-negligible for $\sigma\ge 0.18$. 
%
\subsubsection{2D Euler characteristic : $\chi_\mathrm{2D}$}
 \label{sec:genus2D}
Let us now investigate the Euler characteristic of the field on a 2D planar section, ${\cal S}$,
 of a 3D z-anisotropic space. 
As redshift distortion occurs along the line of sight, this statistics thus depends on
 the angle $\theta_{\cal S}$ between the plane ${\cal S}$ and that line of sight. 
The 2D field restricted to the plane ${\cal S}$  is denoted $\tilde x(s_1,s_2)$, where $s_{1}$ and $s_{2}$
 are two cartesian coordinates on the plane,
chosen so that $s_1$ spans the direction perpendicular to the line of sight.
From equation~(\ref{eq:2DEuler}), the Euler characteristic  on the plane (${\cal S}$) reads
\begin{equation}
\chi_{2\textrm{D}}=\frac{1}{\tilde \sigma_{1\perp}\tilde \sigma_{1\parallel}}\int  \dd \tilde x \dd\tilde x_{1}\dd\tilde x_{2}  \dd\tilde \zeta \dd\tilde \xi \dd\tilde x_{12}\tilde I_{2}\delta_\mathrm{ D}(\tilde x_1)\delta_\mathrm{ D}(\tilde x_2)P(\tilde x,\tilde x_{1},\tilde x_{2},\tilde \zeta, \tilde\xi,\tilde x_{12}), \label{eq:genusformal}
\end{equation}
where  $\tilde I_{2}=\tilde \sigma_{2\perp}\tilde\sigma_{2\parallel} \left(\tilde x_{11}\tilde x_{22}-\tilde\gamma_{2}\tilde x_{12}^{2}\right)$ is the Hessian determinant on the plane.
Rewriting $\tilde I_{2}$ in terms of $\tilde x$, $\tilde\zeta$, $\tilde\xi$ and $\tilde x_{12}$ and using the orthogonality of Hermite polynomials, 
one obtains, after some algebra, an all order expansion
\begin{multline}
\chi_{2\textrm{D}}(\nu)
=\frac {e^{- \nu^{2}/2} } {(2\pi)^{3/2}} \frac{\tilde\sigma_{2\perp}\tilde\sigma_{2\parallel}}{\tilde\sigma_{1\perp}\tilde\sigma_{1\parallel}}\times
\Bigg(\tilde \gamma_{\parallel}\tilde \gamma_{\perp} H_{1}( \nu)
+\tilde \gamma_{\parallel}\tilde \gamma_{\perp} \sum_{n=3}^\infty
\sum_{\sigma_{n}} \!\frac{(-1)^{k+j}}{i!\; j!\;k!\;2^{k}2^{j}}  \left\langle \tilde x^i
\tilde x_{1}^{2j}\tilde
x_{2}^{2k}\right\rangle_{\!\mathrm{GC}}\!\!\left(H_{i+1}\left(\nu\right)+2iH_
{i-1}\left(\nu\right)+i(i-1)H_{i-3}\left(\nu\right)\right)\\
-\sum_{n=3}^\infty\sum_{\sigma_{n-1}}
\!\!\!\frac{(-1)^{k+j}(H_i\!\left(\nu\right)\!+\!iH_{i-2}\!\left(\nu\right))}{
i!\; j!\;k!\;2^{k}2^{j}\sqrt{1 - \tilde\gamma_\perp^2}}\!\!\left[
\tilde\gamma_{\perp}\!\sqrt{1 - \tilde\gamma_\perp^2 -\tilde
\gamma_{\parallel}^2 -\tilde \gamma_2^2 + 2 \tilde\gamma_\perp
\tilde\gamma_{\parallel}\tilde\gamma_2}\left\langle \tilde x^i \tilde x_{1}^{2j}\tilde
x_{2}^{2k}\tilde\xi\right\rangle_\mathrm{ \!GC}\!\!\!\!
+\!\left(\!\tilde\gamma_{\perp}\!(\tilde\gamma_2 -\tilde
\gamma_{\parallel}\tilde \gamma_{\perp})\!+\!\tilde
\gamma_{\parallel}(1-\tilde\gamma_{\perp}^2)\!\right)\!\!\left\langle x^i \tilde
x_{1}^{2j}\tilde x_{2}^{2k}\tilde\zeta\right\rangle_\mathrm{ \!GC}\right] 
\\
+\left. \sum_{n=3}^\infty \sum_{\sigma_{n-2}} \!\!\frac{(-1)^{k+j}}{i!\; j!\;k!\;2^{k}2^{j}} H_{i-1}\!\left(\nu\right)\!\left[
(\tilde\gamma_2 -\tilde \gamma_{\parallel}\tilde \gamma_{\perp})\!\left\langle  \tilde
x^i \tilde x_{1}^{2j}\tilde x_{2}^{2k}\tilde \zeta ^{2}
\right\rangle_{\!\mathrm{GC}}\!\!\!\!-  \tilde\gamma_{2}\! \left\langle  \tilde x^i
\tilde x_{1}^{2j}\tilde x_{2}^{2k}\tilde
x_{12}^{2}\right\rangle_{\!\mathrm{GC}}\!\!\!\!
+\!\sqrt{1 - \tilde\gamma_\perp^2 -\tilde \gamma_{\parallel}^2 -\tilde
\gamma_2^2 + 2 \tilde\gamma_\perp
\tilde\gamma_{\parallel}\tilde\gamma_2}\left\langle  \tilde x^i \tilde x_{1}^{2j}\tilde
x_{2}^{2k}\tilde \zeta \tilde\xi \right\rangle_{\!\mathrm{GC}}\!
\right]
\right)
, 
 \label{eq:chi2D}
\end{multline}
where $\sigma_{n}=\{(i,j,k)\in \mathbb{N}^{3}|i+2 j+2k =n\}$ and 
$H_{-1}(\nu)=\sqrt{\pi/2}\,\textrm{Erfc}\!\left(\nu/\sqrt 2\right)$.
In the Gaussian limit, the Euler characteristic of a 2D plane then reads
\begin{equation}
\chi_{2\textrm{D}}^{(0)}(\nu, \theta_{\cal S})=
\frac {e^{-\nu^{2}/2}} { (2\pi)^{3/2}}
\frac{{\tilde \sigma_{1\perp}}{\tilde \sigma_{1\parallel}}}{\sigma^{2}} H_{1}(\nu) =
\frac {e^{-\nu^{2}/2}} {(2\pi)^{3/2}}
\frac{\sigma_{1\perp}\sigma_{1\parallel}}{\sqrt{2} \sigma^2}
\sqrt{1-\beta_\sigma \sin^2\theta_{\cal S}} \; H_{1}(\nu) \,,
\end{equation}
where $\beta_\sigma$ is defined in equation~(\ref{eq:defbeta}).
This is in agreement with  \cite{Matsubara96}. Note that 
 the amplitude of this Gaussian term is overestimated when assuming isotropy.
The $n=3$ term (i.e the first correction from Gaussianity) in the expansion 
 gives, in ``on plane'' variables
 \begin{equation}
 \chi_{2\textrm{D}}^{(1)}(\nu,\theta_{\cal S})
=\frac {e^{-\nu^{2}/2}} {(2\pi)^{3/2}}
\frac{{\tilde \sigma_{1\perp}}{\tilde \sigma_{1\parallel}}}{\sigma^{2}}
\times
\left[
\frac 1 {3!} H_4\left(\nu\right)  \left\langle  \tilde x^3\right\rangle
 +\frac 1 2  H_{2}(\nu)\left\langle  \tilde x \left({\tilde x_{1}}^{2}+{\tilde x_{2}}^{2}   \right)\right\rangle
-\frac 1 2 {\tilde \gamma_{\perp}^{-1}}\left\langle {\tilde x_{11}} {\tilde x_{2}}^{2}\right\rangle
\right]
\,,
 \end{equation}
or, using the cumulants of the 3D field, 
 \begin{multline}
\chi_{2\textrm{D}}^{(1)}(\nu,\theta_{\cal S})
=\frac {e^{-\nu^{2}/2}} {(2\pi)^{3/2}}\frac{\sigma_{1\perp}\sigma_{1\parallel}}{\sqrt{2} \sigma^2}
\sqrt{1-\beta_\sigma \sin^2\theta_{\cal S}}
\times
\left[ \frac 1 {3!}H_4\left(\nu\right)  \left\langle x^3\right\rangle \right.  
+ H_{2}(\nu) \left(\left\langle x q_{\perp}^2\right\rangle
+ \frac{1}{2} \frac{\cos^{2}\!\theta_{\cal S}}{ 1 - \beta_\sigma \sin^{2}\theta_{\cal S}}
 \left( \left\langle x x_{3}^{2}\right\rangle  - \left\langle x q_{\perp}^2\right\rangle\right)\right)\\
-\left. \frac{1}{\gamma_{\perp}}
\left( \left\langle  q_{\perp}^{2}J_{1\perp}\right\rangle +
\frac{1}{2} \frac{\cos^{2}\!\theta_{\cal S} }{ 1 - \beta_\sigma \sin^{2}\theta_{\cal S} }
\left( \left\langle  x_{3}^{2}J_{1\perp}\right\rangle - 2 \left\langle  q_{\perp}^{2}J_{1\perp}\right\rangle \right)
\right)\right]\qquad
\,, 
\label{eq:2DNGgenus}
 \end{multline}
 where $\sigma_{1\perp}^{2}q_{\perp}^{2}=\sigma^{2}
({\nabla}_{\!\perp}x\cdot\! {\nabla}_{\!\perp}x)$,
$\sigma_{1\parallel}^{2}x_{3}^{2}=\sigma^{2} ({\nabla}_{\parallel}x\cdot\!
{\nabla}_{\parallel}x)$ and $\sigma_{2\perp}J_{1\perp}=\sigma\,\Delta x$.
For the particular case when the cut is done through isotropic 3D field, the 2D Euler characteristic to first order in non-Gaussianity 
is
\begin{equation}
\label{eq:chibar2D}
 \chi_{2\textrm{D}}^{\textrm{iso}}(\nu)
= \frac {e^{-\nu^{2}/2}} {(2\pi)^{3/2}}\frac{\sigma_{1}^{2}}{3\sigma^{2}}\left[H_{1}(\nu)+\frac 1 {3!}H_4\left(\nu\right)  \left\langle x^3\right\rangle
+\frac{\sigma^{2}}{\sigma_{1}^{2}}H_{2}(\nu) \left\langle x\left({\nabla}x\cdot {\nabla}x\right)\right\rangle
-\frac{3\sigma^{4}}{4\sigma_{1}^{4}} \left\langle \Delta x \left( {\nabla}x\cdot  {\nabla}x\right)\right\rangle+{\cal O}(\sigma^{2})\right],
\end{equation}
which is in agreement with \cite{Gay12} (again relying on Appendix \ref{sec:momentsrelation} for some relations between the cumulants).
\begin{figure}
 \begin{center}
   \includegraphics[width=0.45\textwidth]{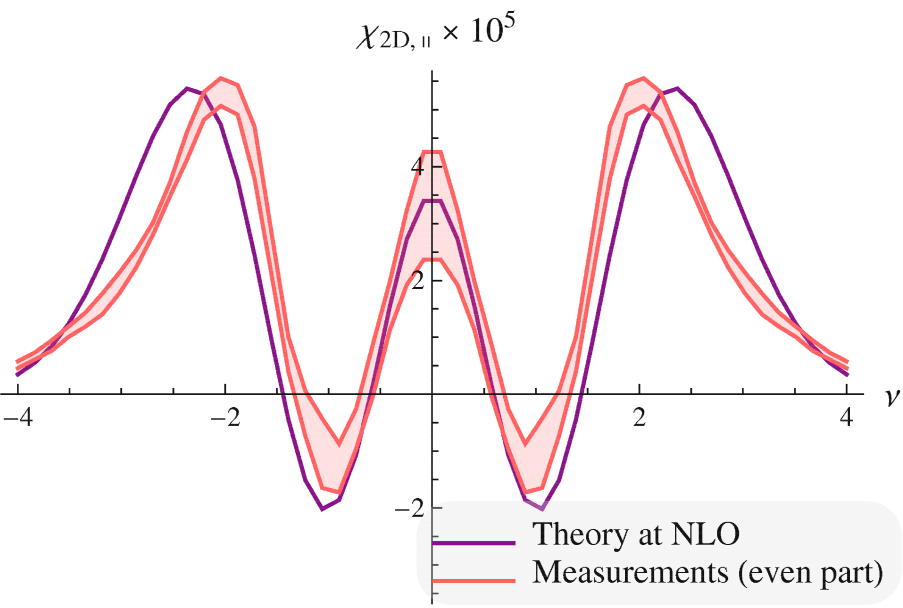}
  \includegraphics[width=0.45\textwidth]{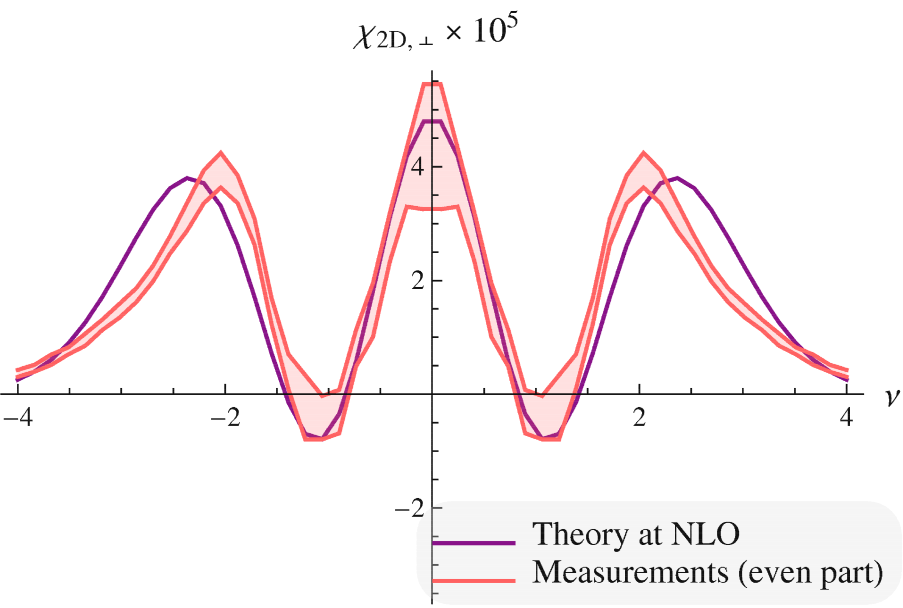}
    \includegraphics[width=0.45\textwidth]{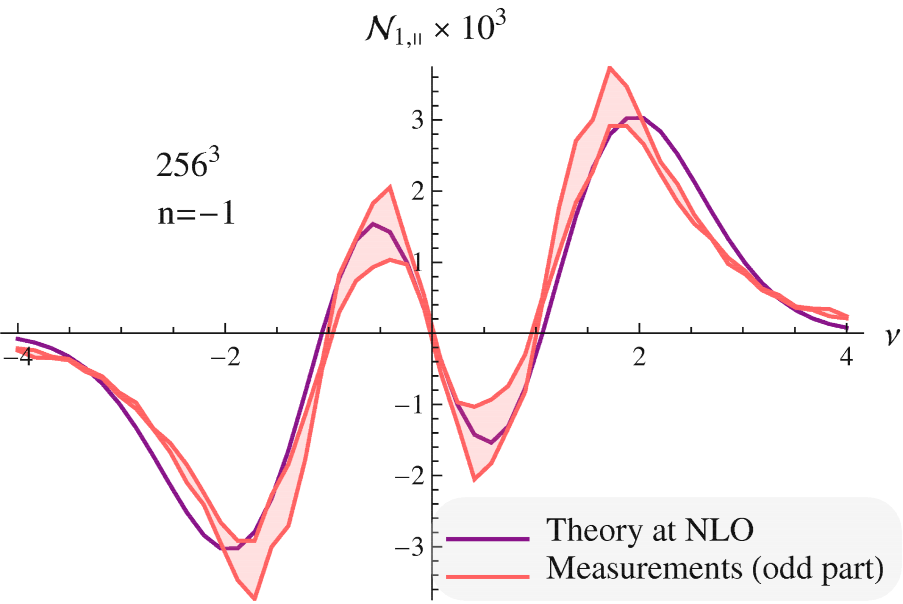}
  \includegraphics[width=0.45\textwidth]{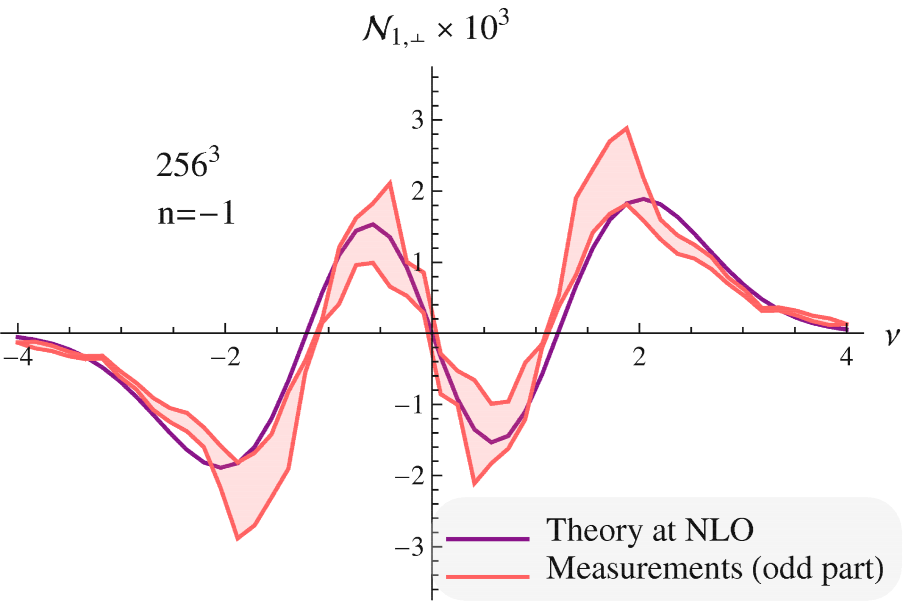}
\includegraphics[width=0.45\textwidth]{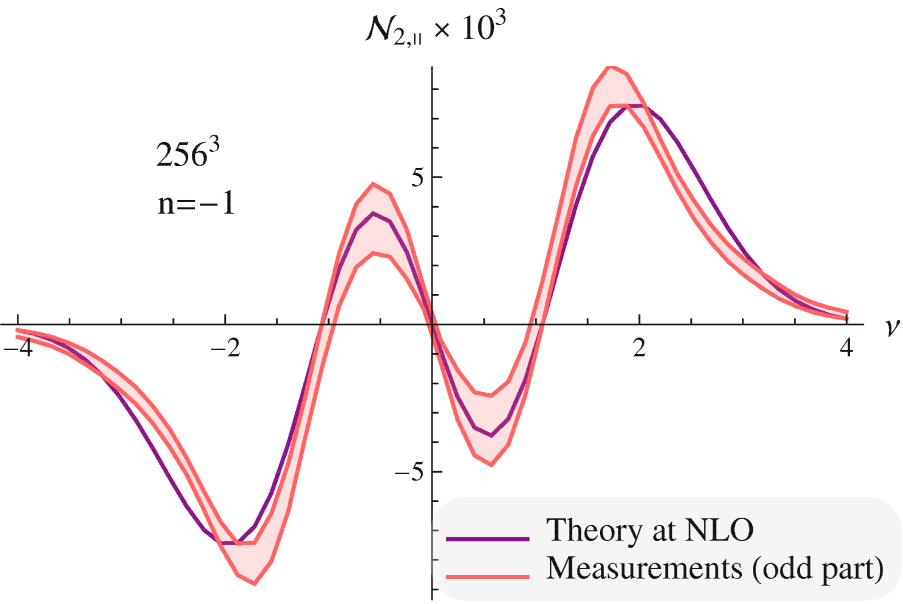}
    \includegraphics[width=0.45\textwidth]{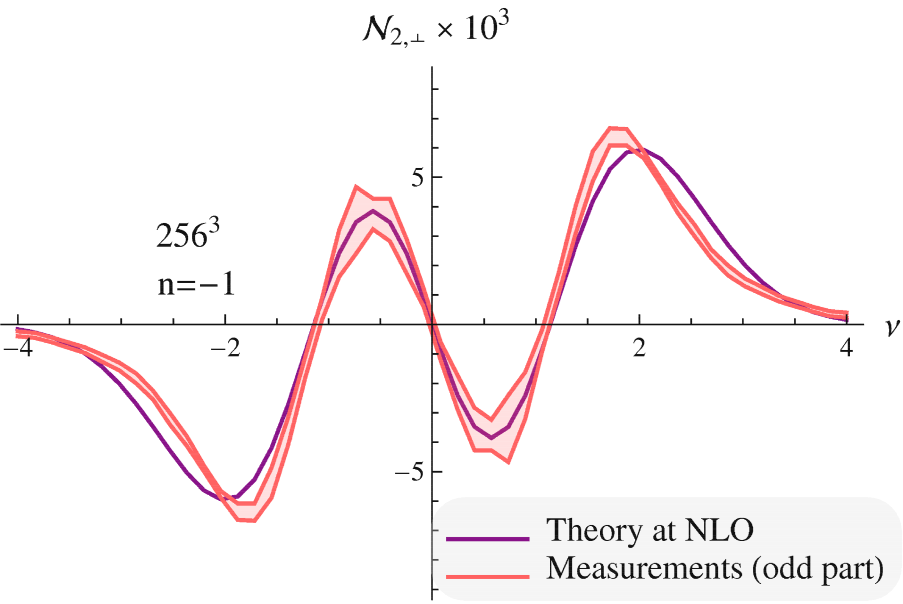}
    \includegraphics[width=0.45\textwidth]{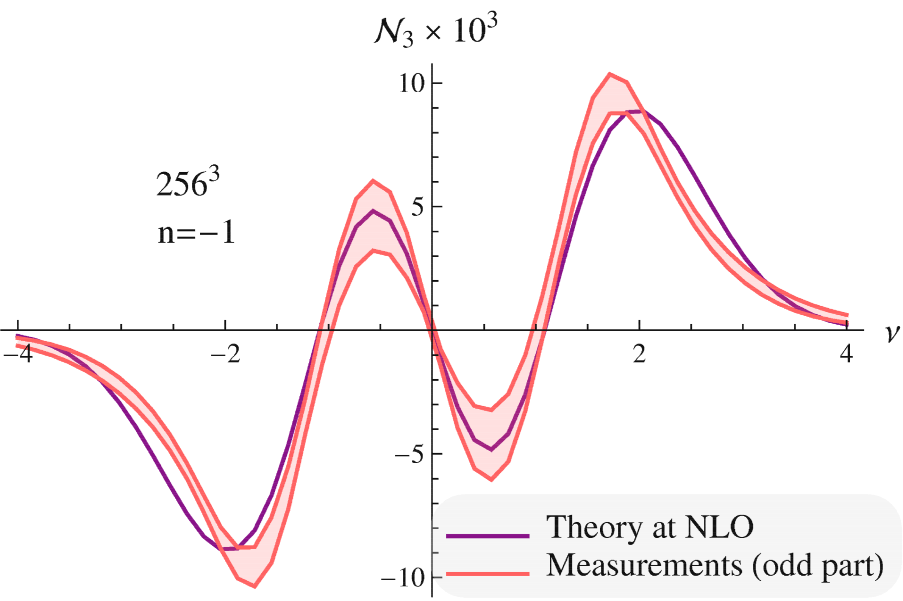}
      \includegraphics[width=0.45\textwidth]{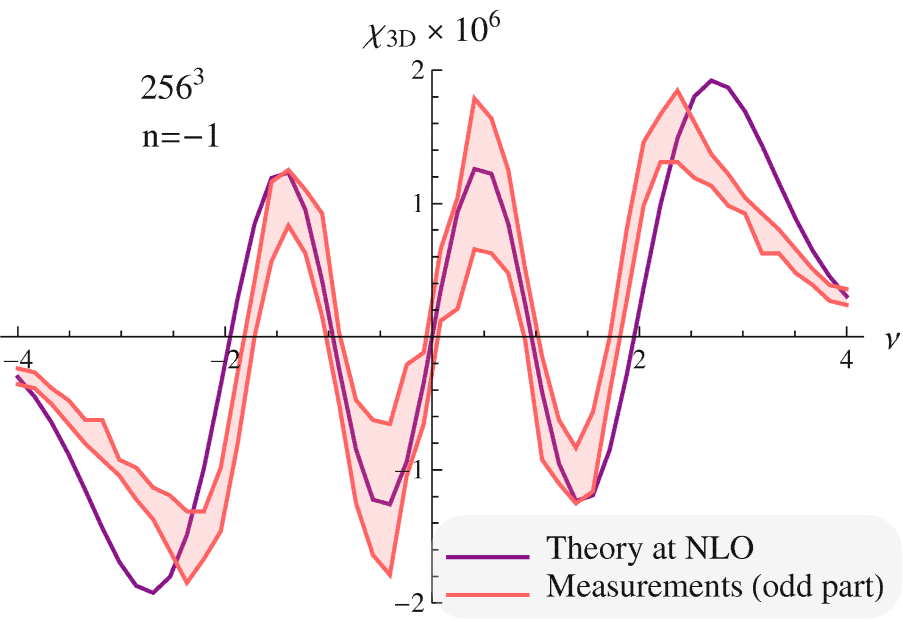}
 \caption{Gravity-induced NLO non-Gaussian corrections for the Minkowski functionals  of a mildly
  non-linear 3D density field ($n=-1$, $\sigma=0.18$) in redshift space ($f=1$). Theoretical predictions
 are displayed with solid purple lines, measurements with shaded one sigma dispersion.
\emph{Top left:} 2D Euler characteristic in planes parallel to the LOS;
\emph{top right:} 2D Euler characteristic in planes perpendicular to the LOS;
\emph{central top left:}  $\mathcal{N}_1$ (contour crossing) in planes parallel to the LOS;
\emph{central top right:}   $\mathcal{N}_1$  in planes perpendicular to the LOS;
\emph{central bottom left:} 
 $\mathcal{N}_2$ (length of iso-contour) in planes parallel to the LOS;
 \emph{central bottom right:} 
 $\mathcal{N}_2$  in planes perpendicular to the LOS;
 \emph{bottom left:} 
 $\mathcal{N}_{3}$ (area of iso-contour);
 \emph{bottom right:} 
 3D Euler characteristic.
{Note that the predicted first order correction fits very well the odd part (resp. even for the 2D Euler characteristic) of the measured correction for intermediate contrast ($-1.5\lesssim\nu\lesssim1.5$).}
\label{fig:minkowski-gravity}
}
 \end{center}
\end{figure}
Figures~\ref{fig:genusGfnl}
 and \ref{fig:genusNGfnl} shows the dependence of the 2D Euler characteristic on $f_{nl}$ and $f$ following equations~(\ref{eq:2DNGgenus}) and (\ref{eq:chibar2D}).
 As expected, the effect of redshift distortion is enhanced in 2D relative to 3D.  This is expected since in 3D two dimensions remain isotropic.
Figure~\ref{fig:genusNGgnl} also shows the evolution of the 2D Euler characteristic as a function of $g_\mathrm{{nl}}$ and $f$.
 It demonstrates that this set of simulations is well fitted by
the  sum of the two contributions from $f_\mathrm{{nl}}$ and $g_\mathrm{{nl}}$. 

For cosmological fields, Figure~\ref{fig:minkowski-gravity}  compares the prediction of equation~(\ref{eq:2DNGgenus})
 for the  ${\chi}_\mathrm{ 2D}$ statistics to scale invariant LSS simulations
in planes parallel and perpendicular to the line of sight. 
Again, this figure shows a very good agreement between measurements and predictions at NLO 
 for intermediate contrasts ($-1.5\lesssim\nu\lesssim1.5$). There is a noticeable angle-dependence
 of the 2D Euler characteristic which suggests that the procedure described in the introduction of
 Section~\ref{sec:Min23} is feasible. This angle dependence is also illustrated in Figure~\ref{fig:2Dgenusangle}
 where one can see how the prediction at next-to-leading order varies with the angle $\theta_{\cal S}$
 between the line of sight and the 2D slice under consideration.
 
\subsection{Other Minkowski functionals }
\label{sec:otherminkowski}
\subsubsection{Area of isodensity contours in 3D : ${\cal N}_3$}
\label{sec:N3}
One of the Minkowski functionals is  ${\cal N}_{3}(\nu)$, the area (per unit volume) of a 3D isosurface of
 the density field at level $\nu$. To compute this functional, it is sufficient to consider the JPDF of the field
 and its first derivatives
\begin{equation}
P(x,q_{\perp}^{2},x_{3})=\frac 1 {2\pi}\exp \left(-\frac {x^{2}}{2}-q_{\perp}^{2}-\frac {x_{3}^{2}}{2}\right)
\left[1+\sum_{n=3}^{\infty}\sum_{\sigma_n}\frac {(-1)^{j}} {i!\,j!\,k!}\left\langle x^{i}q_{\perp}^{2j}x_{3}^{k}\right\rangle_{\textrm{GC}}H_{i}(x)L_{j}(q_{\perp}^{2})H_{k}(x_{3})\right].
\end{equation} 
From equation~(\ref{eq:N1}), the area of a 3D isosurface  is
\begin{equation}
{\cal N}_{3}(\nu)=\frac 1 {\sigma} \int \dd q_{\perp}^{2}\dd x_{3} P(\nu,q_{\perp}^{2},x_{3})\sqrt{\sigma_{1\perp}^{2}q_{\perp}^{2}+\sigma_{1\parallel}^{2}x_{3}^{2}}\,.
\label{eq:N3_ani}
\end{equation}
Computing the integrals, we express the results using the anisotropy parameter 
$\beta_{\sigma} = 1 - {\sigma_{1\perp}^2}/{2 \sigma_{1\parallel}^2}$. 
At Gaussian order, expression (\ref{eq:N3_ani}) yields the result, consistent with \cite{Matsubara96}, 
\begin{equation}
{\cal N}_{3}^{(0)}(\nu)
= \frac {2 }{\pi} \frac{\sigma_{1}}{\sqrt{3} \sigma}  \frac{1 - A(\beta_\sigma)}{\sqrt{1-2 \beta_\sigma/3}} e^{-\nu^{2}/2}, \quad \textrm{with}\quad
A(\beta_\sigma) \equiv \frac{1}{2} \left( \vphantom{\frac{T}{\beta_\sigma}}  \beta_\sigma - T \left(\beta_\sigma\right)  + \beta_\sigma T \left(\beta_\sigma\right) \right), 
\end{equation}
where the function $T$ is defined as $T(\beta_\sigma)=1/\sqrt{\beta_\sigma} \tanh ^{-1}\left(\sqrt{\beta_\sigma}\right) - 1$
for $\beta_\sigma \ge 0$ 
and 
$T(\beta_\sigma)=1/\sqrt{|\beta_\sigma|}\tan ^{-1}\left(\sqrt{|\beta_\sigma|}\right)-1 $ for $\beta_\sigma<0$. Under this definition
$A(\beta_\sigma)$ describes a $\sim {\beta_\sigma}/{3} + \beta_\sigma^2/15 + \ldots $ correction at small anisotropy  $\beta_\sigma \to 0$.
We see that in the Gaussian limit, the anisotropy has a very little effect on ${\cal N}_3$. The amplitude
deviates from unity by less that 1\% in the range $-1 < \beta_\sigma < 0.5$, as its series expansion
$ \propto 1 - \beta_\sigma^2/90 \ldots $ attests. Even at extreme anisotropies, it
changes just to $\approx 0.92$ at $\beta_\sigma \to -\infty$ and $\approx 0.87$ at $\beta_\sigma=1$. 

The computation of the $n=3$ term corresponding to the first non-Gaussian correction is also straightforward
\begin{equation}
{\cal N}_{3}^{(1)}(\nu)
=  {\cal N}_{3}^{(0)}(\nu) \Bigg[
\frac {1}{3!}\left\langle x^{3}\right\rangle H_{3}(\nu) 
+\frac{1}{2}\left( \vphantom{\frac{T}{\beta_\sigma}}
  \left\langle  xq_{\perp}^{2}\right\rangle + 
  \frac{A(\beta_\sigma)/\beta_\sigma}{ 1 - A(\beta_\sigma)} 
\left( \left\langle xx_{3}^{2}\right\rangle -\left\langle  xq_{\perp}^{2}\right\rangle  \right) \right) H_1(\nu)\Bigg]\,.
\label{eq:N3}
\end{equation}
To first order in $\beta_\sigma$ (the anisotropy parameter), one gets the following explicit expression
\begin{equation}
{\cal N}_{3}^{(1)}(\nu)
= \frac {2e^{-\nu^{2}/2}\sigma_{1}}{ \sqrt {3}\pi\sigma}\Bigg[
1+
\frac {1}{3!}\left\langle x^{3}\right\rangle H_{3}(\nu)
+\frac{1}{2} \left(\frac{2}{3} \left\langle x q_\perp^{2}\right\rangle \left[ 1  - \frac{4}{15}\beta_\sigma \right] + 
\frac{1}{3} \left\langle x x_{3}^2 \right\rangle \left[ 1 + \frac{8}{15}\beta_\sigma \right] \right)H_{1}(\nu) 
\Bigg] + {\cal O}(\beta_\sigma^2)\,, \label{eq:N3final}
\end{equation}
from which the isotropic limit of \cite{Gay12} is readily recovered by setting  $\beta_\sigma=0$ and $\left\langle x q_\perp^{2}\right\rangle = \left\langle x x_{3}^2 \right\rangle 
= \left\langle x q^2 \right\rangle$.
We note that anisotropy effects in the ${\cal N}_3$ statistics are almost exclusively  concentrated in the gradient terms 
$\propto H_1(\nu)$.  This suggests, for example, that the recovery of the skewness  $\langle x^3 \rangle$ by fitting the $H_3(\nu)$ mode 
to  ${\cal N}_3 $  will be practically unaffected by redshift distortions. In contrast,  one must focus on the $H_1(\nu)$ mode to measure
anisotropic effects.

Figure~\ref{fig:minkowski-gravity}   also compares the prediction of equation~(\ref{eq:N3final}) for the  ${\cal N}_3$ statistics to scale invariant LSS simulations ($256^{3}$, $n=-1$). Once again, in this mildly non-linear regime ($\sigma=0.18$ in real space), the prediction at NLO matches very well the measurement for intermediate contrasts ($-2 \lesssim\nu\lesssim 2$) . 

\subsubsection{Length of isodensity contours in 2D planes : ${\cal N}_2$}
\label{sec:N2}
Let us consider the length (per unit volume) of isodensity contours in 2D slices of the density field, ${\cal N}_2$. 
This functional  
is the 2D version of the Minkowski functional ${\cal N}_3$ for 3D field. Here the 2D slice is defined by the 
angle, $\theta_{\cal S}$, it makes with the $z$-axis and the statistical properties of the field depend on this angle.
 Again, let us start with the JPDF of the field and its gradient on a 2D plane. 
Using the same variables as in Section \ref{sec:genus2D},
\begin{equation}
P( \tilde x,{\tilde x_{1}},{\tilde x_{2}})=\frac 1 {(2\pi)^{3/2}}
\exp \left(-\frac { \tilde x^{2}}{2}-\frac{{\tilde x_{1}}^{2}}{2}-\frac{{\tilde x_{2}}^{2}}{2}\right)
\left[1+\sum_{n=3}^{\infty}\sum_{\sigma_n}\frac 1 {i!\,j!\,k!}\left\langle  \tilde x^{i}{\tilde x_{1}}^{j}{\tilde x_{2}}^{k}\right\rangle_{\textrm{GC}}H_{i}( \tilde x)H_{j}({\tilde x_{1}})H_{k}({\tilde x_{2}})\right] \; ,
\end{equation}
where $\sigma_{n}=\{(i,j,k)\in \mathbb{N}^{3}|i+2j+k=n\}$.
From equation~(\ref{eq:N1}), the length of isodensity contours in 2D planes is now
\begin{equation}
{\cal N}_{2}(\nu,\theta_{\cal S})=\frac 1 {\sigma} \int \dd \tilde x_{1}\dd \tilde x_{2} P(\nu,\tilde x_{1},\tilde x_{2})\sqrt{{\tilde \sigma_{1\perp}}^{2}\tilde x_{1}^{2}+{\tilde \sigma_{1\parallel}}^{2}\tilde x_{2}^{2}}\,.
\label{eq:N2_ani}
\end{equation}
We shall proceed similarly to the 3D case, defining the 2D anisotropy parameter 
$\tilde \beta_\sigma(\theta_{\cal S}) $
which  depends on the 
orientation of the plane
\begin{equation}
\label{eq:defbeta2D}
 \tilde \beta_\sigma(\theta_{\cal S}) \equiv 1 - \frac{\tilde \sigma_{1\perp}^2}{\tilde \sigma_{1\parallel}^2} = 
\frac{\beta_\sigma \cos^2\theta_{\cal S}}{1 - \beta_\sigma \sin^2\theta_{\cal S}}\,.
\end{equation}
At Gaussian order, the evaluation of equation~({\ref{eq:N2_ani})  yields, in agreement with \cite{Matsubara96}
\begin{equation}
{\cal N}_{2}^{(0)}(\nu,\theta_{\cal S})=\frac {\tilde\sigma_{1\parallel}} {\pi \sigma} e^{-\nu^{2}/2}E\left(\tilde\beta_\sigma\right) 
= \frac {\sigma_{1\perp}} {\sqrt{2} \pi \sigma} e^{-\nu^{2}/2}\frac{ E\left(\tilde\beta_\sigma\right)}{ \sqrt{1-\tilde\beta_\sigma}}\,,
\end{equation}
where $E$ is the complete elliptic integral of the second kind. The amplitude behaves as
$ E(\tilde\beta_\sigma)/\sqrt{1-\tilde\beta_\sigma} \sim {\pi}/{2} + {\pi} \tilde \beta_\sigma/{8}$ at small $\tilde \beta_\sigma$ and is thus strongly dependent on the anisotropy parameter.  This is distinct from the behaviour of ${\cal N}_3$.

Then, the first non-Gaussian correction (corresponding to $n=3$) reads
\begin{equation}
{\cal N}_{2}^{(1)}(\nu,\theta_{\cal S})
= {\cal N}_2^{(0)}
\Bigg[\!\frac {H_{3}(\nu)} {3!}\left\langle  \tilde x^{3} \right\rangle 
\!+\!\frac{1}{2} H_{1}(\nu)\! \left( \! \left\langle \tilde  x \tilde x_{2}^{2}\right\rangle 
\!+\! B(\tilde \beta_\sigma) \left( \left\langle \tilde  x \tilde x_1^{2}\right\rangle\! -\! 
\left\langle x \tilde x_{2}^{2}\right\rangle \right) \right)\!
\Bigg] ,
 \,\textrm{with}\; B(\tilde\beta_\sigma) =  2(\tilde \beta_\sigma - 1 ) \frac{\dd \ln E(\tilde \beta_\sigma)}{\dd \tilde \beta_\sigma}\,.
\label{eq:N2_ani_n3}
\end{equation}
Here on-plane cumulants are related to 3D ones by $\left\langle \tilde x{\tilde x_{1}}^{2}\right\rangle
=\left\langle x q_{\perp}^{2}\right\rangle$ and
 $\left\langle  \tilde x{\tilde x_{2}}^{2}\right\rangle
 =\left(1-\tilde\beta_{\sigma}/\beta_{\sigma}\right)\left\langle xq_{\perp}^{2}\right\rangle+\tilde\beta_{\sigma}/\beta_{\sigma}\left\langle xx_{3}^{2}\right\rangle$.
In the small anisotropy limit, equation (\ref{eq:N2_ani_n3}) becomes 
\begin{equation}
{\cal N}_{2}^{(1)}(\nu,\theta_{\cal S})
= \frac {\sigma_{1\perp}} {2 \sqrt{2} \sigma} e^{-\nu^{2}/2} \left(1+\frac{1}{4} \tilde\beta_\sigma\right)
\Bigg[\frac {H_{3}(\nu)} {3!}\left\langle \tilde x^{3} \right\rangle 
+\frac{1}{4} H_{1}(\nu) \left( \left\langle \tilde x \tilde x_1^{2}\right\rangle 
+\left\langle \tilde x \tilde x_{2}^{2}\right\rangle 
- \frac{3}{8}\tilde\beta_\sigma \left( \left\langle \tilde x \tilde x_{1}^{2}\right\rangle 
- \left\langle \tilde  x \tilde x_2^{2}\right\rangle \right) \right)
\Bigg] + {\mathcal O}\left(\tilde\beta_\sigma^2\right)\, ,
\label{eq:N2_series}
\end{equation}
where the isotropic limit of \cite{Gay12} is readily recognized at $\tilde\beta_\sigma=0$, 
$\left\langle  \tilde x \tilde x_1^{2}\right\rangle=\left\langle \tilde x \tilde x_{2}^{2}\right\rangle=
\left\langle x q^2 \right\rangle$ and $\sigma_{1\perp} = \sqrt{2/3} \sigma_1$.
We should stress that anisotropic effects in equation~(\ref{eq:N2_series}) are contained not only in the
$\tilde \beta_\sigma$ factors, but in the deviation of cumulants from their isotropic values as well. Note that, with the 3D variables, equation~(\ref{eq:N2_series}) becomes
\begin{multline}
{\cal N}_{2}^{(1)}(\nu,\theta_{\cal S})
= \frac {\sigma_{1\perp}} {2 \sqrt{2} \sigma} e^{-\nu^{2}/2} \left(1+\beta_\sigma\frac{\cos^{2}\theta_{\cal S}}{4}\right)
\Bigg[\frac {H_{3}(\nu)} {3!}\left\langle x^{3} \right\rangle \\
+\frac{1}{4} H_{1}(\nu) \left( \left\langle x q_{\perp}^{2}\right\rangle 
+\left\langle x  x_{3}^{2}\right\rangle 
+\left( \sin^{2}\theta_{\cal S}-\beta_\sigma\frac{\cos^{2}\theta_{\cal S}(3+5\sin^{2}\theta_{\cal S})}{8}\right) \left( \left\langle x  q_{\perp}^{2}\right\rangle 
- \left\langle x  x_3^{2}\right\rangle \right) \right)
\Bigg] + {\mathcal O}\left(\beta_{\sigma}^2\right)\, .
\end{multline}

The most interesting case for the ${\cal N}_2$ statistics is when the slice is passing through the observer, and, 
correspondingly, contains the line of sight. This is the case for 2D slices in observational catalogues such as
SDSS. This setup corresponds to $\theta_{\cal S}=0$ and $\tilde \beta_\sigma = \beta_\sigma$. Indeed, the measurement of 
${\cal N}_2$ in such slices gives direct access to the 3D anisotropy parameter $\beta_\sigma$, and, by extention, $\Omega_m^{0.6}/b$,
even when full 3D data is not available. The study of 2D slices at different angles $\theta_{\cal S}$ is possible if the 3D
cube of data is available and offers alternative way of analysing such cubes. Varying $\theta_{\cal S}$ allows
to introduce functional dependence of statistics on the parameter, which gives additional access to $\tilde\beta_\sigma$
through variation of the amplitude of the statistics even when the normalization scale $\sigma_{1\perp}/\sigma$ is poorly determined. 

Figure~\ref{fig:minkowski-gravity}  compares the prediction of equation~(\ref{eq:N2_ani_n3}) for the  ${\cal N}_2$ statistics to scale invariant LSS simulations
in planes parallel and perpendicular to the line of sight. In both cases, measurements are well fitted by the prediction at the first non-Gaussian order  for intermediate contrasts. One can easily notice that this statistics varies macroscopically with the angle $\theta_{\cal S}$. This dependence is also shown in Figure~\ref{fig:2Dgenusangle} for the NLO prediction.
Indeed, Figure~\ref{fig:2Dgenusangle} displays the variation of the non-Gaussian contribution to the  ${\cal N}_2$ statistics as a function of the orientation of the 2D slices relative to the LOS. Note that anisotropy
affects the $H_1(\nu)$ harmonic of this dependence and can be detected by a linear fit to 
${\cal N}_{2}^{(1)}(\nu,\theta_{\cal S})$ at different $\theta_{\cal S}$.
\begin{figure}
 \begin{center}
 \includegraphics[width=0.45\textwidth]{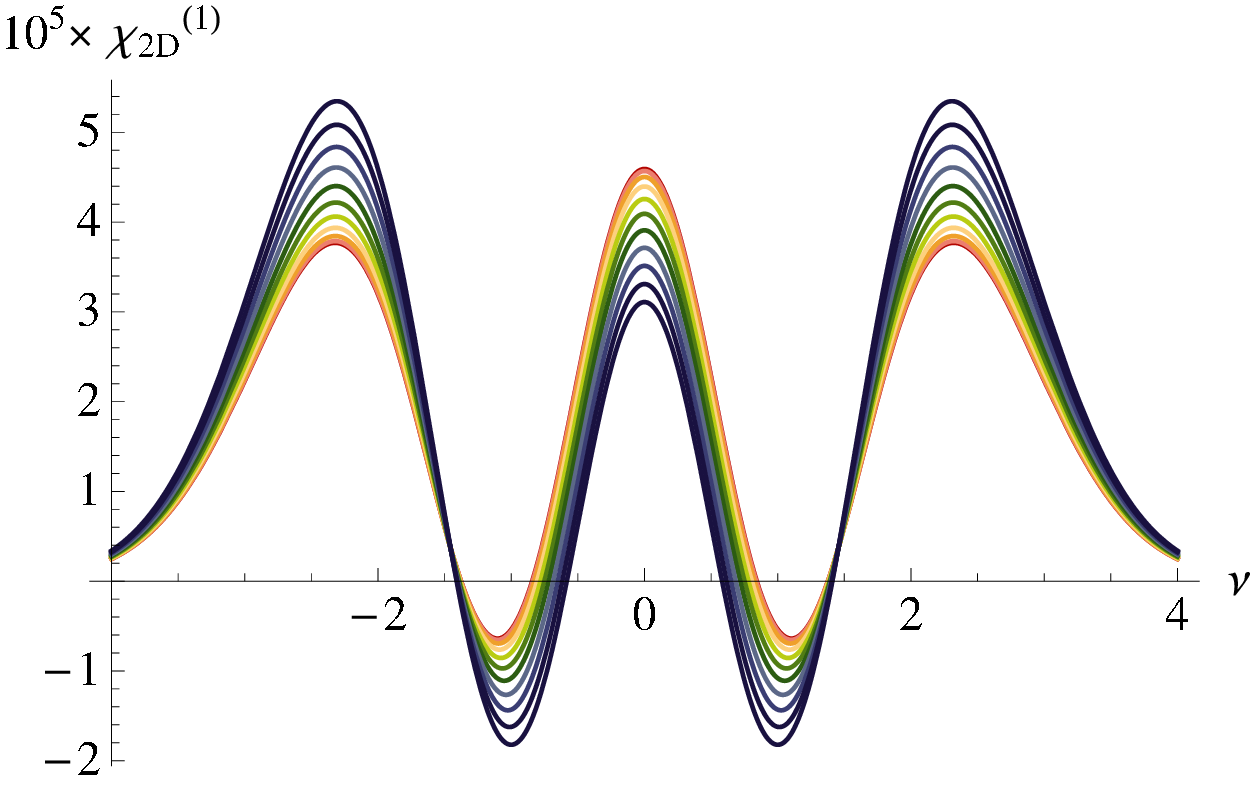}
 \includegraphics[width=0.45\textwidth]{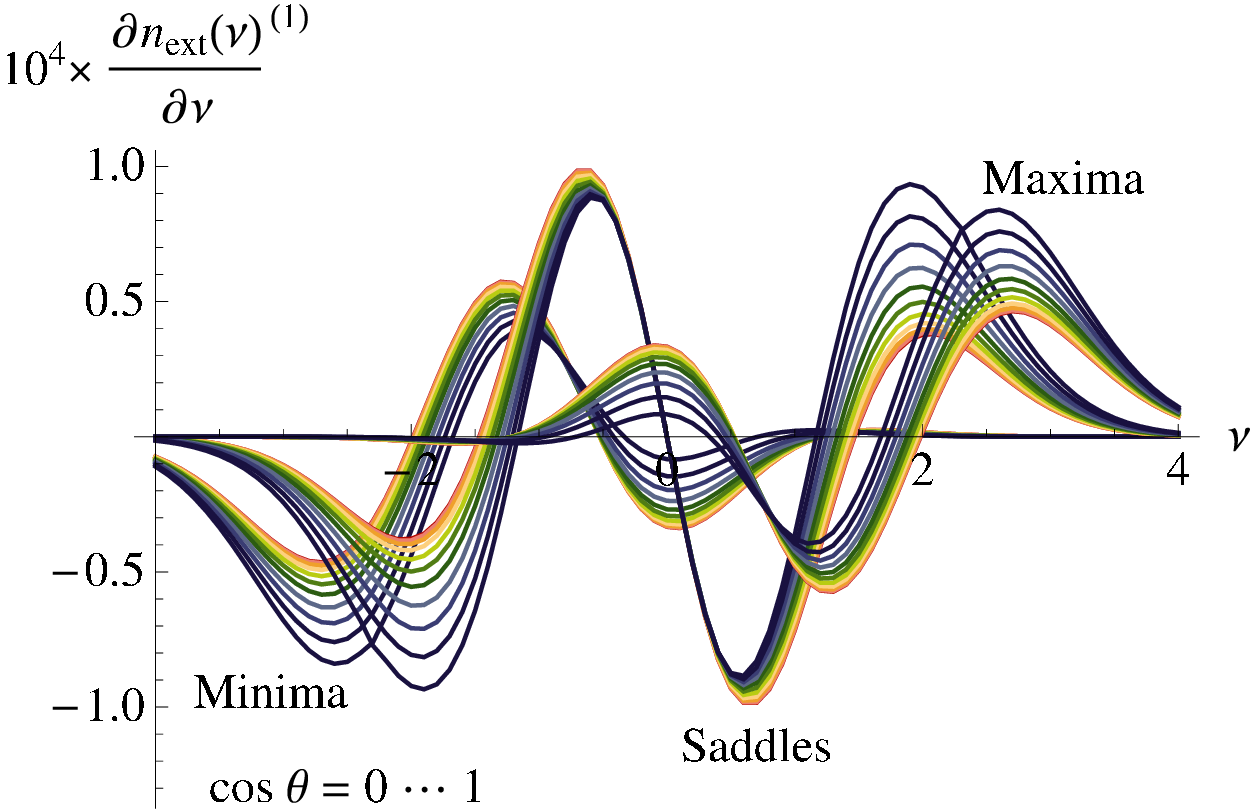}
  \includegraphics[width=0.45\textwidth]{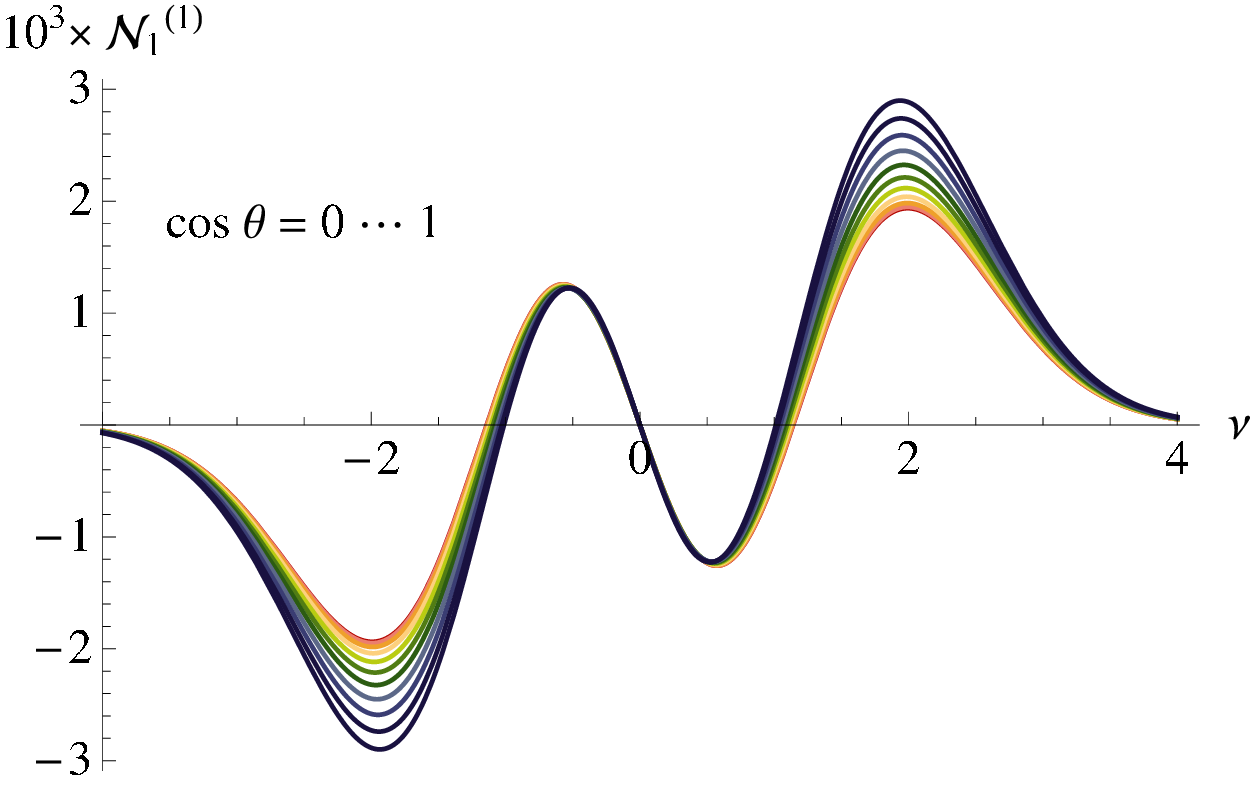} 
    \includegraphics[width=0.45\textwidth]{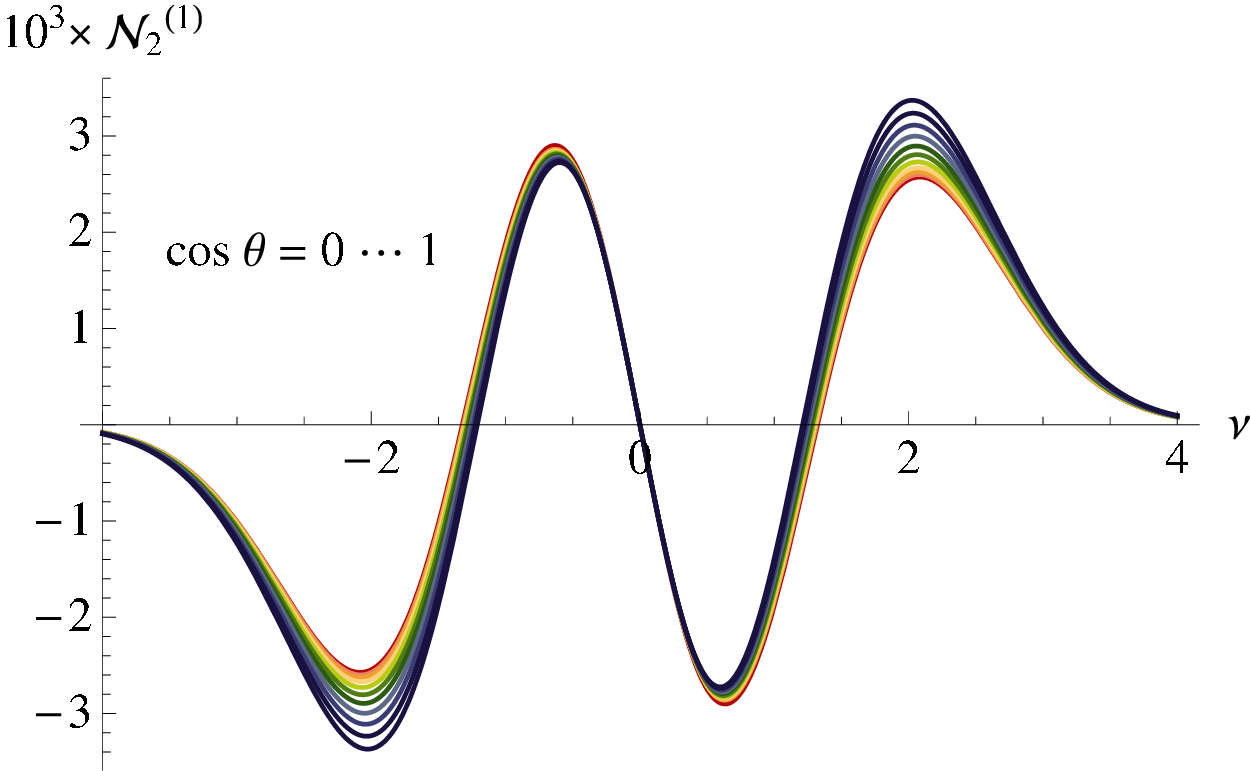} 
\caption{
 2D Euler characteristic (\emph{top left panel}),  2D extrema counts (\emph{top right panel}), 
 length of 2D iso-contour (\emph{bottom left panel}) and  contour crossing (\emph{bottom right panel})
  at first non-Gaussian order  as a function of $\cos \theta_{\cal S}$ from 0 (\emph{red}) to 1 (\emph{dark blue}).
 Cumulants used in equations~(\ref{eq:2DNGgenus})-(\ref{eq:N2_ani_n3})-(\ref{eq:N1NG}) are measured in a scale-invariant
 LSS simulation with $\sigma=0.18$  The prediction for 2D extrema statistics was obtained by numerical integration
 following Section~\ref{sec:ext}. The noticeable dependence of these functionals on the angle $\theta_{\cal S}$
 gives us a mean to measure $\beta=f/b_{1}$ as suggested in Section \ref{sec:measuringf}.
 \label{fig:2Dgenusangle}
}
 \end{center}
\end{figure}
\subsubsection{Contour crossings : ${\cal N}_1$}
\label{sec:N1}
Contour-crossing statistics  measures the average number of times  a
given line crosses the isocontours of a field, ${\cal N}_1$.  It is closely related to the
average area of the isocontours per unit volume, ${\cal N}_3$. It is equivalent
to ${\cal N}_3$ when averaged over all possible orientations of the chosen line
(indeed, the surface area per unit volume ${\cal N}_3$ can be understood as an
average ``hyperflux'' of isosurfaces, i.e how many time per unit length they
cross a random line), but has distinct dependence on line orientation if the
field is anisotropic.  A special advantage of ${\cal N}_1$ is that it can be
applied when only a ``pencil beam'' data is available, as for example in
$Ly_\alpha$ lines, although in this case one is limited to use only
the line-of-sight direction of the field.  In general, the behaviour of ${\cal
N}_1$ along lines with arbitrary orientation relative to the LOS contains additional
information.
 
To compute contour-crossing statistics, we assume (without loss of generality) that the line intersecting the
field lies in the ($\mathbf{u_1},\mathbf{u_3}$) plane with a direction defined by the unit vector
$\hat l= (\sin\theta_{\cal L},0,\cos\theta_{\cal L})$. 
Then the mean number of intersections between a line $(\cal L)$ and the isodensity
contours is simply the average of the field gradient projection $| \nabla x
\cdot \hat l |=\left |\sigma_{1\perp}x_{1}\sin \theta_{L}/\sqrt{2}+\sigma_{1\parallel}x_{3}\cos \theta_{L}\right |$.
In contrast to the 2D statistics case, we shall not introduce
tilde variables on the line,  and write the statistics immediately in terms of
the 3D variables. 
Starting from  the JPDF of the
field and two components of its first derivatives
\begin{equation}
P(x,x_{1},x_{3})=\frac 1 {(2\pi)^{3/2}}\exp \left(-\frac{x^{2}}{2} -\frac{x_{1}^{2}}{2}-\frac {x_{3}^{2}}{2}\right)\left[1+\sum_{n=3}^{\infty}\sum_{i,j,k=0}^{i+j+k=n}\frac 1 {i!\,j!\,k!}\left\langle x^{i}x_{1}^{j}x_{3}^{k}\right\rangle_{\textrm{GC}}H_{i}(x)H_{j}(x_{1})H_{k}(x_{3})\right]\,,
\end{equation} 
from equation~(\ref{eq:N1}), we write
\begin{equation}
{\cal N}_{1}(\nu,\theta_{\cal L})=\frac 1 {\sigma} \int \dd x_{1}\dd x_{3} P(\nu,x_{1},x_{3})
\left |\frac{\sigma_{1\perp}}{\sqrt 2}x_{1}\sin \theta_{\cal L}+\sigma_{1\parallel}x_{3}\cos \theta_{\cal L}\right |.
\end{equation}
At Gaussian order, using again $\beta_\sigma = 1 - {\sigma_{1\perp}^2}/{2 \sigma_{1\parallel}^2}$, we find, in agreement with \cite{Matsubara96}:
\begin{equation}
 \label{eq:N1G}
{\cal N}_{1}^{(0)}(\nu,\theta_{\cal L})=
\frac{\sigma_{1\parallel}}{\pi \sigma} e^{-\nu^{2}/2}
\sqrt{1 - \beta_\sigma \sin^{2} \theta_{ L } }
,
\end{equation}
Focusing on the next
order term ($n=3$) and using the relationships $\langle  xx_{1}x_{3}\rangle  =\langle  x^{2}x_{i}\rangle
 =0$, one  gets
\begin{equation}
{\cal N}_{1}^{(1)}(\nu,\theta_{\cal L})
={\cal N}_{1}^{(0)}(\theta_{\cal L},\nu)\Bigg[
\frac 1 {3!}\left\langle x^{3}\right\rangle H_{3}(\nu) 
+\frac 1 {2}H_{1}(\nu)
\left( \left\langle x q_{\perp}^{2}\right\rangle +
\frac{\cos^{2}\theta_{\cal L}}{1-\beta_\sigma \sin^{2}\theta_{\cal L}}
\left(\left\langle x x_{3}^{2}\right\rangle - \left\langle x q_{\perp}^{2}\right\rangle \right) \right)
\Bigg]
 \label{eq:N1NG}
.
\end{equation}
In the isotropic limit where $\beta_\sigma=0$ and $\left\langle x
q_{\perp}^{2}\right\rangle=\left\langle x x_{3}^{2}\right\rangle=\left\langle x
 q^2\right\rangle$, equations~(\ref{eq:N1G}) and (\ref{eq:N1NG}) together becomes
\begin{equation}
{\cal N}_{1}^{\textrm{iso}}(\nu)
=\frac {e^{-\nu^{2}/2}\sigma_{1}} {\sqrt 3 \pi\sigma}\Bigg[1+\frac 1 {3!}H_{3}(\nu)\left\langle x^{3}\right\rangle  +
\frac 1 {2}H_{1}(\nu) \left\langle x q^{2}\right\rangle+{\cal O}(\sigma^{2})
\Bigg]
.
\end{equation}
Figure~\ref{fig:minkowski-gravity}  compares the prediction of equation~(\ref{eq:N1NG}) for the  ${\cal N}_1$ statistics to scale invariant LSS simulations
in planes parallel and perpendicular to the line of sight. The agreement between the prediction at NLO and measurements
 is very good for contrasts in the range $-2 \lesssim\nu\lesssim 2$. The tails of the distribution are more sensitive to higher-order terms in the GC expansion. Planes parallel and perpendicular to the line of sight give rise to a noticeable difference in the first order correction for the contour crossing statistics. 
Figure~\ref{fig:2Dgenusangle} also displays the variation of ${\cal N}_1$  as a
function of the orientation of the 2D slices relative to the LOS.
\subsection{Extrema counts} \label{sec:ext}
%
Extrema counts are given by averaging the absolute value of the determinant of the Hessian ($I_3$ and $I_2$ in 3D and 2D respectively)
under the condition of zero gradient over the range of Hessian eigenvalues space that maintains the correspondent
signature of the sorted eigenvalues, namely, in 3D, $\lambda_1 > \lambda_2 > \lambda_3 > 0$ for minima, 
$\lambda_1 > \lambda_2 > 0 > \lambda_3$ for pancake-like saddle points, $\lambda_1 > 0> \lambda_2 > \lambda_3$ for
filamentary saddle points and $0 > \lambda_1 > \lambda_2 > \lambda_3$ for maxima.
The integral to perform is similar to that for the Euler characterstic, e.g.~equation~(\ref{eq:defchi3D}) in 3D
\begin{equation}
n_\mathrm{ext,3D} =\frac 1 {\sigma_{1\perp}^{2}\sigma_{1\parallel}} 
\int_{\rm fixed~\lambda_i~signs} \dd x\, \dd \zeta\, \dd J_{2\perp} \dd \xi\, \dd Q^2\dd \Upsilon P_{\textrm{ext}}
(x,\zeta,J_{2\perp}\xi,Q^{2},\Upsilon) | I_{3} |\,,
\label{eq:defext3D}
\end{equation}
the principal difference being in the limits of integration. 
The signature of the eigenvalues set changes where $I_3$ changes sign. In terms of the invariant variables, $I_3$ is given by
equation~(\ref{eq:I3}). The equation $I_3=0$ that follows from equation~(\ref{eq:I3})
is more instructive if one  writes it
in the form that uses the full 3D Hessian rotation invariants, $J_1$, $J_2$, $J_3$ 
\begin{equation}
J_{1}^{3}-3J_2\left(y,J_{2\perp},Q^2,\Upsilon\right)J_{1}-2J_3\left(y,J_{2\perp}
,Q^2,\Upsilon\right)=0\,,
\label{eq:I3roots}
\end{equation}
where $J_{1}=J_{1\perp}+x_{33}$, $J_{2}=y^{2}+ 3 J_{2\perp}/4 +3 Q^{2}$,
$J_{3}=-y^{3}+9 y\left({ J_{2\perp}}-2Q^{2}\right) /4+ {27}\Upsilon/{4}$ with $y=J_{1\perp}/2-x_{33}$ 
where for this equation only, for the sake of simplicity, variables are not rescaled by their variance.
It can be shown that for any values of $Q^2,J_{2\perp},\Upsilon$ and $y$ in their unrestricted allowed range
there exist three real roots for this cubic polynomial in $J_1$  
that split the integration over $J_1$ in four regions corresponding
to different extrema types. No restriction on other variables arise besides choosing the threshold of the field
value for differential counts.

The integral required to predict extrema counts in 2D is given by 
equation~(\ref{eq:genusformal}) (where $P_{2\textrm{D}}(\tilde x,\tilde x_{1},\tilde x_{2},\tilde\zeta,\tilde \xi,\tilde x_{12})$ is defined in equation~(\ref{eq:PDF2D})) 
and is carried in terms of the field variables, $\tilde x_{11},\tilde x_{12},\tilde x_{22} $,
subject to  constraints on the signs of the  eigenvalues of the Hessian. 

Extrema counts in anisotropic  2 and 3D  spaces do not have closed form expressions (indeed, differential 3D extrema
counts do not lead to analytic results even in the Gaussian limit). Therefore we shall not present here intermediate
expressions for formal expansion, and instead  will be performing the averaging  numerically.
Recall however, that in the rare event limit, $\nu \gg 1$ or $\nu \ll -1$, $n_\mathrm{ max/min}(\nu)\simeq  \chi(\nu)$,
so that in this limit, equations~(\ref{eq:chi2D})  and~(\ref{eq:chi3d}) provide an all-order expansion of the extrema
counts in redshift space in 2 and 3D respectively. 

We can also establish the following general symmetry relations
between extrema counts of different types, valid in a space of arbitrary dimension $N$. Let us label
the extrema type by its signature $S$ which is the sum of the signs of the  eigenvalues of the Hessian that define its type.
This way, $S=N$ for minima ($N$ positive eigenvalues), $S=-N$ for maxima ($N$ negative eigenvalues) and $S$ changes with
a step of two between $2-N$ to $N-2$ for saddle points of different types. Then, if we denote by
$\partial_{\nu} n_{\textrm{S}}^{(n)}$ the contribution to the differential
number count of extrema of type $S$  of order $n$ (where $n=2$ corresponds to the Gaussian term), the following holds
\begin{equation}
\partial_{\nu} n_{\textrm{S}}^{(n)}(\nu) = (-1)^n\partial_{\nu} n_{\textrm{-S}}^{(n)}(-\nu) ~~ .
\end{equation}
These relations allow us to predict the expected behaviour of extrema counts of type $S$ from the measurements of their 
``conjugate'' type $-S$. In particular, the minima counts Gram-Charlier terms are equal to the reflected ($\nu \to -\nu$) maxima counts
for even $n$ (including the Gaussian term), and to minus the reflected maxima counts for   odd $n$ 
(including the first non-Gaussian correction).  
The same relation holds between the ``pancake-like'' ($S=1$) and the ``filament-like'' ($S=-1$)
saddle points in 3D.

Figure~\ref{fig:extrema} illustrates the corresponding extrema distribution for  a set of anisotropic fields 
(Gaussian and  first order non-Gaussian correction) in 2D and 3D.
A comparison with extrema counts measured from random realisations of the fields (scale invariant $n=-1$
 field sampled on $4096^2$ pixels) is also presented there.
  Figure~\ref{fig:2Dgenusangle} demonstrates the angular dependence of the predicted extrema counts
 at NLO in a scale-invariant LSS simulation ($n=-1$, $\sigma=0.18$ in real space).
 
\subsection{Invariance of critical sets}
\subsubsection{ Summary of the Minkowski functionals as functions of $\nu_f$ }\label{sec:nuf}
To put the mathematical results derived in Sections \ref{sec:genus3D}-\ref{sec:N1} on a practical footing in cosmology, let us collect them expressed as
functions of the observable threshold variable $\nu_f$. 
The $\nu\rightarrow \nu_f$ remapping is astrophysically motivated by the fact that typically,
 the amplitude of the field is not known, as it may depend on e.g. the bias factor, whereas  
$\nu_f$ can be measured\footnote{Note that the choice of $\nu_f$ as an invariant parametrization of the  Minkowski functionals is not unique. One could plot   Minkowski as a function of e.g. $\nu_L$,
the threshold corresponding to a given fraction of the total skeleton length as an alternative construction.}.
 The value of $\nu_f$ is obtained by  inverting equation~(\ref{eq:degauss}) for the filling factor, $f_V$,
and using it instead of the difficult-to-determine $\nu$, to effectively ``Gaussianize'' the PDF of the field. Indeed,
when the transformation $\nu \to \nu_f$ in equation~(\ref{eq:nufromnuf}) is applied to perturbative results,
the most oscillatory  $\nu$ modes, that are proportional solely to the cumulants of the field, are eliminated. 

In analogy with the skewness parameter $S_3 = \langle x^3 \rangle / \sigma$, we introduce two other scaled cumulants that
involve the derivatives of the field. For isotropic fields,  they are $T_3 = \langle x q^2 \rangle/\sigma$ and 
$U_3 = -\langle J_1 q^2\rangle /(\gamma\sigma)$.  For anisotropic fields, we use their partial versions $T_{3\perp} = \langle x q_\perp^2 \rangle/\sigma$, 
$T_{3\parallel} = \langle x q_\parallel^2 \rangle/\sigma$ and 
$U_{3\perp} = -\langle J_{1\perp} q_\perp^2\rangle/(\gamma_\perp\sigma)$, 
$U_{3\parallel} = -\langle J_{1\perp} q_\parallel^2\rangle/(\gamma_\perp\sigma)$.
In the isotropic limit these partial contributions are summed according to the following rules
$T_3 = \frac{1}{3} \left( 2 T_{3\perp} + T_{3\parallel}\right) $ and 
$U_3 = \frac{4}{9} \left( U_{3\perp} + U_{3\parallel} \right)$ \footnote{To see this, we use the relations between
isotropic and anisotropic variances and correlation parameters from Appendix~\ref{sec:variance_relations} as well
as the following identities: $\gamma_\perp \langle J_{1\parallel} q_\perp^2 \rangle =   \gamma_\parallel \langle J_{1\perp} q_\parallel^2 \rangle$
and $\langle J_{1\parallel}q_\parallel^2 \rangle = 0$}.
It is then straightforward to rewrite our statistics, e.g. the 3D Euler characteristic, as a function of $\nu_f$
\begin{equation}
\label{eq:chi3D-nuf}
\chi_\mathrm{{ 3D}}(\nu_{f})=\frac {\sigma_{1\parallel}\sigma_{1\perp}^{2}}
{\sigma^{3}}\frac {e^{-\nu_{f}^{2}/2}} {8\pi^{2}}
 \left(H_{2}(\nu_{f})+\sigma\left[ \left( T_{3\perp} + \frac{1}{2} T_{3\parallel} - S_3 \right) H_{3}(\nu_{f})
+ \left( U_{3\perp} + U_{3\parallel} - S_3 \right) H_{1}(\nu_{f})\right]
+\cdots\right)\,.
\end{equation}
Note that in contrast to the isotropic case,  in redshift space the coefficient of $H_1$, $ \tilde S^{(1)}_{z}= U_{3\perp} + U_{3\parallel} - S_3$, is  non-zero for scale-invariant
power spectra.
For the other statistics, we get:\\
 the area of 3D isocontours
 \begin{equation}
 \label{N3-nuf}
{\cal N}_{3}(\nu_{f})
=  {\cal N}_{3}^{(0)}(\nu_{f}) \left(
1
+\sigma H_1(\nu_{f})\left[
 \frac{1}{2} \vphantom{\frac{T}{\beta_\sigma}}
 \left(1 - \frac{A(\beta_\sigma)/\beta_\sigma}{ 1 - A(\beta_\sigma)} \right) 
T_{3\perp}
+  \frac 12 
 \left( \frac{A(\beta_\sigma)/\beta_\sigma}{ 1 - A(\beta_\sigma)} \right)
T_{3\parallel}
 -S_{3}\right]
 +\cdots
 \right)\,,
\end{equation}
the 2D Euler characteristic
 \begin{equation}
\chi_{2\textrm{D}}(\nu_{f},\theta_{\cal S})
=\frac{\tilde \sigma_{1\perp}\tilde\sigma_{1\parallel}}{\sigma^{2}}\frac {e^{-\nu_{f}^{2}/2}} {(2\pi)^{3/2}}
\left(H_1\left(\nu_{f}\right) 
 +  \sigma\left[H_{2}(\nu_{f})\left(\frac 1 2\frac{\left\langle \tilde x \left[{\tilde x_{1}}^{2}+{\tilde x_{2}}^{2}   \right]\right\rangle}{\sigma}-S_{3}\right)
+\left(-\frac 1 {2 {\tilde \gamma_{\perp}}}\frac{\left\langle {\tilde x_{11}} {\tilde x_{2}}^{2}\right\rangle}{\sigma}-S_{3}\right)\right]
+\cdots
\right)\,,
 \end{equation}
the length of 2D isocontours
\begin{equation}
 \label{N2-nuf}
{\cal N}_{2}(\nu_{f},\theta_{\cal S})
= {\cal N}_2^{(0)}(\nu_{f})
\left(
1
+\sigma H_{1}(\nu_{f}) \left[  \frac{1}{2}\frac{ \left\langle \tilde x \tilde x_1^{2}\right\rangle}{\sigma} B(\tilde \beta_\sigma)
+ \frac{1}{2}\frac{ \left\langle \tilde x \tilde x_{2}^{2}\right\rangle}{\sigma} \left( 1 - B(\tilde \beta_\sigma) \right) -S_{3}\right]
+\cdots
\right)
\,,
\end{equation}
and the frequency of contour crossings
\begin{equation}
 \label{eq:N1-nuf}
{\cal N}_{1}(\nu_{f},\theta_{\cal L})
={\cal N}_1^{(0)}(\nu_{f})\left(
1+\sigma
H_{1}(\nu_{f}) \left[\frac{1}{2}\frac{(1-\beta_\sigma)\sin^{2}\theta_{\cal L}T_{3\perp}+\cos^{2}\theta_{\cal L}
T_{3\parallel}}{1 - \beta_\sigma \sin^{2}\theta_{\cal L}}-S_{3}
\right]
+\cdots
\right)
\,.
\end{equation}
Note that $ \left\langle \tilde x \tilde x_1^{2}\right\rangle$, $ \left\langle \tilde x \tilde x_1^{2}\right\rangle$ and $\left\langle\tilde x_{11} \tilde x_2^{2}\right\rangle$ are functions of $\theta_{\cal S/\cal L}$, $T_{3\perp}$, $T_{3\parallel}$, $U_{3\perp}$ and $U_{3\parallel}$.
Using  $\nu_f$ as a threshold variable makes explicit  the invariance of Minkowski functionals and extrema statistics under
monotonic transformation of the field.  Indeed, since the filling factor $f_V$ is one of the Minkowski functionals itself,
$\nu_f$ is strictly unchanged under monotonic transformation. Thus the  other statistics, described by functions of $\nu_f$, are invariant.  
\begin{figure}
 \begin{center}
    \includegraphics[width=0.45\textwidth]{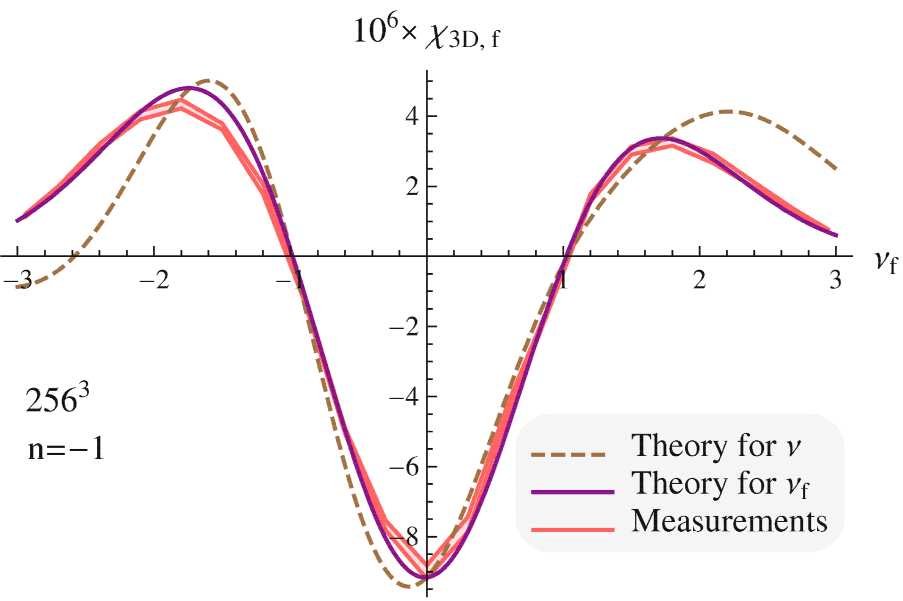}
  \includegraphics[width=0.45\textwidth]{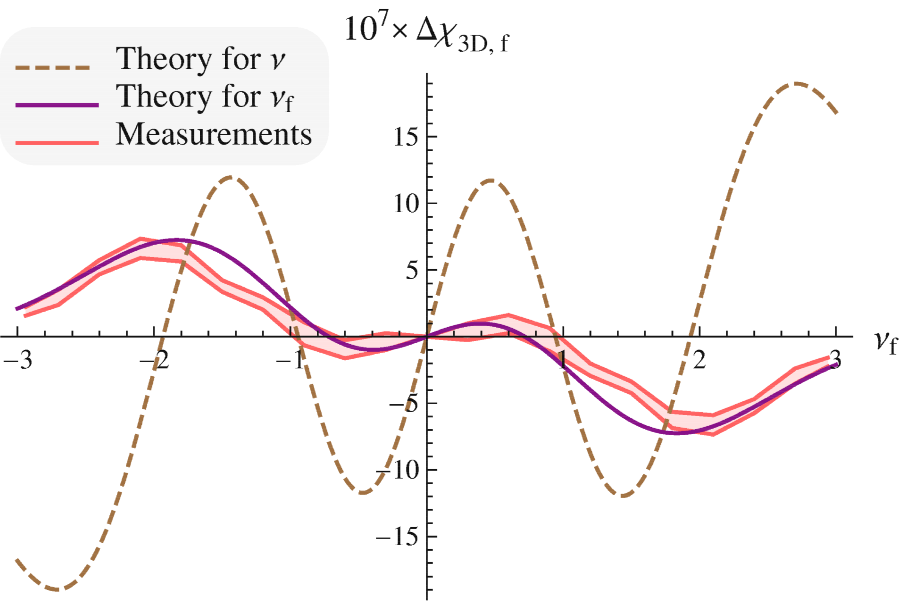}
   \caption{
\emph{Left panel} : 3D Euler characteristic as a function of $\nu_f$ in 19 scale-invariant LSS simulation ($n=-1, 256^{3}, \sigma=0.18$).
    Prediction to first order in non-Gaussianity is displayed in solid purple line.  The displayed error is the error on the median of the measurements (red shaded area). 
    For comparison, the 3D Euler characteristic as a function of $\nu$ is displayed in dashed brown line.
 \emph{Right panel}: same as left panel for the first non-Gaussian correction only compared to the odd part of the measurements (including the first order correction). 
    In this mildly non-linear regime, the agreement between measurements and theory at leading and next-to-leading order is very good.  Note how the use of $\nu_{f}$ instead of $\nu$ ``gaussianizes'' the \genus. 
    \label{fig:minkowski-fillingfactor}
}
 \end{center}
\end{figure}

Figure~\ref{fig:minkowski-fillingfactor} reproduces the 3D Euler characteristic  of Figure~\ref{fig:CGvsRDS} but as a function of the filling factor threshold, $\nu_f$,
defined in Section~\ref{sec:filling}. Agreement between prediction truncated at NLO  and measurements is very good for contrasts $\nu\gtrsim -1.5$. The effect of using $\nu_{f}$ is to gaussianize the PDF as seen on Figure~\ref{fig:minkowski-fillingfactor} : denser regions are brought to the center whereas wider regions are pushed to the outside.

\subsubsection{ Formal invariance of  Minkowski functionals  w.r.t. monotonic transformation} 
\label{sec:invariance}
From Section~\ref{sec:minkowski}, it is straightforward to see that any Minkowski functional as well as extrema counts
are invariant under a local monotonic transformation. This property should also be encoded in the Gram-Charlier expansions
of these Minkowski functionals given in Section~\ref{sec:nuf}.  Indeed the combinations
of cumulants which appear at  first order in the Gram-Charlier expansion are invariant under local monotonic
transformation taken to the same first order in $\sigma$.
Let us illustrate this on the 3D Euler characteristic (given that the same proof can be developed for any Minkowski functionals). Following \cite{Matsubara},
let us henceforth study how cumulants evolve under the transformation: $y\rightarrow b \,x + {b_{2}}/{2}\left(x^{2}-\left\langle x^{2}\right\rangle\right)$,  which represents the local representation of  any analytic function of the density field when Taylor expanded.
Let us first show how $\left\langle x^{3}\right\rangle$ evolve under such transformation
\begin{eqnarray}
\left\langle y^{3}\right\rangle&=&b^{3}\left\langle x^{3}\right\rangle+\frac 3 2 b^{2}b_{2}\left(\left\langle x^{4}\right\rangle-\left\langle x^{2}\right\rangle^{2}\right)+\frac 3 4 b\, b_{2}^{2}\left(\left\langle x^{5}\right\rangle-2\left\langle x^{3}\right\rangle\left\langle x^{2}\right\rangle  \right)
\,,\nonumber\\&=&b^{3}\left\langle x^{3}\right\rangle_{c}+\frac 3 2 b^{2}b_{2}\left(\left\langle x^{4}\right\rangle_{c}+2\left\langle x^{2}\right\rangle_{c}^{2}\right)+\frac 3 4 b\, b_{2}^{2}\left(\left\langle x^{5}\right\rangle_{c}+8\left\langle x^{3}\right\rangle_{c}\left\langle x^{2}\right\rangle_{c}  \right)\,.\label{eq:y3int}
\end{eqnarray}
Selecting only the first order term in the PT expansion, the classical expression for the skewness follows from Equation (\ref{eq:y3int})
\begin{equation}
\left\langle y^{3}\right\rangle
=b^{3}\left\langle x^{3}\right\rangle_{c}+3  b^{2}b_{2}\left\langle x^{2}\right\rangle_{c}^{2}+\textrm{higher order terms }\,.
\label{eq:y3}
\end{equation}
The same construction for other moments  and cumulants leads to the following relationships
\begin{subequations}
\begin{align}
\left\langle y^{2}\right\rangle&=b^{2}\left\langle x^{2}\right\rangle+\textrm{higher order terms }\,,\\
\left\langle \nabla_{\perp} y\cdot \nabla_{\perp} y \right\rangle&=b^{2}\left\langle  \nabla_{\perp} x\cdot \nabla_{\perp} x\right\rangle+\textrm{higher order terms }\,,\\
\left\langle \nabla_{\parallel} y\cdot \nabla_{\parallel} y \right\rangle&=b^{2}\left\langle  \nabla_{\parallel} x\cdot \nabla_{\parallel} x\right\rangle+\textrm{higher order terms }\,,\\
\left\langle y (\nabla_{\perp} y)^{2}\right\rangle&=b^{3}\left\langle x (\nabla_{\perp} x)^{2}\right\rangle+2b^{2}b_{2}\left\langle x^{2}\right\rangle\left\langle( \nabla_{\perp} x)^{2} \right\rangle+\textrm{higher order terms }\,,\\
\left\langle y (\nabla_{\parallel} y)^{2}\right\rangle&=b^{3}\left\langle x (\nabla_{\parallel} x)^{2}\right\rangle+2b^{2}b_{2}\left\langle x^{2}\right\rangle\left\langle( \nabla_{\parallel} x)^{2} \right\rangle+\textrm{higher order terms }\,,\\
\left\langle( \nabla_{\perp} y\cdot \nabla_{\perp} y )\Delta_{\perp} y\right\rangle&=b^{3}\left\langle( \nabla_{\perp} x\cdot \nabla_{\perp} x )\Delta_{\perp} y\right\rangle-b^{2}b_{2}\left\langle( \nabla_{\perp} x)^{2}\right\rangle^{2}+\textrm{higher order terms }\,,\\
\left\langle( \nabla_{\parallel} y\cdot \nabla_{\parallel} y )\Delta_{\perp} y\right\rangle&=
b^{3}\left\langle( \nabla_{\parallel} x\cdot \nabla_{\parallel} x )\Delta_{\perp} y\right\rangle
-2b^{2}b_{2}\left\langle( \nabla_{\perp} x)^{2}\right\rangle\left\langle( \nabla_{\parallel} x)^{2}\right\rangle+\textrm{higher order terms }\,,
\end{align}
\end{subequations}
so that for $b\neq 0 $ at leading order
\begin{eqnarray}
\tilde S^{(1)}_{\chi_{\textrm{3D}},z}(y)&\equiv&\frac{1}{\left\langle y^{2}\right\rangle}\left(
\frac{\left\langle y (\nabla_{\perp} y)^{2}\right\rangle}{\left\langle(\nabla_{\perp} y)^{2} \right\rangle}
+\frac{\left\langle y (\nabla_{\parallel} y)^{2}\right\rangle}{2\left\langle(\nabla_{\parallel} y)^{2} \right\rangle}-
\frac{\left\langle y^{3}\right\rangle}{\left\langle y^{2}\right\rangle}\right)=\tilde S^{(1)}_{\chi_{\textrm{3D}},z}(x)\,,
\\
\tilde S^{(-1)}_{\chi_{\textrm{3D}},z}(y)&\equiv&\frac{1}{\left\langle y^{2}\right\rangle}\left(
 -\frac{\left\langle\Delta_{\perp} y (\nabla_{\perp} y)^{2}\right\rangle}{\left\langle(\nabla_{\perp} y)^{2} \right\rangle}-
 \frac{\left\langle \Delta_{\perp}y (\nabla_{\parallel} y)^{2}\right\rangle}{\left\langle(\nabla_{\parallel} y)^{2} \right\rangle}
 -\frac{\left\langle y^{3}\right\rangle}{\left\langle y^{2}\right\rangle}
  \right)=\tilde S^{(-1)}_{\chi_{\textrm{3D}},z}(x)\,,
\end{eqnarray}
where we denote $\tilde S^{(i)}_{\chi_{\textrm{3D}},z}$ the coefficients in front of the Hermite polynomials in equation (\ref{eq:chi3D-nuf}) (see also equation~(\ref{eq:2Dstatform})). 
 This shows that the combinations of cumulants in front of the Hermite polynomials are invariant under local monotonic transformation and so are the Minkowski functionals. In particular, it demonstrates formally why  if the bias is local and monotonic, the Minkowski functionals are bias-independent.

\section{Application to large-scale structure in redshift space} \label{sec:cosmo}
In the context of cosmology, it is of interest to understand what kind of constraints on the cosmological parameters can be drawn from
the prediction of critical sets in redshift space. Redshift space distortion can be viewed as a nuisance,  but in fact potentially opens new prospects given the broken 
symmetry induced by the kinematics. 
\subsection{Estimating cumulants of the field} \label{sec:5.1}
Table~\ref{tab:cumulants} summarizes the cubic cumulants that determine  the Minkowski functionals studied in this paper
to first non-Gaussian order.
In turn, the combinations of these cumulants is what can be measured by fitting the correspondent functionals with low-order
Hermite modes.
\begin{table}
\begin{displaymath}
\begin{array}{c|c|c|c|c}
&H_{0}&H_{1}&H_{2}&H_{3}\\
\hline \chi_\mathrm{{3D},f}&-&\left\langle  J_{1\perp}{q^2_{\perp}}\right\rangle,\left\langle  J_{1\perp}x_{3}^{2} \right\rangle,\left\langle  x^{3}\right\rangle&-&\left\langle x q_{\perp}^{2}\right\rangle,\left\langle x x_{3}^{2}\right\rangle,\left\langle x^{3}\right\rangle\\
\chi_\mathrm{{2D},f}&\left\langle  J_{1\perp}{q^2_{\perp}}\right\rangle,\left\langle  J_{1\perp}x_{3}^{2} \right\rangle,\left\langle  x^{3}\right\rangle&-&\left\langle x q_{\perp}^{2}\right\rangle,\left\langle x x_{3}^{2}\right\rangle,\left\langle x^{3}\right\rangle&-\\
{\cal N}_{3,f}&-&\left\langle x q_{\perp}^{2}\right\rangle,\left\langle x x_{3}^{2}\right\rangle,\left\langle x^{3}\right\rangle&-&-\\
{\cal N}_{2,f}&-&\left\langle x q_{\perp}^{2}\right\rangle,\left\langle x x_{3}^{2}\right\rangle,\left\langle x^{3}\right\rangle&-&-\\
{\cal N}_{1,f}&-&\left\langle x q_{\perp}^{2}\right\rangle,\left\langle x x_{3}^{2}\right\rangle,\left\langle x^{3}\right\rangle&-&-
\end{array}
\end{displaymath}
\caption{ The cubic cumulants that determine  the Minkowski functionals studied in this paper
to first non-Gaussian order.\label{tab:cumulants}
}
\end{table}
Two main group of cumulants that Minkowski functionals (to first order) give access to are
$\left\langle x q_{\perp}^{2}\right\rangle,\left\langle x x_{3}^{2}\right\rangle$, which relate the field to its 
gradient, and $\left\langle  J_{1\perp}{q^2_{\perp}}\right\rangle,\left\langle  J_{1\perp}x_{3}^{2} \right\rangle$ which
relate the gradient to the Hessian. Note that $\langle x^3 \rangle$ is not accessible as the amplitude of an 
independent Hermite mode if Minkowski functionals are studied as functions of the filling factor threshold $\nu_f$.
Rather it offsets the modes defined by the other two groups of cumulants. 
Redshift space analysis is in principle capable
of mining more information  than real space analysis.
Indeed, in redshift space there is a  qualitative difference between cumulants that involve the LOS direction and those that
involve directions orthogonal to the LOS in the plane of the sky. These differences encode information about  velocities, and
reflect the mechanism of how these velocities originated. In principle, estimating the anisotropic part of such 
cumulants can be used to test the theory of gravity    in the context  of  large-scale structures perturbation
theory. 3D geometrical statistics, such as $\chi_\mathrm{ 3D}$ and ${\cal N}_3$ do not allow by themselves to
determine separately the LOS and sky cumulants. To separate anisotropic contributions one must analyse  {\sl slices} of
3D volume at different angles $\theta_{\cal S}$ to the LOS\footnote{Another related approach would be to study the  2D slices orthogonal to LOS field  with  variable LOS thickness.
This latter technique was successfully used to study ISM turbulence \citep{LazPog00}.}.
For instance, measuring the length of the isocontours, ${\cal N}_2$,
(with possible cross-check from $\chi_\mathrm{ 2D}$) yields a separate handle on
 $\left\langle x q_{\perp}^{2}\right\rangle$ and $\left\langle x x_{3}^{2}\right\rangle$,
while the additional analysis of the 2D Euler characteristic $\chi_\mathrm{ 2D}(\nu_f)$, as a function  of $\theta_{\cal S}$  (via the 
baseline offset $\propto H_0$, see figure~\ref{fig:2Dgenusangle})   allows us to measure
$\left\langle  J_{1\perp}{q^2_{\perp}}\right\rangle$ and $\left\langle  J_{1\perp}{x_{3}^{2}}\right\rangle$. This procedure is further discussed 
in Section~\ref{sec:measuringf} below.

\subsection{Which modes of the bispectrum are geometrical statistics probing?}
\label{sec:bispectrum}
This paper is concerned with geometric probes operating in configuration space. On the other hand, a fair amount of theoretical predictions (such as PT) for the growth of structure are best described in  Fourier space. It is therefore of interest to relate the two and characterize which feature of the multi-spectra these probes constrain. For instance, it is straightforward to 
show via Fourier transform that the third order cumulant, $ \left\langle x^{3}\right \rangle$, can be expressed as a double sum over the anisotropic bispectrum, $B_z(k_1,k_2,k_3)$ via 
\[  \left\langle x^{3}\right \rangle=  \displaystyle\frac{1}{\sigma^{3}}\int\frac{ \dd^{3}k_{1}}{(2\pi)^{3}}\frac{\dd^{3}k_{2}}{(2\pi)^{3}}B(k_{1},k_{2},|\mathbf {k_{1}+k_{2}} |)\,.
\] Measuring $\left\langle x^{3}\right \rangle$ amounts to constraining the monopole of the bispectrum.
 Similarly,  other geometric cumulants involve different weights (see Table~\ref{table:Stilde}), while   higher order cumulants will involve $k$ integrals of multi-spectra.
For instance, the Euler characteristic to all order given in equation~(\ref{eq:chi3Dfinalallorder}), involves $n^{\rm th}= i+2j+2m$  moments, $\left\langle  x^{i} q^{2j}_{\perp}  x_3^{2m}   \right\rangle$ which can be re-expressed via the $n^{\rm th}$ order multi-spectrum, $B^{n}_z(k_{1},\cdots ,k_{n-1},|\mathbf{k}_{1}+\cdots \mathbf{k}_{n-1} |)$  as 
\[\left\langle  x^{i} q^{2j}_{\perp}  x_3^{2m}   \right\rangle=
  \displaystyle\frac{1}{\sigma^{i} \sigma^{2j}_{1\perp}\sigma^{2m}_{1\parallel}}\int\frac{ \dd^{3}k_{1}}{(2\pi)^{3}}\cdots \frac{\dd^{3}k_{n-1}}{(2\pi)^{3}}B^{n}_z(k_{1},\cdots ,k_{n-1},|\mathbf{k}_{1}+\cdots \mathbf{k}_{n-1} |)
 \prod_{j_1\le j} |\mathbf {k}_{j_1\perp} |^{2} 
 \prod_{j<m_1\le m+j} |\mathbf {k}_{m_1\parallel} |^{2}
  \,.\]

 Let us first show explicitly how the first order corrections of the 3D Euler characteristic can be re-expressed in terms of the  underlying   bispectrum in redshift space, $B_z$. 
 Equation (\ref{eq:chi3D-nuf}) shows that it only depends at first order in non-Gaussianity on two numbers (the coefficients in front of the two Hermite polynomials):
$\sigma\tilde S^{(1)}_{\chi_{\textrm{3D}},z}=\left\langle x(q_{\perp}^{2}+x_{3}^{2}/2)\right\rangle-\left\langle x^{3}\right\rangle$ and $\sigma \tilde S^{(-1)}_{\chi_{\textrm{3D}},z}=-\left\langle J_{1\perp}(q_{\perp}^{2}+x_{3}^{2})\right\rangle/\gamma_{\perp}-\left\langle x^{3}\right\rangle$. These quantities can in turn be expressed as special combinations of the underlying bispectrum using Table \ref{table:Stilde}
\begin{eqnarray}
\label{eq:tS1}
\tilde S^{(1)}_{\chi_{\textrm{3D}},z}&=&\frac{1}{\sigma^{4}}\int\frac{ \dd^{3}k_{1}}{(2\pi)^{3}}\frac{\dd^{3}k_{2}}{(2\pi)^{3}}\left(\frac{\sigma^{2}|\mathbf {k_{1\perp}+k_{2\perp}} |^{2}}{2\sigma_{1\perp}^{2}}+\frac{\sigma^{2}|\mathbf {k_{1\parallel}+k_{2\parallel}} |^{2}}{2\sigma_{1\parallel}^{2}}-1\right)B_z(k_{1},k_{2},|\mathbf {k_{1}+k_{2}} |)\,,
\\
\label{eq:tS2}
\tilde S^{(-1)}_{\chi_{\textrm{3D}},z}&=&-\frac{1}{\sigma^{4}}\int\frac{ \dd^{3}k_{1}}{(2\pi)^{3}}\frac{\dd^{3}k_{2}}{(2\pi)^{3}}\left(\frac{\sigma^{4}|\mathbf {k_{1\perp}+k_{2\perp}} |^{2}(\mathbf k_{1\perp}\cdot \mathbf k_{2\perp})}{\sigma_{1\perp}^{4}}+\frac{\sigma^4|\mathbf {k_{1\perp}+k_{2\perp}} |^{2}(\mathbf k_{1\parallel}\cdot \mathbf k_{2\parallel})}{\sigma_{1\perp}^{2}\sigma_{1\parallel}^{2}}+1\right)B_z(k_{1},k_{2},|\mathbf {k_{1}+k_{2}} |)\,.
\end{eqnarray}
The parenthesis in equations~(\ref{eq:tS1})-(\ref{eq:tS2}) define  ``projectors'' for the bispectrum, $(\tilde S^{(1)}_{\chi_{\textrm{3D}},z})_B$
and $(\tilde S^{(-1)}_{\chi_{\textrm{3D}},z})_B$, so that
  \[\tilde S^{(i)}_{\chi_{\textrm{3D}},z}=\frac{1}{\sigma^{4}}\int\frac{ \dd^{3}k_{1}}{(2\pi)^{3}}\frac{\dd^{3}k_{2}}{(2\pi)^{3}}\ B_z(k_{1},k_{2},|\mathbf {k_{1}+k_{2}} |)\left(\tilde S^{(i)}_{\chi_{\textrm{3D}},z}\right)_B . \]
\begin{table}
\begin{center}
$\begin{array}{|c|c|}
\hline
  &\\
    \left\langle x^{3}\right \rangle & \displaystyle \displaystyle\frac{1}{\sigma^{3}}\int\frac{ \dd^{3}k_{1}}{(2\pi)^{3}}\frac{\dd^{3}k_{2}}{(2\pi)^{3}}B(k_{1},k_{2},|\mathbf {k_{1}+k_{2}} |) \\
         \left\langle x q_{\perp}^{2}\right \rangle &\displaystyle \frac{1}{2\sigma\sigma_{1\perp}^{2}}\int\frac{ \dd^{3}k_{1}}{(2\pi)^{3}}\frac{\dd^{3}k_{2}}{(2\pi)^{3}}|\mathbf {k_{1\perp}+k_{2\perp}} |^{2}B(k_{1},k_{2},|\mathbf {k_{1}+k_{2}} |) \\
         \left\langle x q_{\parallel}^{2}\right \rangle &\displaystyle\frac{1}{2\sigma\sigma_{1\parallel}^{2}}\int\frac{ \dd^{3}k_{1}}{(2\pi)^{3}}\frac{\dd^{3}k_{2}}{(2\pi)^{3}}|\mathbf {k_{1\parallel}+k_{2\parallel}} |^{2}B(k_{1},k_{2},|\mathbf {k_{1}+k_{2}} |) \\  
         \left\langle J_{1\perp} q_{\perp}^{2}\right \rangle  &\displaystyle-\frac{1}{\sigma_{2\perp}\sigma_{1\perp}^{2}}\int\frac{ \dd^{3}k_{1}}{(2\pi)^{3}}\frac{\dd^{3}k_{2}}{(2\pi)^{3}}|\mathbf {k_{1\perp}+k_{2\perp}} |^{2}(\mathbf k_{1\perp}\cdot \mathbf k_{2\perp})B(k_{1},k_{2},|\mathbf {k_{1}+k_{2}} |)\\
 \left\langle J_{1\perp} q_{\parallel}^{2}\right \rangle  &\displaystyle-\frac{1}{\sigma_{2\perp}\sigma_{1\parallel}^{2}}\int\frac{ \dd^{3}k_{1}}{(2\pi)^{3}}\frac{\dd^{3}k_{2}}{(2\pi)^{3}}|\mathbf {k_{1\perp}+k_{2\perp}} |^{2}(\mathbf k_{1\parallel}\cdot \mathbf k_{2\parallel})B(k_{1},k_{2},|\mathbf {k_{1}+k_{2}} |)\\
   &\\ \hline
\end{array}$
\end{center}
  \caption{\label{table:Stilde} Rescaled third order cumulants written in terms of the bispectrum.}
\end{table}
In the isotropic limit, theses  projectors becomes respectively \citep{Matsubara}
\begin{equation}
\left(\tilde S^{(1)}_{\chi_{\textrm{3D}},z}\right)_B=\left(\frac 3 2 \frac{\sigma^{2}|\mathbf {k_{1}+k_{2}} |^{2}}{2\sigma_{1}^{2}}-1\right)\,,
\quad \textrm{and} \quad \left(\tilde S^{(-1)}_{\chi_{\textrm{3D}},z}\right)_B=
\left(\frac 9 4\frac{\sigma^{4}|\mathbf {k_{1}+k_{2}} |^{2}(\mathbf k_{1}\cdot \mathbf k_{2})}{\sigma_{1}^{4}}+1\right)\,. \label{eq:defprojiso}
\end{equation}
A multipole expansion of $B_z$ and equation~(\ref{eq:defprojiso}) w.r.t. $\mu=\mathbf k_{1}\cdot \mathbf k_{2}/(k_1k_2) $ shows for instance that neither $\tilde S^{(1)}_{\chi_{\textrm{3D}}}$ nor $\tilde S^{(-1)}_{\chi_{\textrm{3D}}}$
would constrain harmonics of $B_z$ larger than three.
For the 2D Euler characteristic, the corresponding projectors read (with $\beta_\sigma$ defined in equation~(\ref{eq:defbeta}))
\begin{eqnarray}
\label{eq:tS1-2D}
\left(\tilde S^{(1)}_{\chi_{\textrm{2D}},z}\right)_B\!\!&=&\!\!\left(1-\frac{\cos^{2}\theta_{\cal S}}{2(1-\beta_{\sigma}\sin^{2}\theta_{\cal S})}\right)\frac{\sigma^{2}|\mathbf {k_{1\perp}+k_{2\perp}} |^{2}}{2\sigma_{1\perp}^{2}}+\frac{\cos^{2}\theta_{\cal S}}{2(1-\beta_{\sigma}\sin^{2}\theta_{\cal S})}\frac{\sigma^{2}|\mathbf {k_{1\parallel}+k_{2\parallel}} |^{2}}{2\sigma_{1\parallel}^{2}}-1\,,
\\
\label{eq:tS2-2D}
\left(\tilde S^{(-1)}_{\chi_{\textrm{2D}},z}\right)_B\!\!&=&\!\!\left(1-\frac{\cos^{2}\theta_{\cal S}}{1-\beta_{\sigma}\sin^{2}\theta_{\cal S}}\right)\frac{\sigma^{4}|\mathbf {k_{1\perp}+k_{2\perp}} |^{2}(\mathbf k_{1\perp}\cdot \mathbf k_{2\perp})}{\sigma_{1\perp}^{4}}+\frac{\cos^{2}\theta_{\cal S}}{2(1-\beta_{\sigma}\sin^{2}\theta_{\cal S})}\frac{\sigma^4|\mathbf {k_{1\perp}+k_{2\perp}} |^{2}(\mathbf k_{1\parallel}\cdot \mathbf k_{2\parallel})}{\sigma_{1\perp}^{2}\sigma_{1\parallel}^{2}}-1\,.
\end{eqnarray}
An interesting feature of equations~(\ref{eq:tS1})-(\ref{eq:tS2}) is that the projectors are now parametric and depend on $\theta_{\cal S}$.
Via  slicing planes  e.g. along and perpendicular to the line of sight it is therefore possible to measure projections of the bispectrum along $|\mathbf {k_{1\perp}+k_{2\perp}} |^{2}$,
$|\mathbf {k_{1\parallel}+k_{2\parallel}} |^{2}$, 
$|\mathbf {k_{1\perp}+k_{2\perp}} |^{2}(\mathbf k_{1\perp}\cdot \mathbf k_{2\perp})$ and $|\mathbf {k_{1\perp}+k_{2\perp}} |^{2}(\mathbf k_{1\parallel}\cdot \mathbf k_{2\parallel})$
independently.

Given equations~(\ref{N3-nuf}),(\ref{N2-nuf}) and (\ref{eq:N1-nuf}), one can  also easily recover the projectors for  ${\cal N}_{1,f}$,  ${\cal N}_{2,f}$ and  ${\cal N}_{3,f}$ at first order in non-Gaussianity as \begin{eqnarray}
\left(\tilde S^{(1)}_{N_{3},z}\right)_B &=&
\frac{\sigma^{2}}{4\sigma_{1\perp}^{2}} \vphantom{\frac{T}{\beta_\sigma}} \left(1 - \frac{A(\beta_\sigma)/\beta_\sigma}{ 1 - A(\beta_\sigma)} \right)|\mathbf {k_{1\perp}+k_{2\perp}} |^{2} + 
 \frac {\sigma^{2}}{4\sigma_{1\parallel}^{2}} \left( \frac{A(\beta_\sigma)/\beta_\sigma}{ 1 - A(\beta_\sigma)} \right)|\mathbf {k_{1\parallel}+k_{2\parallel}} |^{2} 
 -1\,, 
 \label{eq:SN3B}
\\
\left(\tilde S^{(1)}_{N_{2},z}(\theta_{\cal S})\right)_B&=& \frac{\sigma^{2}}{4\sigma_{1\perp}^{2}}B(\tilde \beta_\sigma)|\mathbf {k_{1\perp}+k_{2\perp}} |^{2} 
+ \frac{1}{4}\sigma^{2}\frac{\sin ^{2}\theta_{\cal S}|\mathbf {k_{1\perp}+k_{2\perp}} |^{2} +2\cos ^{2}\theta_{\cal S}|\mathbf {k_{1\parallel}+k_{2\parallel}} |^{2} }{\sin ^{2}\theta_{\cal S}\sigma_{1\perp}^{2}+2\cos ^{2}\theta_{\cal S}\sigma_{1\parallel}^{2}} \left( 1 - B(\tilde \beta_\sigma) \right) -1\,,  \label{eq:SN2B}
\\
\left(\tilde S^{(1)}_{N_{1},z}(\theta_{\cal L})\right)_B&=&\frac{1}{4}\frac{(1-\beta_\sigma)\sin^{2}\theta_{\cal L}|\mathbf {k_{1\perp}+k_{2\perp}} |^{2}\sigma^{2}/\sigma_{1\perp}^{2}+\cos^{2}\theta_{\cal L}|\mathbf {k_{1\parallel}+k_{2\parallel}} |^{2}\sigma^{2}/\sigma_{1\parallel}^{2}}{1 - \beta_\sigma \sin^{2}\theta_{\cal L}}-1
\,, \label{eq:SN1B}
\end{eqnarray}
where the last two are also parametric in $\theta_{\cal L}$ and $\theta_{\cal S}$ resp. Note that equations~(\ref{eq:SN3B})-(\ref{eq:SN1B})  formally yield no new 
projector, compared to equation~(\ref{eq:tS1-2D}), though the weighting  differs and might be more favorable for noisy datasets.
As no closed form for the extrema counts was found it is not possible to extend this analysis to their cumulants.
\subsection{Predicting cumulants using gravitational  Perturbation Theory}
\label{sec:PT}
In the previous section, no assumption was made on the shape of the anisotropic bispectrum, $B_z$. Let us now turn to the context of gravitational clustering in redshift space and 
start with a rapid overview of the relevant theory.
The fully non-linear expression (generalizing equation~(\ref{eq:Kaiser_mass}))
 for the Fourier transform of the density in redshift space is \citep{Scoccimarro99,Bernardeau}
\begin{equation}
\label{eq:zdensity} \hat
\delta_{s}(\mathbf{k})=\int \frac{\dd^{3}\mathbf{x}}{(2\pi)^{3}}e^{-i\mathbf{k}\cdot \mathbf{x}}e^{ifk_{z}v_{z}(\mathbf{x})}\left[\delta^{( r )}(\mathbf{x})+ f\nabla_{z}v_{z}(\mathbf{x})  \right]\,, 
\end{equation}  with $v_z$ the peculiar velocity along the LOS,  $f= \dd \log D/\dd \log a $,
while    assuming  the plane-parallel approximation and  that only $f\nabla_{z}v_{z}(\mathbf{x}) <1$ terms
contribute\footnote{Note that equation~(\ref{eq:zdensity}) could be more accurately replaced by
$\hat \delta_{s}(\mathbf{k})=-\delta_\mathrm{ D}(\mathbf{k})+1/(2\pi)^{3}\int {\dd^{3}\mathbf{x}}{}e^{-i\mathbf{k}\cdot \mathbf{x}}e^{ifk_{z}v_{z}(\mathbf{x})}\left[\delta^{( r )}(\mathbf{x})+1 \right]$
where no assumption about the amplitude of the radial velocity is made. This gives exactly the same perturbation theory as expected.
}.
\subsubsection{Derivation of geometrical cumulants for standard gravitational clustering}
Expanding the exponential in equation~(\ref{eq:zdensity}),  leads to, using the kernels $Z_{n}$,  the following expression for the density field in redshift space \citep{Verde98,Scoccimarro99,Bernardeau}
\begin{equation}
\hat\delta_{s}(\mathbf{k},\tau)=\sum_{n=1}^{\infty}D_{1}^{n}(\tau)\int\dd^{3}\mathbf{k}_{1}\cdots\int\dd^{3}\mathbf{k}_{n}\delta_\mathrm{ D}(\mathbf{k}-\mathbf{k}_{1}-\cdots-\mathbf{k}_{n})Z_{n}(\mathbf{k}_{1},\cdots,\mathbf{k}_{n})\hat \delta_{l}(\mathbf{k}_{1})\cdots\hat\delta_{l}(\mathbf{k}_{n}),
\end{equation}
where  $\mu_{i}$ is the cosine of the angle between $\mathbf{k}_{i}$ and the line of sight, $\mathbf{k}=\mathbf{k}_{1}+\mathbf{k}_{2}$. The first kernels 
are given by (assuming a quadratic local bias model involving $b_1$ and $b_2$)
\begin{eqnarray}
Z_{1}(\mathbf{k})&=&(b_{1}+f\mu^{2})\,,\,\, \mathrm{ and}\,\,\,
Z_{2}(\mathbf{k}_{1},\mathbf{k}_{2})=b_{1}F_{2}(\mathbf{k}_{1},\mathbf{k}_{2})+f\mu^{2}G_{2}(\mathbf{k}_{1},\mathbf{k}_{2})+\frac{f\mu k}{2}\left[ \frac{\mu_{1}}{k_{1}}(b_{1}+f\mu_{2}^{2})+\frac{\mu_{2}}{k_{2}}(b_{1}+f\mu_{1}^{2})\right]+\frac{b_{2}}{2}\,, \label{eq:defZ1Z2}
\end{eqnarray} with  $f= \Omega_{m}^{\gamma_m}$, $\gamma_m \simeq 6/11$, $\epsilon\simeq3/7\, \Omega_{m}^{-2/63}$ and
\begin{eqnarray}
F_{2}(\mathbf{k}_{1},\mathbf{k}_{2})&=&\frac 1 2 (1+\epsilon)+\frac 1 2\frac{\mathbf{k}_{1}\cdot\mathbf{k}_{2}}{k_{1}k_{2}}\left(\frac{k_{1}}{k_{2}}+\frac{k_{2}}{k_{1}}\right)+\frac 1 2 (1-\epsilon) \left(\frac{\mathbf{k}_{1}\cdot \mathbf{k}_{2}}{k_{1}k_{2}}\right)^{2}\,,\\
G_{2}(\mathbf{k}_{1},\mathbf{k}_{2})&=&\epsilon+\frac 1 2\frac{\mathbf{k}_{1}\cdot\mathbf{k}_{2}}{k_{1}k_{2}}\left(\frac{k_{1}}{k_{2}}+\frac{k_{2}}{k_{1}}\right)+(1-\epsilon) \left(\frac{\mathbf{k}_{1}\cdot \mathbf{k}_{2}}{k_{1}k_{2}}\right)^{2}\,. \label{eq:defG2F2}
\end{eqnarray} 
In particular
\begin{eqnarray}
\hat \delta_{s}^{(1)}(\mathbf{k})&=&D_{1}(\tau)\int\dd^{3}\mathbf{k}_{1}\delta_\mathrm{ D}(\mathbf{k}-\mathbf{k}_{1})Z_{1}(\mathbf{k}_{1})\hat \delta_{l}(\mathbf{k}_{1})=D_{1}(\tau)Z_{1}(\mathbf{k})\hat \delta_{l}(\mathbf{k})\,,\\
\hat \delta_{s}^{(2)}(\mathbf{k})&=&D_{1}^{2}(\tau)\int\dd^{3}\mathbf{k}_{1}\int\dd^{3}\mathbf{k}_{2}\delta_\mathrm{ D}(\mathbf{k}-\mathbf{k}_{1}-\mathbf{k}_{2})Z_{2}(\mathbf{k}_{1},\mathbf{k}_{2})\hat\delta_{l}(\mathbf{k}_{1})\hat\delta_{l}(\mathbf{k}_{2}).
\end{eqnarray}
 Given this expansion, cumulants can be computed at tree order. For conciseness, we  denote simply $\int_{W}$ the weighted 6D integration $\int \dd^{3}\mathbf{k}_{i} \dd^{3}\mathbf{k}_{j} [\qquad] W(k_{i}R)W(k_{j}R)W(|\mathbf{k}_{i}+\mathbf{k}_{j}|R)$. With this notation the cumulant of $\delta^{3}$ reads for instance
\begin{eqnarray*}
\left\langle\delta^{3}\right\rangle
&\simeq&3\left\langle(\delta^{(1)})^{2}\delta^{(2)}\right\rangle
= 3D_{1}^{4}(\tau)\int_{W} Z_{1}(\mathbf{k}_{1})Z_{1}(\mathbf{k}_{2})Z_{2}(\mathbf{k}_{3},\mathbf{k}_{4})\left\langle\hat \delta_{l}(\mathbf{k}_{1})\hat \delta_{l}(\mathbf{k}_{2})\hat\delta_{l}(\mathbf{k}_{3})\hat \delta_{l}(\mathbf{k}_{4})\right\rangle\,,\\
&=& 3D_{1}^{4}(\tau)\int_{W} Z_{1}(\mathbf{k}_{1})Z_{1}(\mathbf{k}_{2})Z_{2}(\mathbf{k}_{3},\mathbf{k}_{4})\left(\left\langle\hat \delta_{l}(\mathbf{k}_{1})\hat \delta_{l}(\mathbf{k}_{2})\right \rangle\left\langle\hat \delta_{l}(\mathbf{k}_{3})\hat \delta_{l}(\mathbf{k}_{4})\right \rangle+2\left\langle\hat \delta_{l}(\mathbf{k}_{1})\hat \delta_{l}(\mathbf{k}_{3})\right \rangle\left\langle
\hat \delta_{l}(\mathbf{k}_{2})\hat \delta_{l}(\mathbf{k}_{4})\right \rangle\right)\,,\\
&=& 3\cdot(2\pi)^{6}D_{1}^{4}(\tau)\left[
\int_{W} Z_{1}(\mathbf{k}_{1})Z_{1}(-\mathbf{k}_{1})Z_{2}(\mathbf{k}_{3},-\mathbf{k}_{3})P(k_{1})P(k_{3})
+2\int_{W} Z_{1}(\mathbf{k}_{1})Z_{1}(\mathbf{k}_{2})Z_{2}(-\mathbf{k}_{1},-\mathbf{k}_{2})P(k_{1})P(k_{2})
\right]\,,
\end{eqnarray*}
where we can note that $Z_{2}(\mathbf{k}_{3},-\mathbf{k}_{3})=b_{2}/2$. This method can be generalized to all relevant cumulants.
Let us sum up these results in Table \ref{table:cums}\footnote{
Note that these expressions (precisely, the second term in $\beta$) can be compared to the results of \cite{Gay12} in real space. 
They are found to be in full agreement. In this table, it is also of interest to notice that $\langle \delta I_{2\perp}\rangle  =-3/4 \langle J_{1\perp}q_{\perp}^{2}\rangle  $ as mentioned in Appendix \ref{sec:momentsrelation}.}  where each cumulant is generically written
 \begin{equation}
\left\langle {\cal Y} \right\rangle =(2\pi)^{6}D_{1}^{4}(\tau)\left[
\int_{W} \!\! Z_{1}(\mathbf{k}_{1})^{2}Z_{2}(\mathbf{k}_{2},-\mathbf{k}_{2})P(k_{1})P(k_{2})\alpha(\mathbf{k}_{1},\mathbf{k}_{2})
+2\int_{W} \!\! Z_{1}(\mathbf{k}_{1})Z_{1}(\mathbf{k}_{2})Z_{2}(\mathbf{k}_{1},\mathbf{k}_{2})P(k_{1})P(k_{2})\beta(\mathbf{k}_{1},\mathbf{k}_{2})
\right], \label{eq:defalphabeta}
\end{equation} 
with
$
 {\cal Y}= \delta^{3}, \delta^{2}J_{1\perp}\, ,
\delta q_{\perp}^{2}\,,
\delta q_{\parallel}^{2}\,,
\delta I_{2\perp}\,,
 J_{1\perp} q_{\perp}^{2}\,, $ and $
J_{1\perp} q_{\parallel}^{2}\,,
$ resp.
%
\begin{table}
\begin{center}
$\begin{array}{|c|c|c|}
\hline
   & \alpha & \beta \\
  \hline
  &&\\
  \left\langle \delta^{3}\right\rangle & 3 & 3 \\
 \left\langle\delta^{2}J_{1\perp}\right\rangle & -2k_{1\perp}^{2} & -(\mathbf{k}_{1\perp}+\mathbf{k}_{2\perp})^{2}-2(k_{2\perp})^{2} \\
  \left\langle\delta q_{\perp}^{2}\right\rangle & k_{1\perp}^{2}& \mathbf{k}_{1\perp}\cdot\mathbf{k}_{2\perp}+2k_{1\perp}^{2}\\
   \left\langle\delta q_{\parallel}^{2}\right\rangle& k_{1\parallel}^{2} & {k}_{1\parallel}{k}_{2\parallel}+2k_{1\parallel}^{2} \\
\left\langle\delta I_{2\perp}\right\rangle & 0 & \frac 3 4 \left( (\mathbf{k}_{1\perp}\times\mathbf{k}_{2\perp})^{2}-(\mathbf{k}_{1\perp}\cdot\mathbf{k}_{2\perp})^{2}+k_{1\perp}^{2}k_{2\perp}^{2}\right)\\
\left\langle J_{1\perp}q_{\perp}^{2}\right\rangle&0& -(\mathbf{k}_{1\perp}\times\mathbf{k}_{2\perp})^{2}+(\mathbf{k}_{1\perp}\cdot\mathbf{k}_{2\perp})^{2}-k_{1\perp}^{2}k_{2\perp}^{2}\\
      \left\langle J_{1\perp}q_{\parallel}^{2}\right\rangle & 0 & k_{1z}k_{2z}(\mathbf{k}_{1\perp}+\mathbf{k}_{2\perp})^{2}-2k_{1\perp}^{2}k_{2z}(k_{1z}+k_{2z})\\
& &\\  \hline
\end{array}$
\end{center}
  \caption{\label{table:cums} $\alpha$ and $\beta$ coefficients defined via equation~(\ref{eq:defalphabeta}) for the relevant cumulants.}
\end{table}
As discussed  in the previous section,
the Minkowski functionals and the critical sets described in the main text yield access to (geometrically weighted by $\alpha$ and $\beta$) averages 
of products of the $Z_1$ and $Z_2$ kernels, which in turn depend on the underlying cosmological parameters via, say, $\epsilon$ in  equations~(\ref{eq:defZ1Z2}) --(\ref{eq:defG2F2}). 
Hence, provided the corresponding components of  the $Z_i$'s do not fall into the null space of the $\alpha$ and $\beta$ projectors, one should expect to 
be able to access the values of some of these cosmic parameters through appropriate combinations of the geometrical sets,
$\langle
 {\cal Y} \rangle=\langle
 {\cal Y} \rangle
 \left[\Omega_m,b_1,b_2,\gamma_m,D(z)\right]
 $.
In practice, all these moments can be integrated using the decomposition of $W_{G}(|\mathbf{k}_{1}+\mathbf{k}_{2}|R)$ in Legendre polynomials and Bessel functions
\begin{equation}
W_{G}(|\mathbf{k}_{1}+\mathbf{k}_{2}|R)=\exp\!\left[-\frac{k_{1}^{2}+k_{2}^{2}}{2}R^{2}\right]\sum_{l=0}^{\infty}(-1)^{l}(2l+1)P_{l}\left(\frac{\mathbf{k_{1}}\cdot\mathbf{k_{2}}}{k_{1}k_{2}}\right)I_{l+1/2}\!\left(k_{1}k_{2}R^{2}\right)\sqrt{\frac{\pi}{2k_{1}k_{2}R^{2}}}\,,
\end{equation}
except one, given by
\begin{equation}
I=\int_{W} Z_{1}(\mathbf{k}_{1})Z_{1}(\mathbf{k}_{2})f\mu^{2}G_{2}(\mathbf{k}_{1},\mathbf{k}_{2})P(k_{1})P(k_{2})\beta(\mathbf{k}_{1},\mathbf{k}_{2}).
\end{equation}
This difficulty was highlighted e.g by \cite{Hivon95} but was not analytically solved until now. For this purpose,
 let us use the following trick: if we introduce three different smoothing length $W(k_{1}R_{1})$, $W(k_{2}R_{2})$ and $W(kR)$  with a Gaussian filter, then one can see that
\begin{equation}
-2\frac{\partial I}{\partial R^{2}}=\int_{W} Z_{1}(\mathbf{k}_{1})Z_{1}(\mathbf{k}_{2})f\mu^{2} k^{2}G_{2}(\mathbf{k}_{1},\mathbf{k}_{2})P(k_{1})P(k_{2})\beta(\mathbf{k}_{1},\mathbf{k}_{2})\,. \label{eq:trick}
\end{equation}
Now the integration of equation~(\ref{eq:trick}) over  $\mu^{2}k^{2}$($=(\mathbf{k}_{1}+\mathbf{k}_{2})\cdot\mathbf{\hat z}$) is straightforward;  the integration  over $R^{2}$ is also straightforward and involves no constant of integration.
Cumulants integrated numerically on $k_{1}$, $k_{2}$ but analytically on the angles as described above are summed up in Table \ref{table:cumulants}.
Here, all variables are rescaled by their respective variance so that for instance $x=\delta/\sigma$, $q_{\perp}^{2}=(\delta_{1}^{2}+\delta_{2}^{2})/\sigma_{1\perp}^{2}$, ... Note that they are computed for $\Omega=1$ (i.e $f=1$), $b_{1}=1$, $P(k)=k^{n}$ in the plane parallel approximation and for a Gaussian filter. Skewness can be compared to \cite{Hivon95}  (where they do not assume plane parallel approximation). These cumulants are also computed for a $\Lambda$CDM power spectrum ($\Omega_{m}=0.27$, $\Omega_{\Lambda}=0.73$, $h=0.7$) smoothed over different scales in Table~\ref{table:cumulantsLCDM} (redshift space) and Table~\ref{table:cumulantsrealspace} (real space).
\begin{table}
\centering
$\begin{array}{|c|c|c|c|c|c|}
\hline
 & \text{n=1} & \text{n=0} & \text{n=-1} & \text{n=-2} \\
  \hline
  \left\langle x^3\right\rangle/\sigma &2.65 &  2.91&3.36  & 4.03\\
 \left\langle x^2 J_{1\bot }\right\rangle / \sigma&-2.62& -2.59&-2.62  &-2.47 \\
 \left.\text{$\langle $ }xq_{\perp}^2\right\rangle/ \sigma &1.87&1.95& 2.15&2.49 \\
  \left.\text{$\langle $}xq_{\parallel}^2\right\rangle/ \sigma &1.66&1.94 &2.36  &2.93\\
 \left.\text{$\langle $ }xI_{2\bot }\right\rangle / \sigma&0.49&0.41  &0.33  &0.22  \\
 \left\langle J_{1\bot }q_{\perp}^2\right\rangle  /\sigma&  -0.94&-0.81  & -0.72&-0.59 \\
 \left\langle J_{1\bot }q_{\parallel}^2\right\rangle  /\sigma& -1.73 &-1.54  &-1.38 &-1.16 \\
 \hline
\end{array}$  
\caption{Predicted $\sigma$'s and 3-point cumulants for power-law power spectra in redshift space using PT.}
\label{table:cumulants}
\end{table}
\begin{table}
\centering
$
\begin{array}{|c|c|c|c|c|c|c|c|c|}
\hline
  \text{R (Mpc/h):}& 8 & 16 & 24 & 32 & 40&48&56&64\\
    \hline
    \sigma & 0.54 & 0.29 & 0.19 & 0.13 & 0.10 & 0.081& 0.066 & 0.055 \\
    \sigma_{1\perp} & 0.047 & 0.014 & 0.0063 & 0.0035 & 0.0022 & 0.0015& 0.0011 & 0.00079 \\
    \sigma_{1\parallel} & 0.038 & 0.011 & 0.0052 & 0.0029 & 0.0018 & 0.0012& 0.00088 & 0.00065 \\
    \sigma_{2\perp} & 0.0065 & 0.0010 & 0.00032 & 0.00014 & 0.000070 & 0.000040& 0.000025 & 0.000016 \\
    \sigma_{2\parallel} & 0.0050 & 0.00078 & 0.00025 & 0.00011 & 0.000053 & 0.000030& 0.000019 & 0.000012 \\
 \left\langle x^3\right\rangle/\sigma  & \text{ 3.68} & \text{ 3.49} & \text{ 3.39} & \text{ 3.34} & \text{ 3.30} & \text{ 3.26} & \text{ 3.22} & \text{ 3.19} \\
 \left\langle x^2 J_{\text{1$\bot $}}\right\rangle  /\sigma& -3.01 & -3.02 & -2.97 & -2.96 & -2.98&\text{-3.00}&\text{-3.00}&\text{-3.00} \\
 \left\langle x q_{\bot }^2\right\rangle/\sigma  & \text{ 2.43} & \text{ 2.32} & \text{ 2.25} & \text{ 2.22} & \text{ 2.20}& \text{ 2.18}& \text{ 2.16}& \text{ 2.14} \\
 \left\langle x q_{\parallel }^2\right\rangle /\sigma & \text{ 2.61} & \text{ 2.45} & \text{ 2.35} & \text{ 2.30} & \text{2.27}& \text{2.24}& \text{2.20}& \text{2.17} \\
 \left\langle x I_{\text{2$\bot $}}\right\rangle/\sigma  & \text{ 0.35} & \text{ 0.39} & \text{ 0.40} & \text{ 0.41} & \text{0.43}& \text{0.44}& \text{0.46}& \text{0.46} \\
 \left\langle J_{\text{1$\bot $}} q_{\bot }^2\right\rangle /\sigma & -0.76 & -0.80 & -0.82 & -0.83 & -0.84&-0.86&-0.87&-0.89 \\
 \left\langle J_{\text{1$\bot $}} q_{\parallel }^2\right\rangle  /\sigma& -1.50 & -1.58 & -1.61 & -1.63 & -1.67 &-1.70&-1.72&-1.74\\
  \hline
\end{array}$
\caption{Predicted $\sigma$'s and 3-point cumulants for $\Lambda$CDM power spectra in redshift space using PT.}
\label{table:cumulantsLCDM}
\end{table}

\begin{table}
\centering
$
\begin{array}{|c|c|c|c|c|c|c|c|c|}
\hline
  \text{R (Mpc/h):}& 8 & 16 & 24 & 32 & 40&48&56&64\\
  \hline
      \sigma & 0.46 & 0.25 & 0.16 & 0.12 & 0.088 & 0.070& 0.057 & 0.047 \\
         \sigma_{1\perp} & 0.042 & 0.013 & 0.0058 & 0.0032 & 0.0020 & 0.0014& 0.00097 & 0.00072 \\
    \sigma_{1\parallel} & 0.030 & 0.0089 & 0.0041 & 0.0023 & 0.0014 & 0.00096& 0.00067 & 0.00051 \\
    \sigma_{2\perp} & 0.0061 & 0.00096 & 0.00030 & 0.00013 & 0.000065 & 0.000037& 0.000023 & 0.000015 \\
    \sigma_{2\parallel} & 0.0038 & 0.00059 & 0.00019 & 0.000079 & 0.000040 & 0.000023& 0.000014 & 0.0000093 \\
 \left\langle x^3\right\rangle/\sigma  & \text{ 3.70} & \text{ 3.52} & \text{ 3.43} & \text{ 3.38} & \text{ 3.34} & \text{ 3.31} & \text{ 3.28} & \text{ 3.25} \\
 \left\langle x^2 J_{\text{1$\bot $}}\right\rangle/\sigma  & -3.21 & -3.20 & -3.15 & -3.13 & -3.16 & -3.17 & -3.17 & -3.16 \\
 \left\langle x q_{\bot }^2\right\rangle/\sigma  & \text{ 2.51} & \text{ 2.39} & \text{ 2.32} & \text{ 2.28} & \text{ 2.25} & \text{ 2.23} & \text{ 2.21} & \text{
   2.19} \\
 \left\langle x q_{\parallel }^2\right\rangle  /\sigma& \text{ 2.51} & \text{ 2.39} & \text{ 2.32} & \text{ 2.28} & \text{ 2.25} & \text{ 2.23} & \text{ 2.21} &
   \text{ 2.19} \\
 \left\langle x J_{\text{2$\bot $}}\right\rangle  /\sigma& \text{ 0.38} & \text{ 0.42} & \text{ 0.43} & \text{ 0.45} & \text{ 0.46} & \text{ 0.48} & \text{ 0.49} &
   \text{ 0.50} \\
 \left\langle J_{\text{1$\bot $}} q_{\bot }^2\right\rangle/\sigma  & -0.79 & -0.83 & -0.85 & -0.86 & -0.88 & -0.90 & -0.91 & -0.92 \\
 \left\langle J_{\text{1$\bot $}} q_{\parallel }^2\right\rangle /\sigma & -1.58 & -1.66 & -1.70 & -1.73 & -1.76 & -1.79 & -1.82 & -1.85\\
 \hline
\end{array}
$
\caption{Predicted $\sigma$'s and 3 pt cumulants for $\Lambda$CDM power spectra in real space using PT.}
\label{table:cumulantsrealspace}
\end{table}

\begin{table}
\centering
$\begin{array}{|c|c|c|c|c|c|c|c|c|}
\hline
  \text{R (Mpc/h):}& 8 & 16 & 24 & 32 & 40&48&56&64\\
  \hline
 \tilde S^{(1)} & 0.075 & 0.062 & 0.044 & 0.033 & 0.036& 0.039 & 0.038 & 0.035 \\
 \tilde S^{(1)}_z & 0.061 & 0.053& 0.037 & 0.028 & 0.033& 0.038& 0.039 & 0.037 \\
 {\tilde S}^{(-1)} & 0.015 & 0.20 & 0.32 & 0.38& 0.43 & 0.48& 0.53 & 0.59 \\
 {\tilde S}^{(-1)}_z & -0.014 & 0.17 & 0.29 & 0.35 & 0.40& 0.46 & 0.51 & 0.57 \\
 \hline
\end{array}
$
\caption{Comparison of the reduced coefficients in front of the Hermite polynomials in the 3D Euler characteristic at first NG order for $\Lambda$CDM power spectra with different smoothings. 
\label{LCDMcoef}
}
\end{table}

In the regime where standard perturbation theory holds in redshift space\footnote{Extensions of  perturbation theory in redshift space using the streaming model  \citep[see][for instance]{Scoccimarro04,Taruya10} would 
allow us to extend the validity of the predicted cumulants to smaller scales, but stands beyond the scope of this paper.}, we therefore may assume that all cumulants entering equations~(\ref{eq:chi3D-nuf})-(\ref{eq:N1-nuf})
can be  {\sl predicted} by a given standard cosmological  model, while the amplitude of the non-Gaussian correction scales like $\sigma$.
For $\Lambda$CDM cosmology, Table~\ref{LCDMcoef} provides the predictions for the combinations of cumulants entering 3D Euler characteristic at first non-Gaussian order, as a function of the smoothing length in real  and in redshift space. The dependence on the linear bias $b_{1}$ of these particular combinations of cumulants can also be computed :  for a $\Lambda$CDM power spectrum smoothed e.g over 32 Mpc/h, it is found that $\tilde S^{(1)}_{\chi_{\textrm{3D}},z}/\tilde S^{(1)}_{\chi_{\textrm{3D}},r}$ is slightly varying around 0.8 while $\tilde S^{(-1)}_{\chi_{\textrm{3D}},z}/\tilde S^{(-1)}_{\chi_{\textrm{3D}},r}$  behaves approximatively  like $1.285-0.36/b_{1}$ for $1<b_{1}<2.5$.
Conversely, one may parametrize these cumulants while exploring alternative theories of gravity and attempt to  fit these cumulants using geometric probes.
\subsubsection{Constraining modified gravity with $\hat\gamma$ models}
Over the last few years  many flavours of  so-called modified gravity (MG) models have been presented \citep[see, e.g. ][ for a review]{Ferreira}. In the context of the upcoming dark energy missions,
it is of interest to understand how topological estimators also allow us to test  such extensions of  general relativity.
Following \cite{BernardeauBrax2011}, let us consider the so-called $\hat\gamma$ model  as an illustrative example of how to 
test  MG theories.
 This model  parametrizes modifications of gravity through a change in the amplitude of  Euler equation's source term.
 For this set of  models, the generalization of equation~(\ref{eq:defG2F2})  becomes parametric:
\begin{eqnarray}
F^{\hat\gamma}_{2}(\vk_{1},\vk_{2})&=&\left(\frac{3\hat\nu_{2}(\hat\gamma)}{4}-\frac{1}{2}\right)+\frac{1}{2}\frac{\vk_{1}.\vk_{2}}{k_{1}^2}+\frac{1}{2}\frac{\vk_{1}.\vk_{2}}{k_{2}^2}
+\left(\frac{3}{2}-\frac{3\hat\nu_{2}(\hat\gamma)}{4}\right)\frac{(\vk_{1}.\vk_{2})^2}{k_{1}^2 k_{2}^2}\,,\label{PrF2Exp}\\
G^{\hat\gamma}_{2}(\vk_{1},\vk_{2})&=&\left(\frac{3\hat\mu_{2}(\hat\gamma)}{4}-\frac{1}{2}\right)+\frac{1}{2}\frac{\vk_{1}.\vk_{2}}{k_{1}^2}+\frac{1}{2}\frac{\vk_{1}.\vk_{2}}{k_{2}^2}
+\left(\frac{3}{2}-\frac{3\hat\mu_{2}(\hat\gamma)}{4}\right)\frac{(\vk_{1}.\vk_{2})^2}{k_{1}^2 k_{2}^2}\,,\label{PrG2Exp}
\end{eqnarray}
where $\nu_{2}(\hat\gamma)$ and $\mu_{2}(\hat\gamma)$ are  given by
\begin{equation}
\hat\nu_{2}(\hat \gamma)=\nu_{2}^\mathrm{GR}-\frac{10}{273}(\hat\gamma-\gamma^\mathrm{GR})(1-\Omega_m)\Omega_m^{\gamma^\mathrm{GR}-1}\,,\quad
\hat\mu_{2}(\hat\gamma)=\mu_{2}^\mathrm{GR}-\frac{50}{273}(\hat\gamma-\gamma^\mathrm{GR})(1-\Omega_m)\Omega_m^{\gamma^\mathrm{GR}-1}\,,\label{mu2fit}
\end{equation}
with
\begin{equation}
\nu_{2}^\mathrm{ GR}=\frac{4}{3}+\frac{2}{7}\Omega_m^{-1/143},\quad
\mu_{2}^\mathrm{ GR}=-\frac{4}{21}+\frac{10}{7}\Omega_m^{-1/143}\,,\quad \gamma_\mathrm{ GR}=\frac{6}{11}\,.
\end{equation}
Then
\begin{equation}
Z^{\hat\gamma}_{2}(\mathbf{k}_{1},\mathbf{k}_{2})=b_{1}F^{\hat\gamma}_{2}(\mathbf{k}_{1},\mathbf{k}_{2})+f\mu^{2}G^{\hat\gamma}_{2}(\mathbf{k}_{1},\mathbf{k}_{2})+\frac{f\mu k}{2}\left[ \frac{\mu_{1}}{k_{1}}(b_{1}+f\mu_{2}^{2})+\frac{\mu_{2}}{k_{2}}(b_{1}+f\mu_{1}^{2})\right]+\frac{b_{2}}{2}\,, \label{eq:defZ1Z2MD}
\end{equation}
 with  $f= \Omega_{m}^{\hat\gamma}$ and $F^{\hat\gamma}_{2}, G^{\hat\gamma}_{2}$ given by equation~(\ref{PrF2Exp}).
Third order cumulants, e.g. $\tilde S^{(1)}_{\chi_{\textrm{3D}},z}$, in equations~(\ref{eq:tS1})-(\ref{eq:tS2}) become functions of $\hat\gamma$  and 
can now be fitted for $\hat \gamma$ to measured  departure to Gaussianity in dark energy surveys. 
We computed numerically these cumulants for $\Lambda$CDM power spectra $(\Omega_{m}=0.27,h_{0}=70,f=\Omega_{m}^{\hat \gamma})$ with different $\hat \gamma$. For instance, the values of the observables, $\sigma \tilde S^{(1)}_{\chi_{\textrm{3D}},z}$ and $\sigma \tilde S^{(-1)}_{\chi_{\textrm{3D}},z}$ 
(that can be accessed by measuring the 3D \genus) 
are typically enhanced by a factor of $1\%$ when $\hat \gamma$ varies from 0.55 (GR)
 to 0.67 (DGP). 
More generally, any parametrization of modified gravity (illustrated here on $\hat\gamma$ models) can be implemented in this framework. A measurement of Minkowski functionals then lead to constraints on the parameters (e.g $\hat \gamma$) of the theory.
 
\subsection{Illustration on $\Lambda$CDM simulations: DM and halo catalogues}
Let us now  illustrate our statistics on the Horizon 4$\pi$ N-body simulation \citep{b4} which contains $4096^3$
 dark matter particles distributed in a 2 $h^{-1}$Gpc periodic box to validate our number count prediction in
 a more realistic framework. This simulation is characterized by the following $\Lambda$CDM cosmology:
 $\Omega_\mathrm{ m}=0.24 $, $\Omega_{\Lambda}=0.76$, $n_s=0.958$, $H_0=73 $ km$\cdot s^{-1} \cdot $Mpc$^{-1}$ and $\sigma _8=0.77$ within one standard deviation of WMAP3 results \citep{Spergeletal03}. 
These initial conditions were evolved non-linearly down to redshift zero using the AMR code RAMSES \citep{b26}, on a $4096^3$ grid. The motion of the particles was followed with a multi grid Particle-Mesh Poisson solver using a Cloud-In-Cell interpolation algorithm to assign these particles to the grid (the refinement strategy of  40 particles as a threshold for refinement allowed us to reach a constant physical resolution of 10 kpc, see the above mentioned two references).
\subsubsection{Minkowski functionals and extrema counts for dark matter}
A measurement of the three points cumulants in the $4\pi$ simulation is displayed in Table \ref{table:measuredcum}.
\begin{table}
\centering
$\begin{array}{|c|c|c|c|c|c|}
\hline
  \text{R (Mpc/h):}& 16 & 23 & 32 & 45 &64\\
   \hline
      \sigma & 0.28 (0.25) & 0.20 (0.17) & 0.14 (0.12) & 0.090 (0.077) & 0.057 (0.049)  \\
 \sigma_{1||} & 0.044 (0.035) & 0.023 (0.018) & 0.012 (0.0092) & 0.0057 (0.0045) & 0.0027 (0.0021)  \\
 \sigma_{1\perp} & 0.054 (0.050) & 0.028 (0.026) & 0.014 (0.013) & 0.0068 (0.0062) & 0.0032 (0.0029)  \\
 \sigma_{2||} & 0.012 (0.0094) & 0.0045 (0.0035) & 0.0017 (0.0013) & 0.00059 (0.00045) & 0.00021 (0.00016)  \\
 \sigma_{2\perp} & 0.016 (0.015) & 0.0061 (0.0057) & 0.0022 (0.0021) & 0.00076 (0.00072) & 0.00026 (0.00024)  \\
  \sigma_{Q} & 0.0089 (0.0076) & 0.0034 (0.0029) & 0.0012 (0.0010) & 0.00043 (0.00036) & 0.00015 (0.00012)  \\
 \hline
 \left\langle x^3\right\rangle /\sigma &3.36 (3.49)&3.27 (3.35)&3.24 (3.28)&3.38 (3.41)&4.03 (4.07)  \\
\left\langle x^2 J_{\text{1$\bot $}}\right\rangle /\sigma &  -2.92 (-3.19)&-2.84 (-3.07)&-2.75 (-2.95)&-2.76 (-2.95)&-3.00 (-3.20) \\
 \left\langle x q_{\bot }^2\right\rangle  /\sigma& 2.27 (2.40)&2.28 (2.27)&2.09 (2.16)&2.05 (2.12)&2.19 (2.26)\\
 \left\langle x q_{\parallel }^2\right\rangle /\sigma & 2.33 (2.41)&2.27 (2.30)&2.24 (2.24)&2.24 (2.21)&2.52 (2.53)\\
 \left\langle x I_{\text{2$\bot $}}\right\rangle  /\sigma&  0.39 (0.43)&0.39 (0.42)& 0.37 (0.41)&0.37 (0.40)&0.42 (0.46)\\
 \left\langle J_{\text{1$\bot $}} q_{\bot }^2\right\rangle  /\sigma&-0.81 (-0.87)&-0.79 (-0.84)&-0.76 (-0.80)&-0.72 (-0.76)&-0.81 (-0.86) \\
 \left\langle J_{\text{1$\bot $}} q_{\parallel }^2\right\rangle /\sigma & -1.57 (- 1.75)&-1.55 (-1.68)&-1.53 (-1.64)&-1.51 (-1.60)&-1.53 (-1.64)\\
  \hline
  \beta_{\sigma}&0.24 (0.0066)&0.26 (0.014)&0.28 (0.030)&0.30 (0.058)&0.33(0.091)\\
  \hline
\end{array}$
\caption{Measured cumulants in redshift space (real space) for different smoothings in the {\sc horizon} $4\pi$ simulation.
{
Note that the ratio of the field dispersion in redshift space versus real space at large scales is $\approx 1.17$ which
corresponds to $\beta \approx 0.5$ as expected.}
}
\label{table:measuredcum}
\end{table}
It shows that redshift distortion has a  small impact on three-point cumulants. We thus expect 3D Minkowski
 functionals to be weakly affected by redshift space distortion. However, let us keep in mind that even if
 this difference is small it should be of great importance to model it properly in the context of high precision cosmology,
 especially since in 2D slices, the effect should be boosted. 

Figure~\ref{fig:4pigen3D}
\begin{figure}
 \begin{center}
\includegraphics[width=0.45\textwidth]{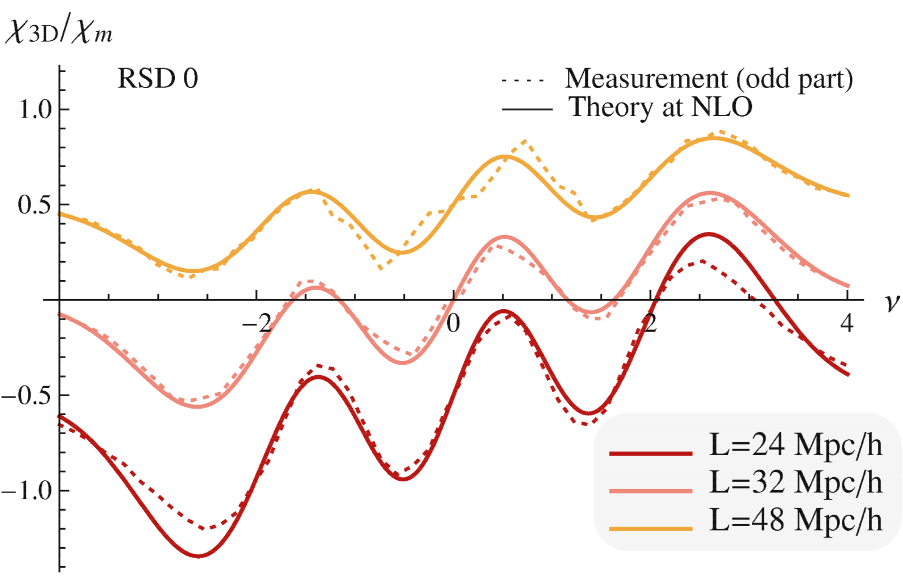}
\includegraphics[width=0.45\textwidth]{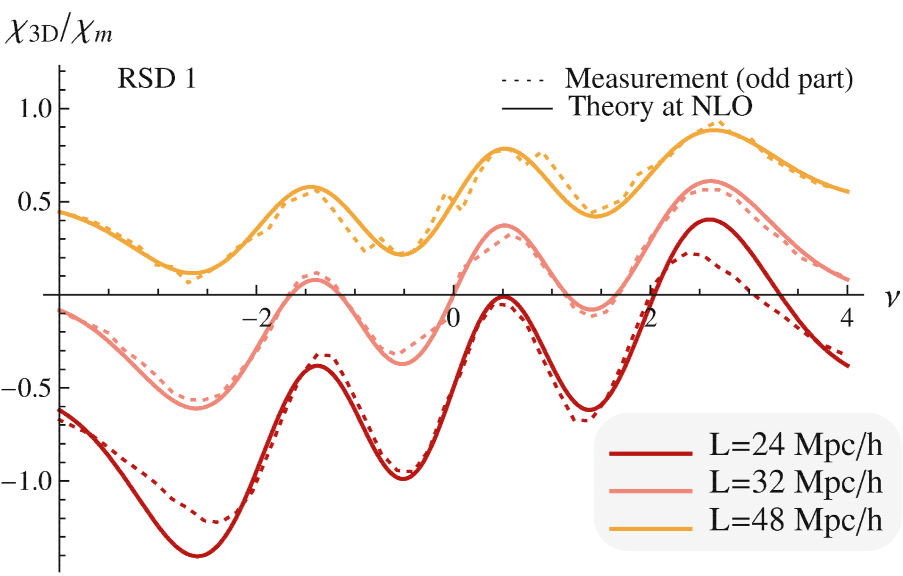}
\includegraphics[width=0.45\textwidth]{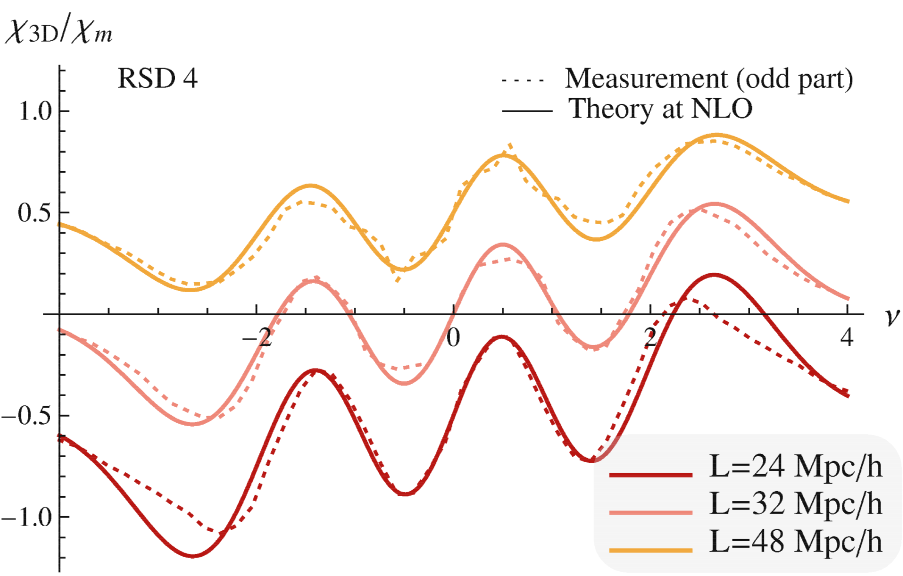}
\includegraphics[width=0.45\textwidth]{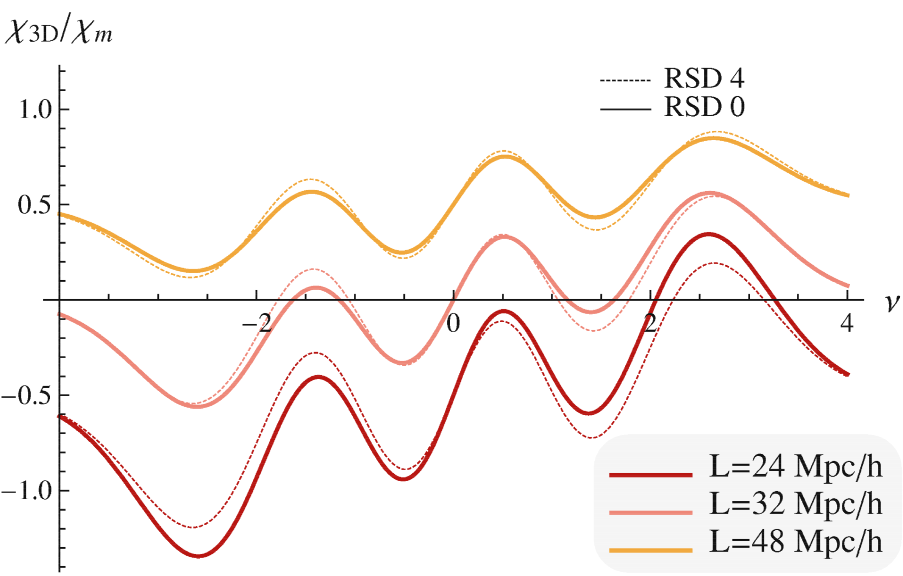}
\caption{
{\sl Top left panel}:  first non-Gaussian correction of the 3D \genus\, as a function of smoothing (as labeled) measured (dotted lines) and predicted (plain line) in \horizon simulation in real space.
Each curve has been normalized by the maximum of the Gaussian component of the \genus. Those corresponding to different smoothing length have been shifted for clarity. 
 {\sl Top right panel}:  same as top left panels, but in  redshift space.
{\sl Bottom left panel}:  same as top right panels, but redshift displacements were boosted by a factor of four. {\sl Bottom right panel}: theoretical prediction in real space and boosted redshift space. For these ranges of smoothing the theory predicts well the  \genus\, to first order in non-Gaussianity, in particular for low-intermediate thresholds. The difference in the \genus\, introduced by redshift distortion (even boosted by a factor of four) is rather small.
\label{fig:4pigen3D}
}
 \end{center}
\end{figure}
illustrates this for the 3D \genus. First, note that the theory mimics very well the measurement.
This figure also shows that the difference between real and redshift space is indeed small for the first correction
from Gaussianity. But in 2D, this difference increases as seen in
Figure~\ref{fig:4pigen2Dth} 
\begin{figure}
 \begin{center}
\includegraphics[width=0.45\textwidth]{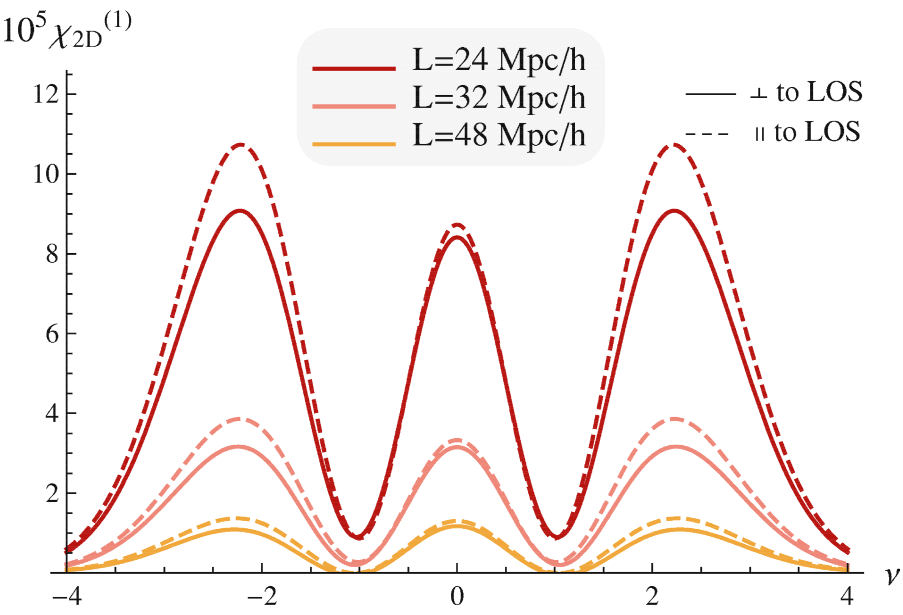}
\caption{
 Prediction for the 2D \genus\, in redshift space along (dashed lines) and perpendicular (solid lines) to the LOS with cumulants measured in the \horizon simulation.
 There is a clear dependence on the angle between the slice and the LOS, especially for smaller smoothing.
\label{fig:4pigen2Dth}
}
 \end{center}
\end{figure}
which shows how the correction from Gaussianity depends on the angle between the LOS and the slice on which the 2D \genus\,
is computed.
\subsubsection{Minkowski functionals for halo catalogues}
The Friend-of-Friend Algorithm \citep[hereafter FOF,][]{b27} was used  over $18^3$ overlapping subsets of the simulation with a linking length of  0.2 times the mean inter-particular distance  to define dark matter halos. 
In the present work, we only consider halos with more than 40 particles, which corresponds to a minimum halo mass of $3 \cdot 10^{11} M_{\odot}$ (the particle mass is $7.7\cdot 10^{9}M_{\odot}$). The mass dynamical range of this simulation spans about 5 decades. 
Overall, the catalog contains 43 million dark halos. The density of dark halos in real space and redshift space
 was re-sampled over a $512^3$ grid with box size $2000~\mathrm{Mpc}/h$ (in redshift space, after shifting the z-ordinate of the halo  by
 its  velocity along that direction divided by the Hubble constant, $H_0$) and smoothed with a Gaussian 
filter of width $\sigma=$ 16, 24... up to $94$ Mpc$/h$. 
%
\begin{figure}
 \begin{center}
\includegraphics[width=0.45\textwidth]{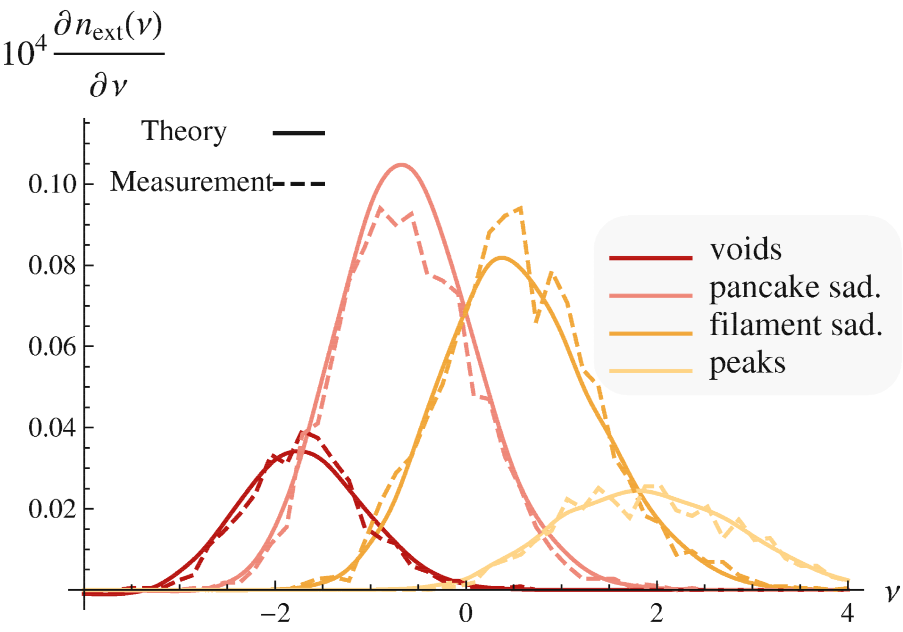}
\includegraphics[width=0.45\textwidth]{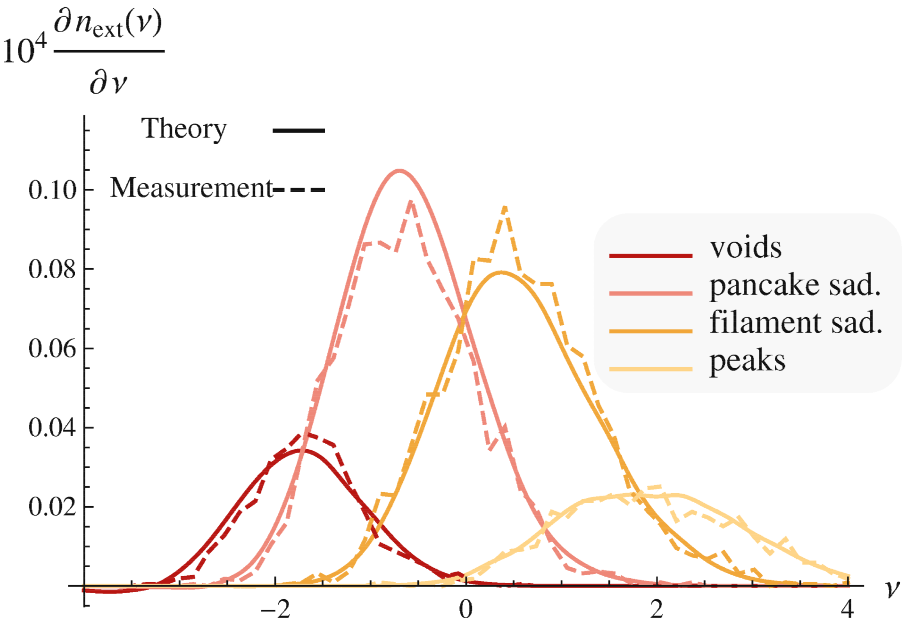}
\caption{
{\sl Left panel}: The distribution of 3D extrema in real space as a function of threshold $\nu$
in real space, as predicted (solid) and measured (dashed) in dark matter {\sl halo catalogue} of the \horizon simulation smoothed over 48 Mpc/h.
 {\sl Right panel}: 
same as left panel but in redshift space.
\label{fig:4piext}
}
 \end{center}
\end{figure}
For each cube in real and redshift space the number density of extrema is computed by
a local quadratic fit to the function profile with control over the double counting  of the neighbouring extrema
(see \citet{PPG2011} for the technique description and \citet{Percolation} for a first application).
The relevant 35 cumulants are computed via fast Fourier transform.
Figure~\ref{fig:4piext} displays the corresponding predicted ({\sl solid line}) and measured ({\sl dashed line})
 number count difference in real space ({\sl left panel}) and  redshift space ({\sl right panel}).
 {Note that redshift and real space extrema counts are almost indistinguishable. Indeed, for this biased population,
 $b_1$ is large ($\sim$2) and  $f/b_1$ is small enough.}
\subsection{A cosmic  fiducial experiment: measuring $\beta$  and $\sigma$ via slicing along and perpendicular to the line of sight}
\label{sec:measuringf}
As mentioned in Section~\ref{sec:Min23}, the angle-dependence of the 2D Minkowski functionals allows us to probe $\beta$. Indeed, equations~(\ref{eq:2DNGgenus})-(\ref{eq:N2_ani_n3})-(\ref{eq:N1NG}) displays a functional dependence on $\theta_{\cal S/L}$. 
At Gaussian order first, it appears that
\begin{equation}
\chi_{2\textrm{D}}^{(0)}(\nu, \theta_{\cal S})\propto
\sqrt{1-\beta_\sigma \sin^2\theta_{\cal S}}  \,,
\end{equation}
so that measuring the Gaussian part of the 2D \genus\, in different slices (with different orientation relative to the LOS) can give access to $\beta_{\sigma}=1-\sigma_{1\perp}^{2}/2\sigma_{1\parallel}^{2}=\frac 4 5 \beta+{\cal O}(\beta^{2})$. For example, in the most favorable case of $\theta_{1}=\pi/2$ and $\theta_{2}=0$, one have directly access to
\begin{equation}
\label{beta-genus}
\frac{\chi_{2\textrm{D}}^{(0)}(\nu, \theta_{1})}{\chi_{2\textrm{D}}^{(0)}(\nu, \theta_{2})}=
\sqrt{\frac{1-\beta_\sigma \sin^2\theta_{1}}{1-\beta_\sigma \sin^2\theta_{2}} }=\sqrt{1-\beta_\sigma} =1-\frac 2 5 \beta +{\cal O}(\beta^{2}) \,.
\end{equation}
Contour crossing statistics have the same angle-dependence as 2D \genus\, at Gaussian order. The ${\cal N}_{2}$ statistics (area of isocontours) leads in turn
\begin{equation}
\label{beta-N2}
\frac{{\cal N}_{2}^{(0)}(\nu, \theta_{1})}{{\cal N}_{2}^{(0)}(\nu, \theta_{2})}=
\frac{E(\tilde \beta_{\sigma}(\theta_{1}))\sqrt{1-\tilde \beta_{\sigma}(\theta_{2})}}{E(\tilde \beta_{\sigma}(\theta_{2}))\sqrt{1-\tilde \beta_{\sigma}(\theta_{1})}}
 =\frac{\pi}{2}\frac{\sqrt{1-\beta_{\sigma}}}{E(\beta_{\sigma})}
 =1-\frac 1 5 \beta +{\cal O}(\beta^{2})\,. 
 \end{equation}
Note that beyond a simple overall amplitude effect (which is enough to measure $\beta$ alone), this angle-dependence also arises in the first non-Gaussian correction as plotted in Figure~\ref{fig:2Dgenusangle}. 

If the value of $\beta$ is important on its own (e.g to study the bias  or to test modifications of gravity), it is also of prime importance to measure $D(z)$ as it allows us to map the dispersion in redshift space, $\sigma$, into its value in real space, $\sigma_{0}\propto D(z)$. 
One way to proceed is to use the Gaussian term to put constraints on $\beta$ 
and then the non-Gaussian correction for $\sigma$ (the amplitude of which is not probed by the Gaussian part).
Indeed, following \cite{Gay12},  the theory presented in Section~\ref{sec:Min23} should allow us to measure the dispersion of the field in redshift space $\sigma$ from the amplitude of the 
departure from non-Gaussianity. As  equation~(\ref{beta-N2}) demonstrates, the comparison of the 2D Minkowski functionals in planes parallel and perpendicular 
to the LOS allows us to measure independently $\beta=f/b$. Hence Minkowski functionals in redshift space yield a geometric estimate of  the real-space field dispersion, $\sigma_0$, via 
\begin{equation}
\label{eq:sigmazr}
 \sigma_0=\sigma/\sqrt{1+{2}\beta/3+{\beta^2}/{5}}.
 \end{equation}
Let us apply this scheme  to measure $\beta$ and $\sigma_0$ in a fiducial experiment.  
Minkowski functionals are measured from our set of 19 scale-invariant (n=-1) $256^{3}$ dark matter simulations smoothed over 15 pixels, corresponding to $\sigma_0=0.18$ and displayed in Figure~\ref{fig:minkowski-gravity}. 
A first step is to use the Gaussian term of the 1D and 2D statistics, while varying the angle of the slices to constrain $\beta$ using equations~(\ref{beta-genus})-(\ref{beta-N2}). 
For that purpose, we extract the even part of ${\cal N}_{2}$ (to get rid of odd parity effects arising from the first NLO correction)
 and restrict ourselves to the intermediate domain $-1.2<\nu<1.2$ where the Gaussian term is dominant.
The resulting constraints on $\beta$ are found to be 
$\hat \beta=1.04\pm 0.05\,, $
 and illustrated on Figure~\ref{fig:recov-beta} (left panel). 
 The same analysis on the other 1D (${\cal N}_{1}$) and 2D statistics (2D genus) leads 
 to similar constraints.
The next step is to fit the first non-Gaussian correction of each statistics with PT predictions in order to constrain $\sigma$.  
The predictions at first order are expressed as a function of the contrast $\nu$ and $\sigma$ only using Table~\ref{table:cumulants}. 
The value of the free parameter $\sigma$ in the model is then constrained by fitting the odd part of the data (which is dominated by the first non-Gaussian correction for intermediate contrasts).
 The result for ${\cal N}_{2}$ is shown in Figure~\ref{fig:recov-beta} (right panel) and yields 
$\hat \sigma=0.22\pm 0.08$.
The other statistics give similar results e.g  for the 3D \genus\,
$\hat\sigma=0.26\pm 0.06. $
Altogether, using $\sigma$ as measured by a 3D probe, $\chi_{\textrm{3D}}$, and $\beta$ by a 2D statistics, ${\cal N}_{2}$,
we finally get 
$\hat \sigma_0 =0.18 \pm 0.04$,
which is fully consistent with the underlying dispersion in our mocks.
The accuracy  on the measurement of $\beta$ and $D(z)$ through $\sigma_0$ can naively be scaled to the expected accuracy for a Euclid-like survey (assuming one quarter of the sky is observed) leading to a relative 0.3\% precision on $\beta$ and 
1.5\% on $D(z)$ at redshift zero. See also  \cite{Gay12}, which
  translates this accuracy in
terms of estimates for the dark energy parameters $w_0, w_a$.

It is worth noting that in this simple fiducial experiment, several assumptions were made : 
first, we assumed we knew the contrast, $\nu$, while in realistic surveys, the accessible quantity is $\nu_{f}$ (see section \ref{sec:nuf});
for the simulation, the cosmology is not $\Lambda$CDM but Einstein-de Sitter with a scale-invariant initial power spectrum; 
 finally the error on the estimated $\sigma$ depends on the accuracy of the theory used to predict the cumulants (Standard Perturbation Theory here which is known to perform 
 somewhat poorly in redshift space). No account of masking, redshift evolution of S/N ratio or finite survey volume, nor comparison with other 
 dark energy probes  was attempted.
Carrying out the road map sketched in this section while addressing these issues should be one of the target of the upcoming  surveys that have been planned specifically
to probe dark energy, either from ground-based facilities (eg BigBOSS, VST-KIDS, %
DES, %
 Pan-STARRS, %
LSST\footnote{\texttt{http://bigboss.lbl.gov},\, 
\texttt{http://wtww.astro-wise.org/projects/KIDS},\, 
\texttt{https://www.darkenergysurvey.org},\,
\texttt{http://pan-starrs.ifa.hawaii.edu},\,
\texttt{http://www.lsst.org}}) 
or space-based observatories (EUCLID\citep{Euclid}, SNAP and JDEM\footnote{\texttt{http://sci.esa.int/euclid},\, 
 \texttt{http://snap.lbl.gov},\, \texttt{http://jdem.lbl.gov}}). 
\begin{figure}
 \begin{center}
  \includegraphics[width=0.45\textwidth]{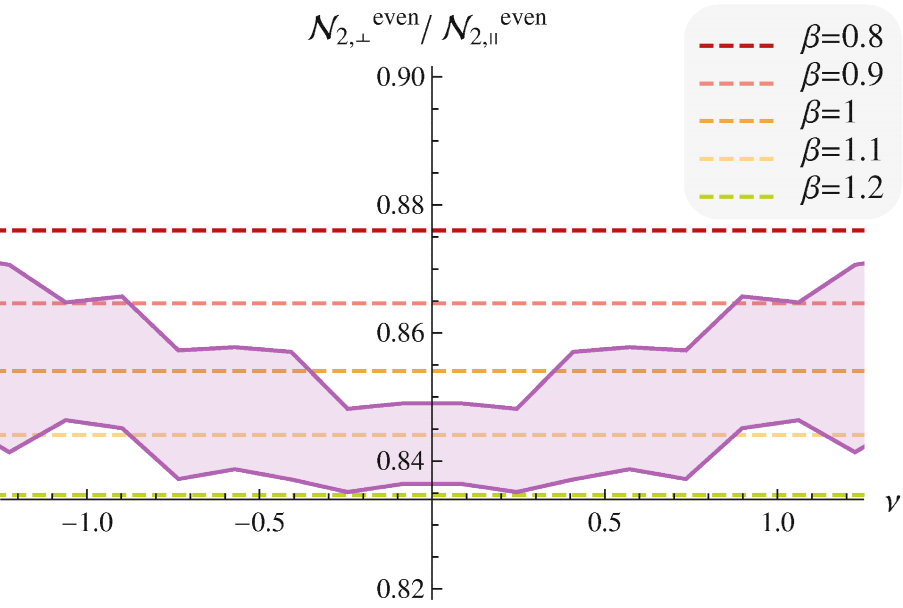}
    \includegraphics[width=0.45\textwidth]{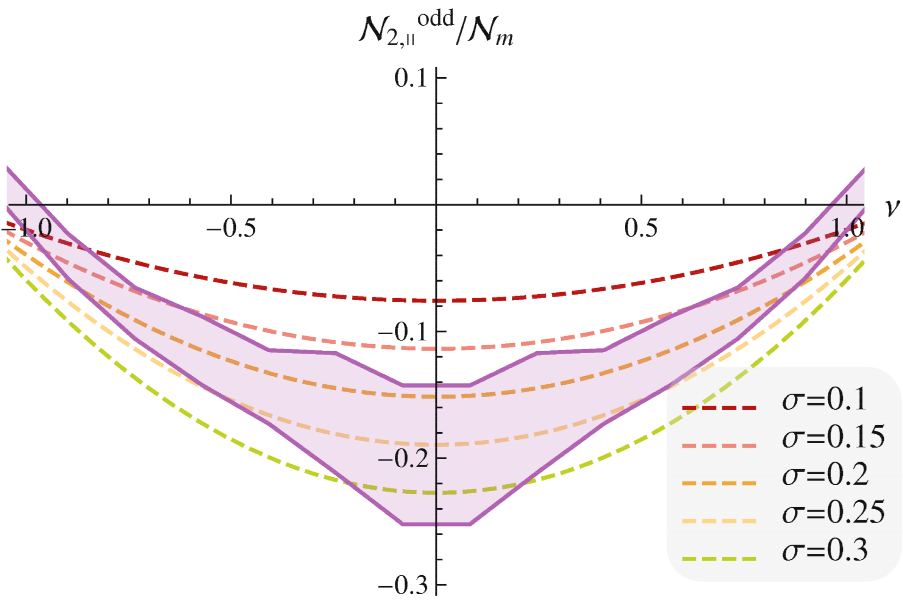}
  \caption{
Reconstructed $\beta$ from the Gaussian part of ${\cal N}_{2}$ (\sl{left panel}) and $\sigma$ from its non-Gaussian correction (\sl{right panel}) . 
\label{fig:recov-beta}
}
 \end{center}
\end{figure}

\section{Conclusion} \label{sec:conclusion}

This paper has placed on a firm footing the  statistical analysis of topological sets in anisotropic spaces.
Specifically, it has presented  extensions of \cite{Matsubara96} and \cite{Gay12} in two directions: it has accounted for anisotropic fields {\sl and } non-Gaussianity for all Minkowski's functionals.
{The main results of this work are:  i) the anisotropic JPDF in 2 and 3D: a new building block for redshift space analysis;
 ii) the analytical Euler characteristic at all order in 2D/3D (and therefore the rare event limit for extrema counts);
 iii) other Minkowski functionals and extrema counts to first order in non-Gaussian correction in anisotropic space;
   iv) and extension of PT for the relevant cumulants.
The theory presented in Sections~\ref{sec:JPDF} and \ref{sec:Min23} makes no assumption about the origin of the anisotropy and could therefore be implemented in contexts 
beyond astrophysics.

In the context  of cosmology, the theoretical expectations were robustly derived both
in invariant and field variables, and checked against Monte Carlo simulations of $f_\mathrm{{nl}}$ models, scale-invariant dark matter simulations, and $\Lambda$CDM simulations (both for DM and halo catalogues). It was shown how to use these predictions to measure $\sigma_\mathrm{ 0}$ and $\beta$ as a function of redshift.
The implication of the invariance of  Minkowski functionals and extrema counts versus  monotonic mapping on the relevant  combinations of cumulants (i.e. projections of the bispectrum) was made explicit. Its relevance for modified gravity probes was discussed. 
The implementation on dark matter halos of $\Lambda$CDM simulations for biased populations with $M>10^{11} M_\odot$, 
allowed us to quantify and formalize the  (weak) effect of redshift distortion on the 3D geometric descriptors of the field.
This weak sensitivity reflects the robustness of topological estimates.
 In contrast, the comparison of 2D
slices perpendicular and parallel to the LOS should allow us to also measure $\sigma(z)$ and correct it for  $\Omega_m^\gamma/b_1$ as demonstrated in a fiducial experiment.
At scales above $\sim$50 Mpc$/h$ it has been found that standard perturbation theory allows us to predict  the theoretical  cumulants at the level of accuracy required to 
match the measured cumulants from simulations within a range of contrast $\nu \in [-2,2]$. As  Minkowski functionnals are configuration space 
probes, we can expect a better convergence of (extended) PT for these contrasts which are less sensitive to the dynamics of very non-linear regions. 

Improvements beyond the scope of this paper include 
i) improving perturbation theory in redshift space (while implementing variations of  the streaming model  \citep[see][for instance]{Scoccimarro04,Taruya10} and/or anisotropic smoothing;
ii) departing from the plane parallel approximation while constructing a full-sky prescription for non-Gaussian Minkowski functionals of realistic catalogues;
iii) extending the prediction to the statistics of the skeleton and walls;
iv) propagating to cosmic parameter estimation the residual mis-fits;
v) extending the prediction of the JPDF to N-point statistics for non-local analysis (e.g. void size and non-linear N-points peak statistics);
vi) exploring alternative expansion to the Gram-Charlier's vii) deriving the statistics of {\sl errors} on one-point statistics such as those presented in this paper;
 viii) implementing the relevant theory on realistic mocks and demonstrating pros and cons of geometrical probes (e.g. in the presence of masks), and contrast those to existing dark energy probes (lensing, SN1a, etc..).
  
In the context of  upcoming 3D spectroscopic surveys such as {\sc Euclid},  the statistical analysis of the geometry of our  redshift-distorted  Universe 
will allow us to robustly measure  weighted moments of the multi-spectra as a function of redshift, and  henceforth quantify the cosmic evolution of the equation of 
state of dark energy and possible departure from GR.

\section*{Acknowledgments}
We thank S. Colombi, T. Nishimichi  and S. Prunet for useful comments during the course of this project, and Lena for her hospitality when  part of this work was completed.
DP thanks the LABEX ``Institut de Lagrange de Paris'' and the ``Programme National de Cosmologie'' for funding. DP research was  supported in part by the National Science Foundation under Grant No. NSF PHY11-25915. SC and CP thank the university of Alberta for funding.
This research is part of the Horizon-UK project and the  ANR Cosmo@NLO (ANR-12-BS05-0002) and Spin(E).
 Special thanks to T. Sousbie for his help in producing the Friend of Friend catalog for the 
{\sc horizon} $4\pi$ simulation  and to
 our collaborators of the  Horizon project ({\tt www.projet-horizon.fr}) for helping us produce the  simulation.  We warmly thank S. Prunet and S. Rouberol for running the \texttt{Horizon} cluster for us. 
Let us finally thank D.~Munro for freely distributing his {\sc \small  Yorick} programming language and opengl interface (available at {\tt yorick.sourceforge.net})
and the community of {\tt mathematica.stackexchange.com} for their help.

\bibliographystyle{mn2e}
\bibliography{DofKin}

\appendix
 \section{Spectral parameters and cumulants}
\label{sec:appendixA}
\subsection{Relations with the parametrization of Matsubara (1996)}
\label{sec:variance_relations}
At linear order, the redshift distortion, equation~(\ref{eq:zdensity}) reduces to \cite{Kaiser87} result in equation~(\ref{eq:Kaiser}).
\cite{Matsubara96} defines the dispersions and angular moments
\begin{equation}
\sigma_{j}^{2}( R )\equiv\frac 1 {2\pi^{2}} \int \dd k\; k^{2+2j}P^{( r )}(k)\, W^{2}(k R )\,,\quad C_{j}(f/b)=\frac 1 2 \int_{-1}^{1} \dd\mu \;\mu^{2j}(1+f b^{-1}\mu^{2})^{2}\,,
\end{equation}
and uses the normalized quantities
\begin{equation}
\alpha\equiv\frac {\delta_{R}^{(s)}}{{\sigma_0 \sqrt{C_0}}},\quad
\beta_{i}\equiv\frac {\partial_{i}\delta_{R}^{(s)}}{{\sigma_0 \sqrt{C_0}}}\quad \textrm{and}\quad
\omega_{ij}\equiv\frac {\partial_{i}\partial_{j}\delta_{R}^{(s)}}{{\sigma_0 \sqrt{C_0}}},
\end{equation}
to describe the smoothed density contrast in redshift space $\delta_{R}^{(s)}$ and its derivatives.
Our variables relate to these as
\begin{equation}
x=\alpha,\quad
x_{I}=\frac{\sqrt{C_{0}}\sigma_{0}}{\sigma_{1\perp}}\beta_{I},\quad
x_{3}=\frac{\sqrt{C_{0}}\sigma_{0}}{\sigma_{1\parallel}}\beta_{3},\quad
x_{IJ}=\frac{\sqrt{C_{0}}\sigma_{0}}{\sigma_{2\perp}}\omega_{IJ} \textrm{ and }
x_{33}=\frac{\sqrt{C_{0}}\sigma_{0}}{\sigma_{2\parallel}}\omega_{33},
\end{equation}
where $I,J\in \{1,2\}$, so that
\begin{equation}
\sigma^{2}=C_{0}\sigma_{0}^{2},\quad
\sigma_{1\perp}^{2}=(C_{0}-C_{1})\sigma_{1}^{2},\quad
\sigma_{1\parallel}^{2}=C_{1}\sigma_{1}^{2},\quad
\sigma_{2\perp}^{2}=(C_{0}-2C_{1}+C_{2})\sigma_{2}^{2},\quad
\sigma_{2\parallel}^{2}=C_{2}\sigma_{2}^{2},
\end{equation}
with
\begin{equation}
C_{0}=1+\frac{2f}{3b}+\frac{f^{2}}{5b^{2}},\quad 
C_{1}=\frac 1 3 +\frac {2f}{5b}+\frac{f^{2}}{7b^{2}},\quad
C_{2}=\frac 1 5 +\frac{2f}{7b}+\frac{f^{2}}{9b^{2}}\,.
\end{equation}
In terms of these variables,   both formulations are in exact agreement.
In the isotropic Gaussian limit (i.e the linear regime with no redshift correction) $C_0 = 1$, $C_1 = 1/3$, $C_2=1/5$.
\subsection{Inter-relations between the moments of random fields  and simplification of Gay  et al. (2012) results}
\label{sec:momentsrelation}
Not all cumulants of the field and its derivatives that appear in Gram-Charlier expansion are independent.
Relations between the cumulants can be established, for instance, by expressing statistical averages as averages over spatial coordinates
using homogeneity of the statistics and utilizing integration by parts.  The details of the results depend on the properties of
the manifold, in this paper we limit ourselves to the theory in the flat infinite Euclidean space. As an example
\begin{equation}
\langle x^2 x_{33} \rangle = \lim_{V \to \infty} \frac{1}{V} \int_V dV x^2 x_{33} = -\gamma_{\parallel} \lim_{V \to \infty} \frac{2}{V} 
\int_V d V x x_3^2 
= -2\gamma_{\parallel} \langle x x_3^2 \rangle\,,
\end{equation}
where the boundary term vanishes as volume $V$ is taken in the infinite limit. A similar procedure shows that,
e.g.,  $ \langle x_{1}^{2}x_{11}\rangle  =0$. 

Proceeding in this manner one can establish for isotropic fields
several useful general relations that involve the rotation invariants $I_s$ of the Hessian matrix $x_{ij}$
\footnote{$I_s$ are the coefficients of the characterisitic polynomial of the matrix $x_{ij}$, namely in $N$ dimensions $I_{1}$ is the trace,
$I_{N}$ is the determinant and for $1 < s < N$, $I_s$ is the sum of 
the minors of order $s$ of $x_{ij}$.}:
\begin{eqnarray}
\label{eq:ipp1}
\langle x^{n}I_{1}\rangle&=&-n \gamma \langle x^{n-1}q^{2}\rangle,
\\
\label{eq:ipp2}
\langle x^{n}I_{2}\rangle&=&-\frac 3 4 n \gamma \langle x^{n-1}q^{2}I_{1}\rangle-\frac 1 4 n(n-1) \gamma^{2} \langle x^{n-2}q^{4}\rangle,
\\
\label{eq:ipp3}
\langle x^{n} I_s \rangle& =& -\frac{1}{s} \sum_{k=1}^{\mathrm{min}(s,n)} \frac{s+k}{(2k)!!} \frac{n!}{(n-k)!}
 \gamma^k \left\langle x^{n-k} q^{2k} I_{s-k} \right\rangle\, , s\geq1\, , I_0=1
.
\end{eqnarray}
These relations are valid in any dimensions, for correspondingly defined $I_s$'s, \footnote{These relations can be generalized
to homogeneous spaces of constant curvature, e.g., 2-sphere, with additional terms that are proportional 
to the curvature of the space appearing during some integrations by parts. } and also hold for the cumulants as well
as the moments.

\subsubsection{Isotropic 3D Euler characteristic}
The relations~(\ref{eq:ipp1}-\ref{eq:ipp3}) allow to simplify the expression for the Euler characteristic
given by \cite{Gay12} for isotropic fields. Indeed, these authors found\footnote{We shall take this opportunity to
correct several unfortunate misprints in  expressions   for the second order corrections $\chi^{(2)}$
as presented in \cite{Gay12}. In this text equations~(\ref{eq:genus3DGay2}) and (\ref{eq:chi2_2D}) are corrected
versions.}
\begin{eqnarray}
&& \chi^{\textrm{iso}(0)}_{3\textrm{D}}(\nu)\label{eq:genus3DGay0}
=\frac{e^{-\nu^{2}/2} }{(2\pi)^{2}R_{\star}^{3}}\frac{\sqrt 3}{9} \gamma^{3}  H_{2}(\nu)\,,\\
 &&\chi^{\textrm{iso}(1)}_{3\textrm{D}}(\nu)\label{eq:genus3DGay1}
=\frac{e^{-\nu^{2}/2} }{(2\pi)^{2}R_{\star}^{3}}\frac{\sqrt 3}{9} \left[ 
\frac{1}{6} \gamma^{3}\left\langle  x^3\right\rangle H_{5}(\nu)
-\frac{3}{2}\left( \gamma^{3}\left\langle xq^{2}\right\rangle+ \gamma^{2} \left\langle x^{2} J_{1} \right\rangle\right) H_{3}(\nu)
+9 \left(
\frac 1 {2} \gamma^{2} \left\langle q^{2}J_{1}\right\rangle
+ \gamma \left\langle x I_{2}\right\rangle \right)  H_{1}(\nu)\right]\,,
 \\
  &&\chi^{\textrm{iso}(2)}_{3\textrm{D}}(\nu)=\frac{e^{-\nu^{2}/2} }{(2\pi)^{2}R_{\star}^{3}}\frac{\sqrt 3}{9}\left[
 - \left( \frac{27}{2} \gamma \langle q^2 I_2 \rangle_\mathrm{c} + 27 \langle x I_3 \rangle_\mathrm{c} \right) H_0(\nu) + \left( \frac{9}{8} \gamma^3 \langle q^4 \rangle_c + \frac{9}{2} \gamma^2 \langle x q^2 J_1 \rangle_\mathrm{c} + \frac{9}{2} \gamma \langle x^2 I_2 \rangle_\mathrm{c} \right) H_2(\nu)\right. \nonumber\\
 &&- \left( \frac{3}{4} \gamma^3 \langle x^2 q^2 \rangle_\mathrm{c} + \frac{1}{2} \gamma^2 \langle x^{3} J_1 \rangle_\mathrm{c} \right) H_4(\nu) + \frac{1}{24} \gamma^3 \langle x^4 \rangle_\mathrm{c} H_6(\nu)
 + \left( \frac{135}{16} \gamma \langle q^2 J_1 \rangle^2 + \frac{27}{2} \langle q^2 J_1 \rangle \langle x I_2 \rangle + \frac{81}{2} \langle x q^2 \rangle \langle I_3 \rangle \right) H_0(\nu)\nonumber \\
 &&- \left( \frac{45}{4} \gamma^2 \langle x q^2 \rangle \langle q^2 J_1 \rangle + \frac{9}{2} \gamma \langle q^2 J_1 \rangle \langle x^2 J_1 \rangle + \frac{27}{2} \gamma \langle x q^2 \rangle \langle x I_2 \rangle + \frac{9}{2} \langle x^2 J_1 \rangle \langle x I_2 \rangle + \frac{9}{2} \langle x^3 { \rangle \langle} I_3 \rangle \right) H_2(\nu)\nonumber \\
 &&+ \left(\frac{15}{8} \gamma^3 \langle x q^2 \rangle^2 + \frac{3}{4} \gamma^2 \langle x^3 \rangle \langle q^2 J_1 \rangle + \frac{9}{4} \gamma^2 \langle x q^2 \rangle \langle x^2 J_1 \rangle + \frac{3}{4} \gamma \langle x^2 J_1 \rangle^2 + \frac{3}{2} \gamma \langle x^3 \rangle \langle x I_2 \rangle \right) H_4(\nu)\nonumber \\
&& \left. - \left( \frac{1}{4} \gamma^3 \langle x q^2 \rangle \langle x^3 \rangle + \frac{1}{4} \gamma^2 \langle x^3 \rangle \langle x^2 J_1 \rangle \right) H_6(\nu) + \frac{1}{72} \gamma^3 \langle x^3 \rangle^2 H_8(\nu)
 \right],
  \label{eq:genus3DGay2}
\end{eqnarray}
where $R_{\star}=\sigma_{1}/\sigma_{2}$.
Using equations~(\ref{eq:ipp1})-(\ref{eq:ipp3}), the equations~(\ref{eq:genus3DGay0})-(\ref{eq:genus3DGay1}) can be rewritten up to  NLO as
\begin{equation}
 \chi^{\textrm{iso}(0+1)}_{3\textrm{D}}(\nu)
=\frac{e^{-\nu^{2}/2} }{(2\pi)^{2}}\frac{\sigma_{1}^{3} }{\sigma^{3}}\frac{\sqrt 3}{9} \left[H_{2}(\nu)+
\frac{1}{6}H_{5}(\nu)\left\langle  x^3\right\rangle
+\frac{3}{2}H_{3}(\nu)\left\langle xq^{2}\right\rangle
-\frac{9}{4} H_{1}(\nu) 
\frac 1 {\gamma} \left\langle q^{2}J_{1}\right\rangle
\right]\,,
\end{equation}
which is the form we quote in equation~(\ref{eq:chibar}). The NNLO, ${{\cal O}(\sigma^2)}$,
term 
can be also simplified from equation~(\ref{eq:genus3DGay2}) to a more compact form
\begin{multline}
\chi_{3\textrm{D}}^{\textrm{iso}(2)}(\nu)=
\frac{e^{-\nu^{2}/2} }{(2\pi)^{2}}\frac{\sigma_{1}^{3}}{\sigma^{3}}\frac{\sqrt 3}{9} \times
\\
\left[ H_{0}(\nu)\left(\frac 9 2 \gamma^{-2} \left\langle q^{2}I_{2}\right\rangle_{c}-\frac{27}{16}\gamma^{-2}
 \left\langle J_{1}q^{2}\right\rangle^{2} \right)
 +H_{2}(\nu)\left(-\frac 9 8  \left\langle q^{4}\right\rangle_{c} - \frac 9 4 \gamma^{-1} 
 \left\langle x q^2 J_1\right\rangle_{c}+\frac 9 8 \gamma^{-1}
  \left\langle x q^{2}\right\rangle\left\langle J_{1}q^{2}\right\rangle\right) \right. \\
\left.
+H_{4}(\nu)\left(\frac 3 4  \left\langle x^{2}q^{2}\right\rangle_{c} +\frac 3 8 \left\langle xq^{2}\right\rangle^{2}
-\frac 3 8 \gamma^{-1} \left\langle x^{3}\right\rangle \left\langle J_{1}q^{2}\right\rangle\right)
 +H_{6}(\nu)\left(\frac 1 {24} \left\langle x^{4}\right\rangle_{c}+\frac 1 4  \left\langle x^{3}\right\rangle \left\langle x q^{2}\right\rangle\right)
+H_{8}(\nu)\frac 1 {72} \left\langle x^{3}\right\rangle^{2}
\right]\,.
\end{multline}
\subsubsection{Isotropic 2D Euler characteristic}
The same procedure holds for the 2D Euler characteristic for which the expression given by \cite{Gay12}
 in terms of a 2D isotropic field $x$
\begin{eqnarray}
&&\chi_{2\textrm{D}}^{\textrm{iso}(0)}(\nu)
=\frac {e^{-\nu^{2}/2}} {(2\pi)^{3/2}}\frac{1}{2R_{\star}^{2}}\gamma^{2}H_{1}(\nu),\\
&&\chi_{2\textrm{D}}^{\textrm{iso}(1)}(\nu)
=\frac {e^{-\nu^{2}/2}} {(2\pi)^{3/2}}\frac{1}{2R_{\star}^{2}}\left[\frac 1 {3!} \gamma^{2}\left\langle x^3\right\rangle H_4\left(\nu\right) 
-\left( \gamma^{2}\left\langle x q^{2}\right\rangle+ {\gamma} \left\langle x^{2}J_{1}\right\rangle\right)H_{2}(\nu)
+\left( {2}{\gamma}\left\langle q^{2} J_{1}\right\rangle+4 \left\langle x I_{2}\right\rangle\right)
\right],\\
&&\chi_{2\textrm{D}}^{\textrm{iso}(2)}(\nu)
=\frac {e^{-\nu^{2}/2}} {(2\pi)^{3/2}}\frac{1}{2R_{\star}^{2}}\left[
\left( \frac{\gamma^{2}}{2} \langle q^4 \rangle_\mathrm{c}
 + 2\gamma \langle x q^2 J_1 \rangle_\mathrm{c} + 2\langle x^2 I_2
\rangle_\mathrm{c}  \right) H_1(\nu) - \left( \frac{1}{2}\gamma^{2} \langle x^2 q^2
\rangle_\mathrm{c}  + \frac{\gamma}{3} \langle x^3 J_1 \rangle_\mathrm{c}
\right) H_3(\nu) + \frac{\gamma^{2}}{24} \langle x^4 \rangle_\mathrm{c} H_5(\nu)
\nonumber\right.
\\&&-\left( 4 \gamma \langle x q^2 \rangle \langle q^2 J_1 \rangle + 
\langle q^2 J_1 \rangle \langle x^2 J_1 \rangle + 4\langle x q^2 \rangle \langle
x I_2 \rangle \right) H_1(\nu)
- \left( \frac{{\gamma^{2}}}{6} \langle x q^2 \rangle \langle x^3 \rangle + 
\frac{\gamma}{6} \langle x^3 \rangle \langle x^2 J_1 \rangle \right) H_5(\nu) +
\frac{\gamma^{2}}{72} \langle x^3 \rangle^2 H_7(\nu) \, \nonumber \\
\label{eq:chi2_2D}
&&
\left.+ \left( \gamma^{2}\langle x q^2 \rangle^2 + \frac{\gamma}{3} \langle x^3
\rangle \langle q^2 J_1 \rangle +
 \gamma \langle x q^2 \rangle \langle x^2 J_1 \rangle + \frac{1}{4} \langle x^2
J_1 \rangle^2 + \frac{2}{3} \langle x^3 \rangle \langle x I_2 \rangle \right)
H_3(\nu)
\right]
\end{eqnarray}
can be simplified up to NLO to
\begin{equation}
\chi_{2\textrm{D}}^{\textrm{iso}(0+1)}(\nu)
=\frac{\sigma_{1}^{2}}{2\sigma^{2}} \frac {e^{-\nu^{2}/2}} {(2\pi)^{3/2}}\left[H_{1}(\nu)+\frac 1 {3!}H_4\left(\nu\right)  \left\langle x^3\right\rangle
+H_{2}(\nu) \left\langle x q^{2}\right\rangle
-\gamma^{-1} \left\langle q^2 J_{1}\right\rangle
+{\cal O}(\sigma^{2})\right],
\end{equation}
which is to be compared with expression~(\ref{eq:chibar2D}) for the Euler characteristic on 2D slices.
The NNLO term adds: 
\begin{multline}
\chi_{2\textrm{D}}^{\textrm{iso}(2)}(\nu)=
\frac{\sigma_{1}^{2}}{2\sigma^{2}} \frac {e^{-\nu^{2}/2}} {(2\pi)^{3/2}}\left[
H_{1}(\nu)\left(-\frac 1 2 \left\langle  q^{4}\right\rangle_{c} - \gamma^{-1} \left\langle x q^2 J_1 \right\rangle_{c}
+\gamma^{-1} \left\langle x q^{2}\right\rangle\left\langle  q^{2}J_{1}\right\rangle\right)\right. \\
\left.
+H_{3}(\nu)\left(\frac 1 2 \left\langle x^{2} q^{2}\right\rangle_{c}-
\frac 1 6 { \gamma^{-1}} \left\langle x^{3}\right\rangle\left\langle q^{2}J_{1}\right\rangle\right)
+H_{5}(\nu)\left(\frac 1 {24} \left\langle x^{4}\right\rangle_{c}
+\frac 1 6 \left\langle x q^{2}\right\rangle\left\langle x^{3}\right\rangle\right)
+H_{7}(\nu)\frac 1 {72} \left\langle x^{3}\right\rangle^{2}
\right].
\end{multline}

\section{  ($f_\mathrm{{nl}}$,$g_\mathrm{{nl}}$) non-Gaussian field  toy models} \label{sec:toys}
\subsection{  $f_\mathrm{{nl}}$  toy model for the field}
Let us briefly describe the toy model used in the main text to validate our predictions for extrema counts and Minkowski functionals in 2D and 3D.
If X is a Gaussian (possibly anisotropic) field with zero mean and variance $\left\langle X^{2}\right\rangle=\sigma_{0}^{2}$, let us define the following non-Gaussian field
\begin{equation}
X^\mathrm{ NG}=X+\frac{f_\mathrm{{nl}}}{\sigma_{0}}\left( X^{2}-\sigma_{0}^{2}\right)+\frac{g_\mathrm{{nl}}}{\sigma_{0}^2}\left( X^{3}-3 \sigma_{0}^{2} X\right)\,.
\end{equation}
For the purpose of this paper, let us define $X$ as a Kaiser transform of a Gaussian isotropic field, $X^\mathrm{ G}$, which can be written in Fourier space, following
equation~(\ref{eq:Kaiser_mass}), as 
\begin{equation}
\hat X_{k}=(1+f\mu^{2}) \hat X^\mathrm{ G}_{k}, \quad\mathrm{ with}\quad \mu=k_{\parallel}/k.
\end{equation}
Then any Minkowski functional or extrema count of this non-Gaussian field can simply be written in terms only of $f$,
 $f_\mathrm{{nl}}$,  $g_\mathrm{{nl}}$ and $\sigma$'s (or $n$ for a scale invariant powerspectrum).
 Indeed, one can easily show that the spectral parameters of the non-Gaussian field are the following in terms
 of the spectral parameters of $X$ (in 2D):
 $\tilde\sigma^{2}=(1+2f_\mathrm{{nl}}^{2})\sigma_{0}^{2}$,
 $\tilde\sigma_{1\perp}^{2}=(1+4f_\mathrm{{nl}}^{2})\sigma_{1\perp}^{2}$,
 $\tilde\sigma_{1\parallel}^{2}=(1+4f_\mathrm{{nl}}^{2})\sigma_{1\parallel}^{2}$,
 $\tilde\sigma_{2\perp}^{2}=(1+4f_\mathrm{{nl}}^{2}+{ 12}f_\mathrm{{nl}}^{2}\gamma_{\perp}^{2})\sigma_{2\perp}^{2}$,
 $\tilde\sigma_{2\parallel}^{2}=(1+4f_\mathrm{{nl}}^{2}+12f_\mathrm{{nl}}^{2}\gamma_{\parallel}^{2})\sigma_{2\parallel}^{2}$
 and $\tilde\sigma_{Q}^{2}=(1+4f_\mathrm{{nl}}^{2}+4f_\mathrm{{nl}}^{2}\gamma_{\perp}\gamma_{\parallel}/\gamma_{2})\sigma_{Q}^{2}$
 where in turn the spectral parameters of $X$ can be expressed as functions of the spectral 
parameters of $X^\mathrm{ G}$ : $\sigma_{0}^{2}=(1+f+3f^{2}/8)\sigma^{2}$, 
$\sigma_{1\parallel}^{2}=1/16(8+12f+5f^{2})\sigma_{1}^{2}$, $\sigma_{1\perp}^{2}=1/16(8+4f+f^{2}
)\sigma_{1}^{2}$, $\sigma_{2\parallel}^{2}=1/128(
48+80f+35f^{2})\sigma_{2}^{2}$, $\sigma_{2\perp}^{2}=1/128(
48+16f+3f^{2})\sigma_{2}^{2}$ and $\sigma_{Q}^{2}=1/128(16+16f+5f^{2})\sigma_{2}^{2}$. In 2D, the three-point cumulants of the non-Gaussian field can also be computed: 
$\left\langle x^{3}\right \rangle=f_\mathrm{{nl}}(6+8f_\mathrm{{nl}}^{2})/(1+2f_\mathrm{{nl}}^{2})^{3/2}$, 
$\left\langle x q_{||,\perp}^{2}\right\rangle=4f_\mathrm{{nl}}\sqrt{1+2f_\mathrm{{nl}}^{2}}/(1+4f_\mathrm{{nl}}^{2})$ and
$\left\langle J_{1\perp} q_{||}^{2}\right\rangle/\gamma_{\perp}=-4 f_\mathrm{{nl}}\sqrt{1+2f_\mathrm{{nl}}^{2}}/(1+4f_\mathrm{{nl}}^{2})$.

In 3D, similar relations hold:
$\tilde\sigma^{2}=(1+2f_\mathrm{{nl}}^{2})\sigma_{0}^{2}$,
 $\tilde\sigma_{1\perp}^{2}=(1+4f_\mathrm{{nl}}^{2})\sigma_{1\perp}^{2}$,
 $\tilde\sigma_{1\parallel}^{2}=(1+4f_\mathrm{{nl}}^{2})\sigma_{1\parallel}^{2}$,
 $\tilde\sigma_{2\perp}^{2}=(1+4f_\mathrm{{nl}}^{2}+{ 8}f_\mathrm{{nl}}^{2}\gamma_{\perp}^{2})\sigma_{2\perp}^{2}$,
 $\tilde\sigma_{2\parallel}^{2}=(1+4f_\mathrm{{nl}}^{2}+12f_\mathrm{{nl}}^{2}\gamma_{\parallel}^{2})\sigma_{2\parallel}^{2}$
 and $\tilde\sigma_{Q}^{2}=(1+4f_\mathrm{{nl}}^{2}+4f_\mathrm{{nl}}^{2}\gamma_{\perp}\gamma_{\parallel}/\gamma_{2})\sigma_{Q}^{2}$
 where again it can be related to the spectral parameters of the isotropic Gaussian field $X$ : $\sigma_{0}^{2}=(1+2f/3+f^{2}/5)\sigma^{2}$, 
$\sigma_{1\parallel}^{2}=(1/3+2f/5+f^{2}/7)\sigma_{1}^{2}$, $\sigma_{1\perp}^{2}=(2/3+4f/15+2f^{2}/35)
)\sigma_{1}^{2}$, $\sigma_{2\parallel}^{2}=(
1/5+2f/7+f^{2}/9)\sigma_{2}^{2}$, $\sigma_{2\perp}^{2}=(8/15+16f/105+8f^{2}/315)\sigma_{2}^{2}$ 
and $\sigma_{Q}^{2}=(2/15+4f/35+2f^{2}/63)\sigma_{2}^{2}$; the three-point cumulants of interest being: 
$\left\langle x^{3}\right \rangle=f_\mathrm{{nl}}(6+8f_\mathrm{{nl}}^{2})/(1+2f_\mathrm{{nl}}^{2})^{3/2}$, 
$\left\langle x q_{||,\perp}^{2}\right\rangle=4f_\mathrm{{nl}}\sqrt{1+2f_\mathrm{{nl}}^{2}}/(1+4f_\mathrm{{nl}}^{2})$,
$\left\langle J_{1\perp} q_{\perp}^{2}\right\rangle/\gamma_{\perp}=-2 f_\mathrm{{nl}}\sqrt{1+2f_\mathrm{{nl}}^{2}}/(1+4f_\mathrm{{nl}}^{2})$ and
$\left\langle J_{1\perp} q_{||}^{2}\right\rangle/\gamma_{\perp}=-4 f_\mathrm{{nl}}\sqrt{1+2f_\mathrm{{nl}}^{2}}/(1+4f_\mathrm{{nl}}^{2})$.

Figures~{\ref{fig:genusGfnl} and {\ref{fig:genusNGfnl} illustrate  the interplay between anisotropy and non-Gaussianity 
that this paper investigates on this example of  the ``$f_{nl}$'' toy model.
It presents the 2D/3D Euler characteristic  of a mildly non-Gaussian anisotropic field 
for different values of the quadratic parameter  $f_\mathrm{{nl}}$ and anisotropy, $f$. Both predictions derived in Sections~(\ref{sec:genus3D}) and (\ref{sec:genus2D}), and measurements from Monte Carlo realizations of the  corresponding field are shown. The agreement between theory and measurements is very good.
As expected, in 2D the effect of anisotropy  is larger than in 3D. 
\begin{figure}
 \begin{center}
 \includegraphics[width=0.45\textwidth]{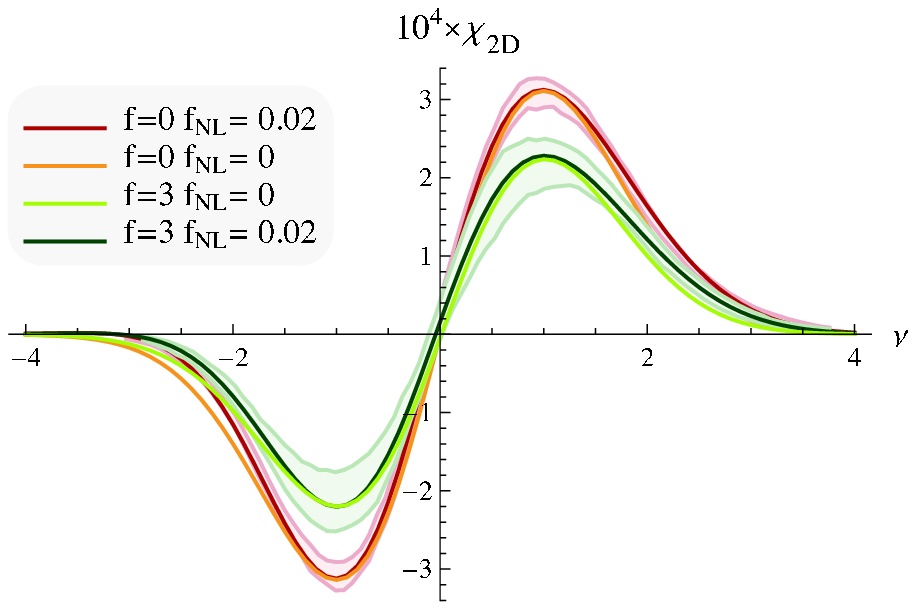}
  \includegraphics[width=0.45\textwidth]{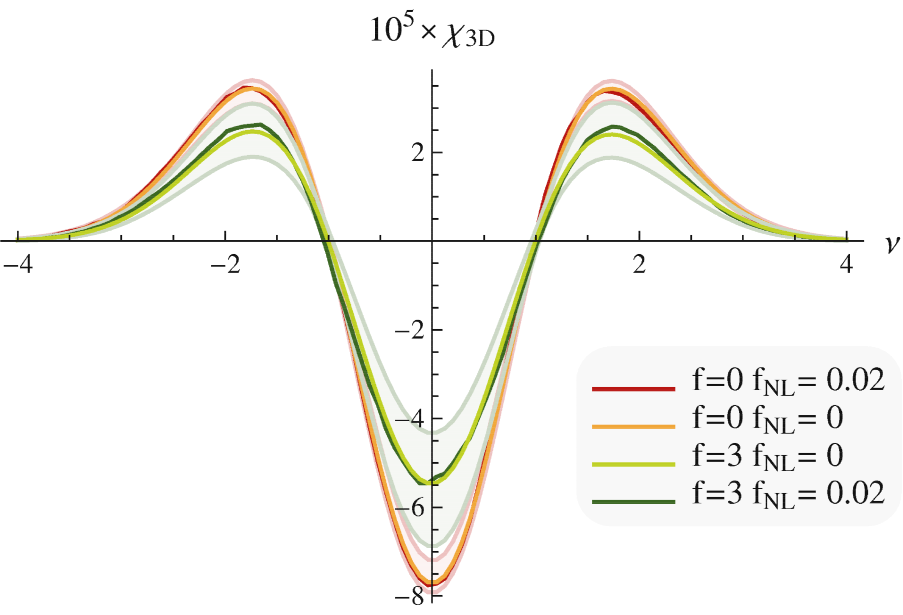}
 \includegraphics[width=0.45\textwidth]{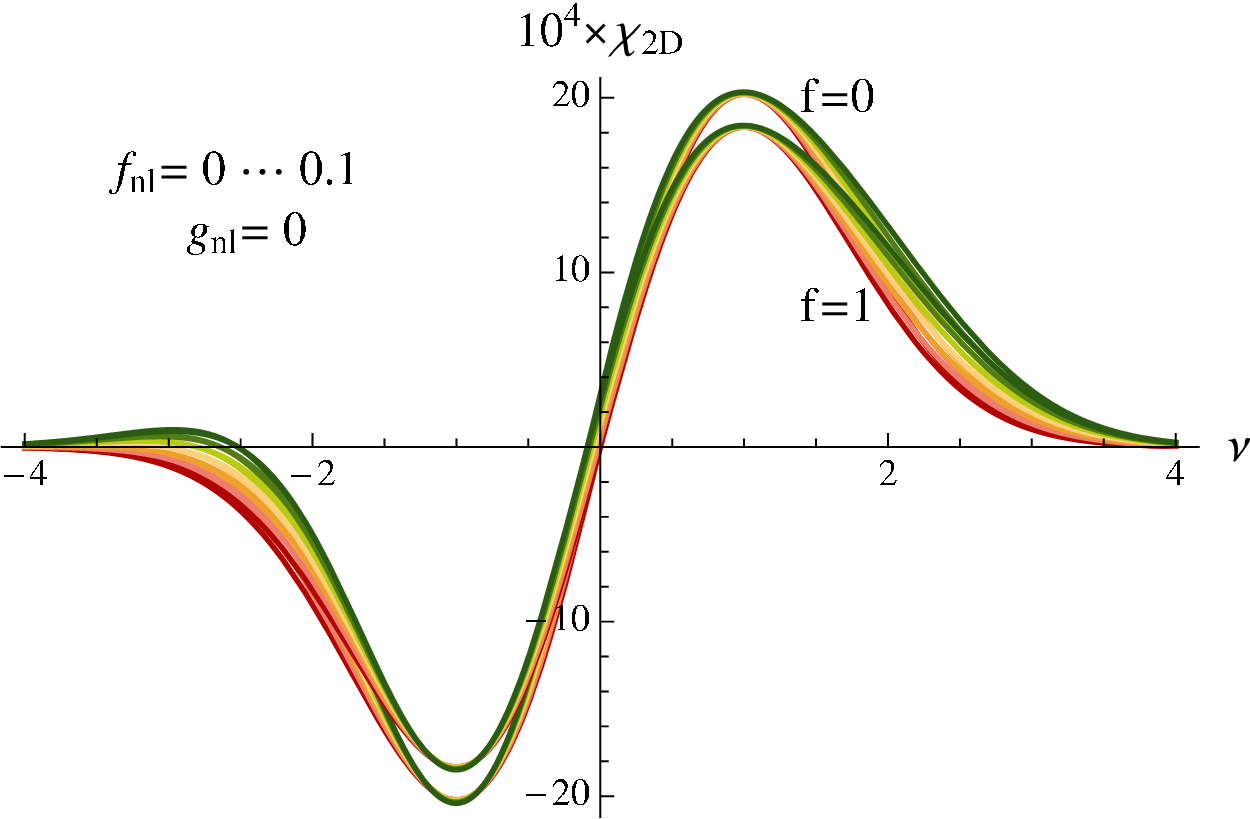}
  \includegraphics[width=0.45\textwidth]{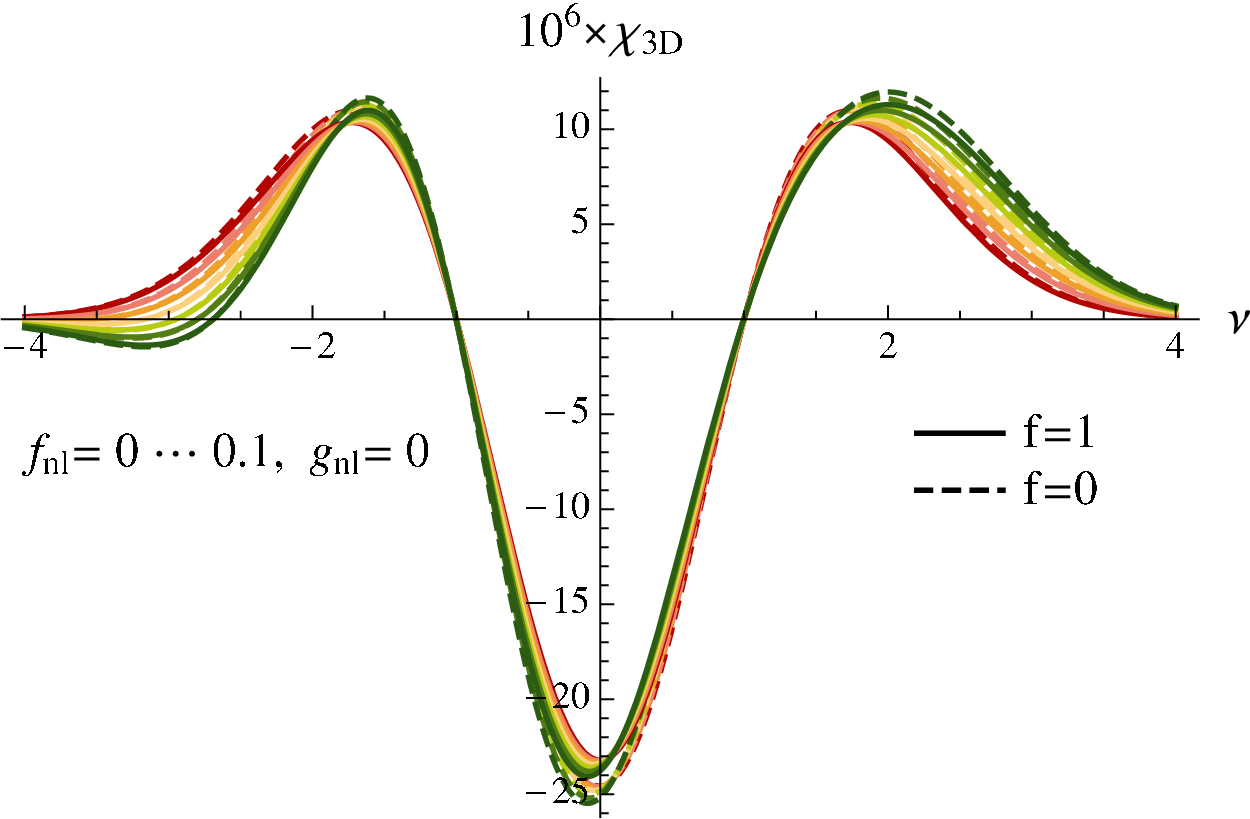}
\caption{2D ({\sl left}) and 3D ({\sl right}) Euler characteristic  of a mildly non-Gaussian anisotropic field in a toy  model.
{\sl Top left:} Comparison with measurements for different values of quadratic parameter  $f_\mathrm{{nl}}$ and anisotropy, $f$ as labeled.
The shaded area corresponds to the  one sigma measurements  error on   realisations of the non-Gaussian fields.
The overall amplitude has been rescaled by $10^4$.
 Note both the change in amplitude with $f$ and the distortion in the tails of the distribution with $f_\mathrm{{nl}}$ which are well fitted by the model. {\sl Bottom left panel:} theoretical evolution of the 2D Euler characteristic as a function of $f_\mathrm{{nl}}$ as labeled for $f=0$ and $f=1$. For this level of anisotropy, the difference in amplitude is less significant.
{\sl Top right } and {\sl bottom right} panels: same as the correspondent left panels in 3D.
\label{fig:genusGfnl}
}
 \end{center}
\end{figure}
\begin{figure}
 \begin{center}
 \includegraphics[width=0.45\textwidth]{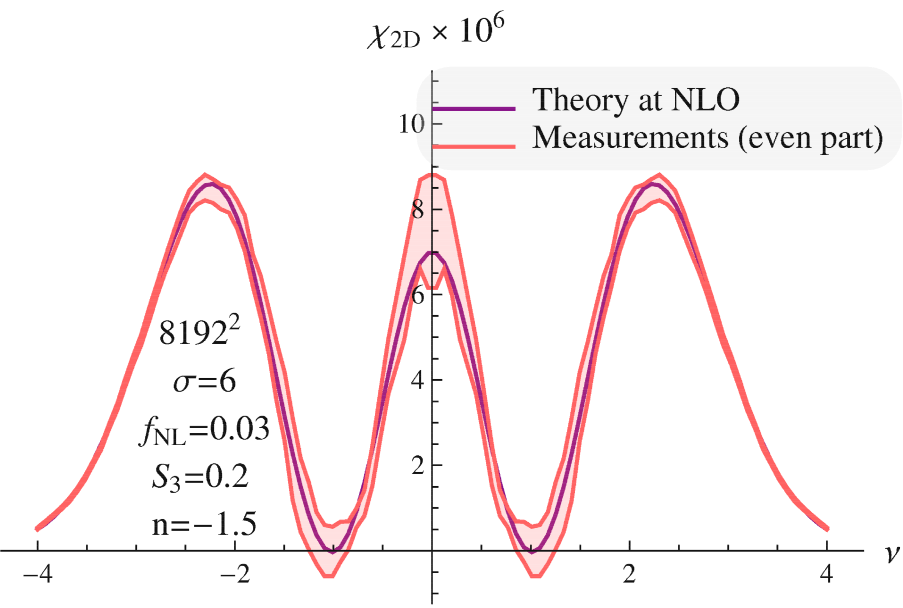}
 \includegraphics[width=0.45\textwidth]{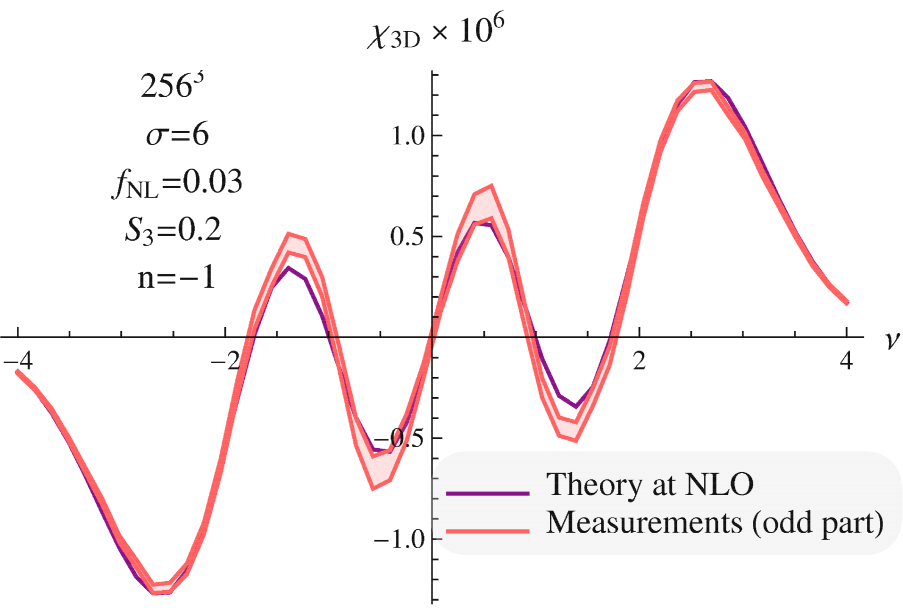}
 \includegraphics[width=0.45\textwidth]{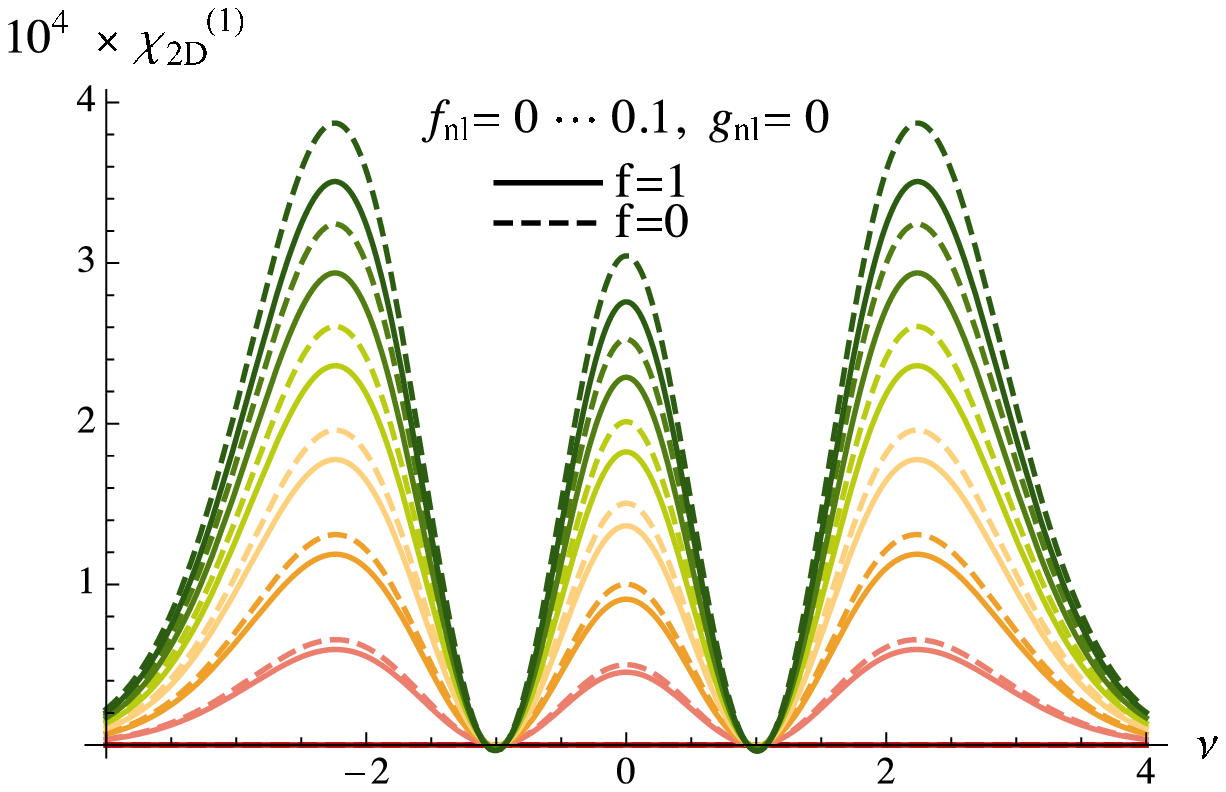}
 \includegraphics[width=0.45\textwidth]{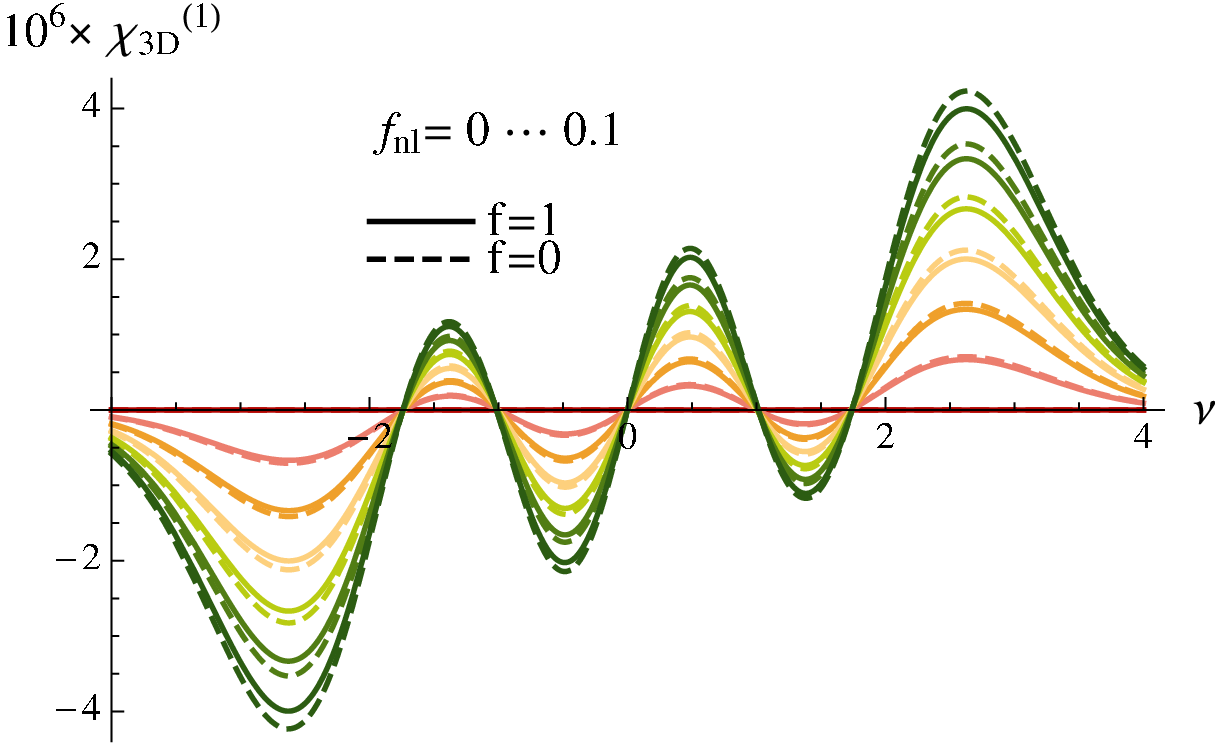}
 \caption{  non-Gaussian part of the Euler characteristic  of a mildly non-Gaussian anisotropic field in 2D (\emph{left panels}) and 3D  (\emph{right panels}).
 The \emph{top panels} compares the theoretical prediction to  the measurements of   realisations of the non-Gaussian fields with  parameters as labeled, while the
   \emph{bottom panels} illustrates the variation of the theory with the amplitude of $f_\mathrm{{ nl}}$.
 The shaded area corresponds to the  one sigma measurements  error. 
 The match to the theory is very good throughout. The effect of redshift distortion is clearly weaker in 3D, as expected. 
\label{fig:genusNGfnl}
}
 \end{center}
\end{figure}

Figure~{\ref{fig:genusNGgnl} shows how $g_{nl}$ acts in turn. For this purpose, the prediction at NNLO ($n=4$ in the Gram-Charlier expansion) is computed from equation~(\ref{eq:chi2D}).
\begin{figure}
 \begin{center}
  \includegraphics[width=0.45\textwidth]{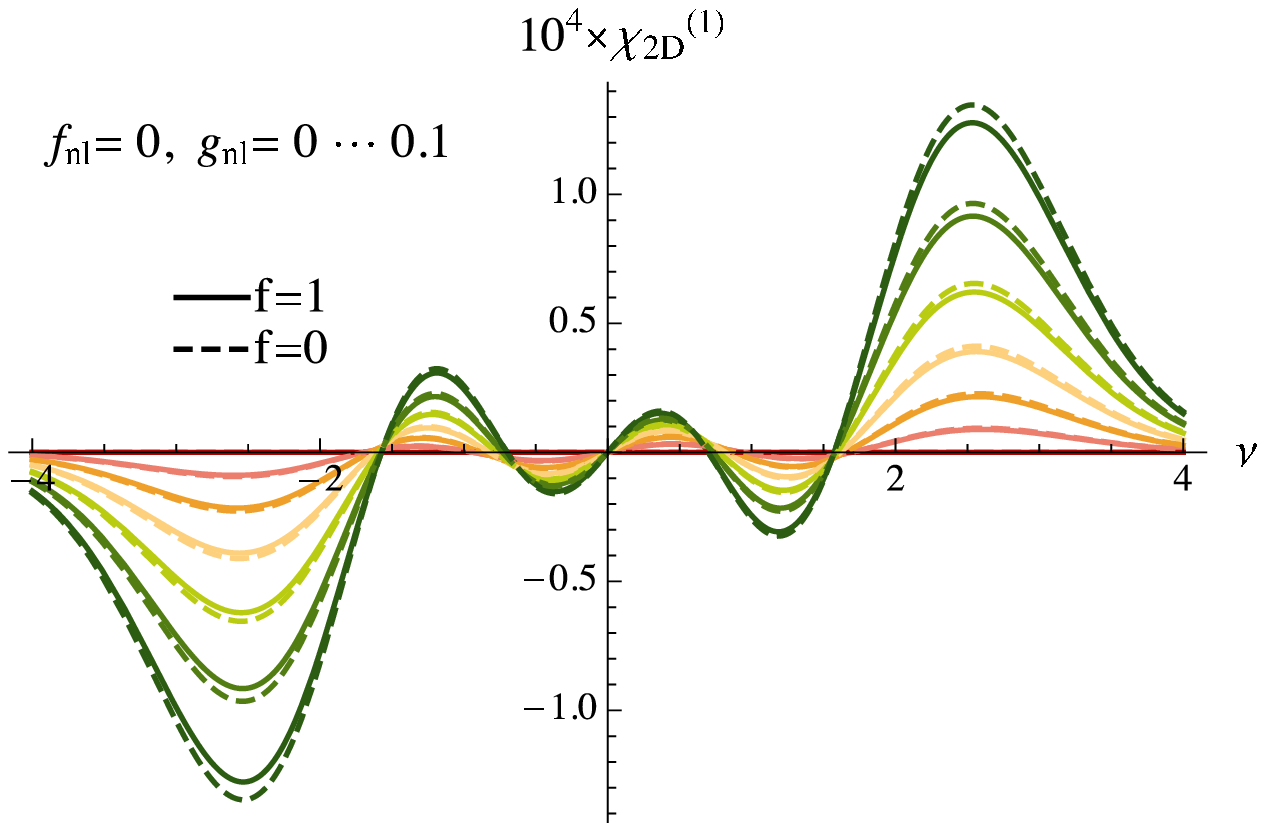}
    \includegraphics[width=0.45\textwidth]{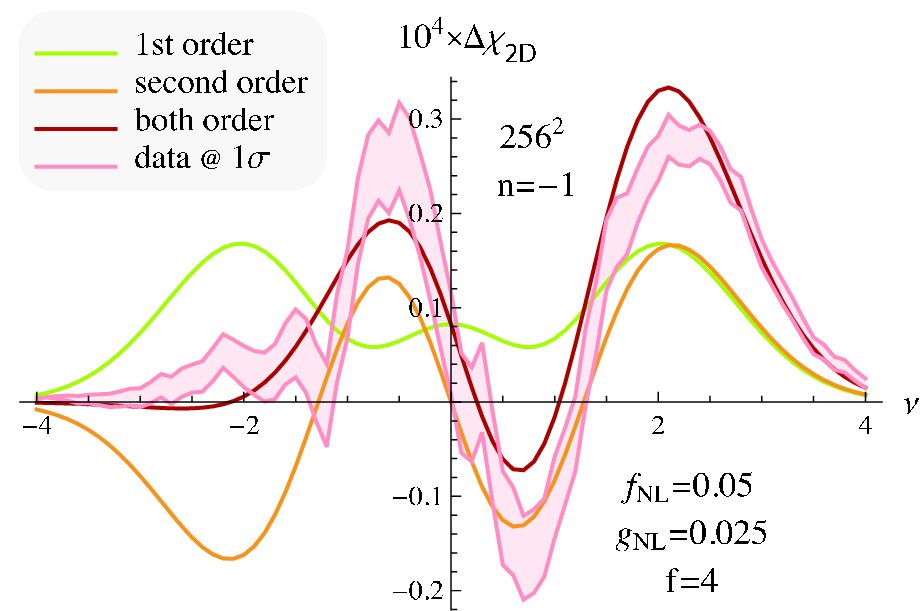}
\caption{ \emph{ Left:}  variation  with the amplitude of $g_\mathrm{{nl}}$ of the predicted Euler characteristic at NLO  of a mildly 2D non-Gaussian anisotropic field.
\emph{ Right:} match between a given realisation with $f_\mathrm{{nl}}$ and $g_\mathrm{{nl}}$ as labeled, and the corresponding model decomposed into first ($n=3$) and second order ($n=4$) corrections. 
\label{fig:genusNGgnl}
}
 \end{center}
\end{figure}
Finally, Figure~\ref{fig:extrema} displays predictions and measurements for differential extrema counts in these kinds of $f_{nl}$ fields. As mentioned in the main text, at even orders in the GC expansion (in particular at Gaussian order), 2D saddle counts is even, peak and void counts are symetric by reflexion across the y-axis meaning $\partial_{\nu} n_{\rm peaks}(\nu)=\partial_{\nu} n_{\rm voids} (-\nu)$ and pancake and filament counts are also symetric by reflection across the y-axis. On the other hand, at odd orders in the GC expansion (in particular at first non-Gaussian order), the 2D saddle count is odd in $\nu$, peak and void counts are symmetric w.r.t. reflection relative to the origin, meaning $\partial_{\nu} n_{\rm peaks}(\nu)=-\partial_{\nu} n_{\rm voids} (-\nu)$ and pancake and filament counts are also symmetric by reflection relative to the origin. Figure~\ref{fig:extrema} thus displays on left panels the even part of the data which should be dominated by the Gaussian term, namely $(\partial_{\
nu}n_{\rm sad}(\nu)
+\partial_{\nu}n_{\rm sad}(-\nu))/2$ for 2D saddle points, $(\partial_{\nu}n_{\rm peaks}(\nu)+\partial_{\nu}n_{\rm voids}(-\nu))/2$ for peaks and $(\partial_{\nu}n_{\rm voids}(\nu)+\partial_{\nu}n_{\rm peaks}(-\nu))/2$ for voids (pancakes and filaments accordingly); and in the right panels, the even part of the measurement, namely $(\partial_{\nu}n_{\rm sad}(\nu)-\partial_{\nu}n_{\rm sad}(-\nu))/2$ for 2D saddle points, $(\partial_{\nu}n_{\rm peaks}(\nu)-\partial_{\nu}n_{\rm voids}(-\nu))/2$ for peaks and $(\partial_{\nu}n_{\rm voids}(\nu)-\partial_{\nu}n_{\rm peaks}(-\nu))/2$ for voids. Agreement between measurement and theory to NLO is very good. The small difference between both is due to NNLO correction and to a bias in the measurement of pancake counts which is yet to be fixed. 
\begin{figure}
 \begin{center}
\includegraphics[width=0.45\textwidth]{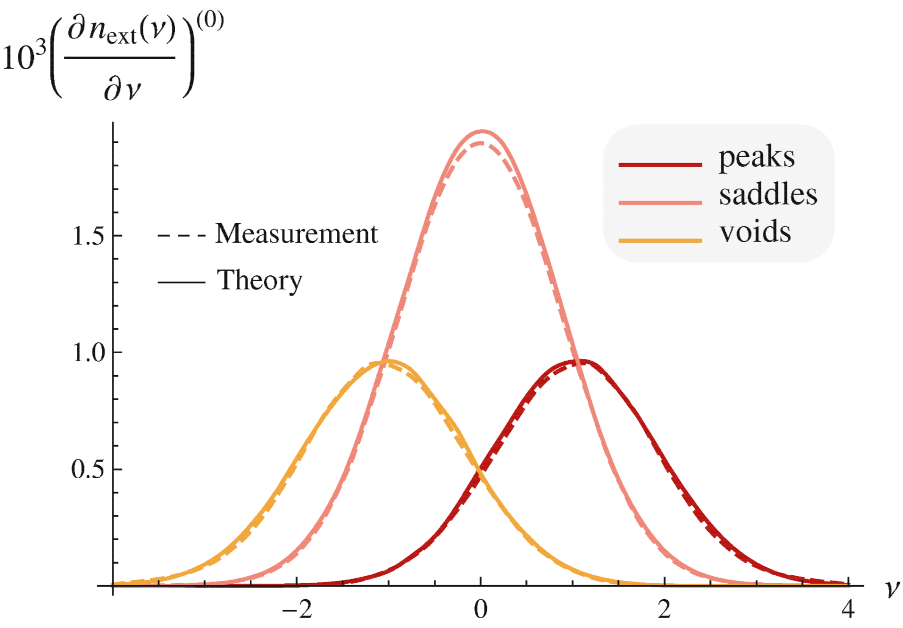}
\includegraphics[width=0.45\textwidth]{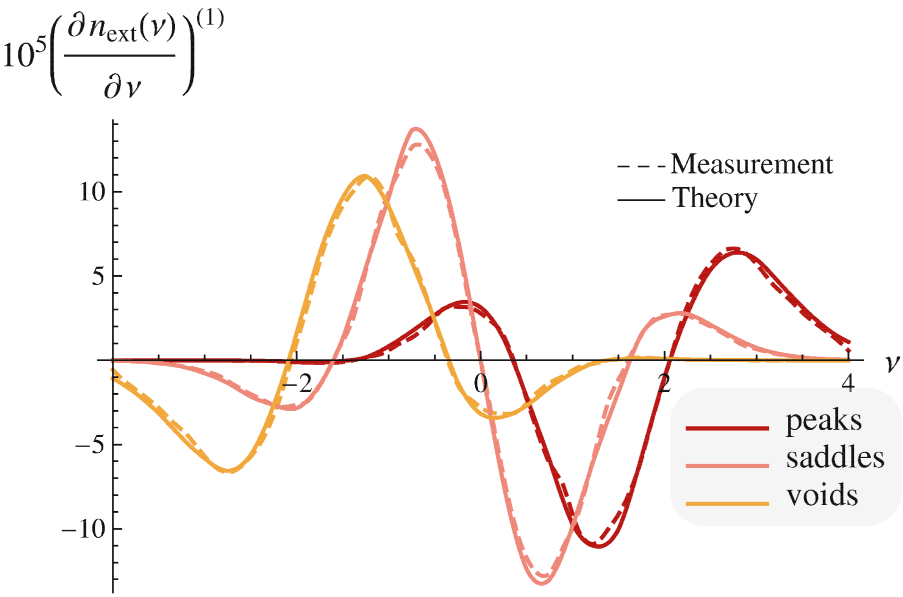}
\includegraphics[width=0.45\textwidth]{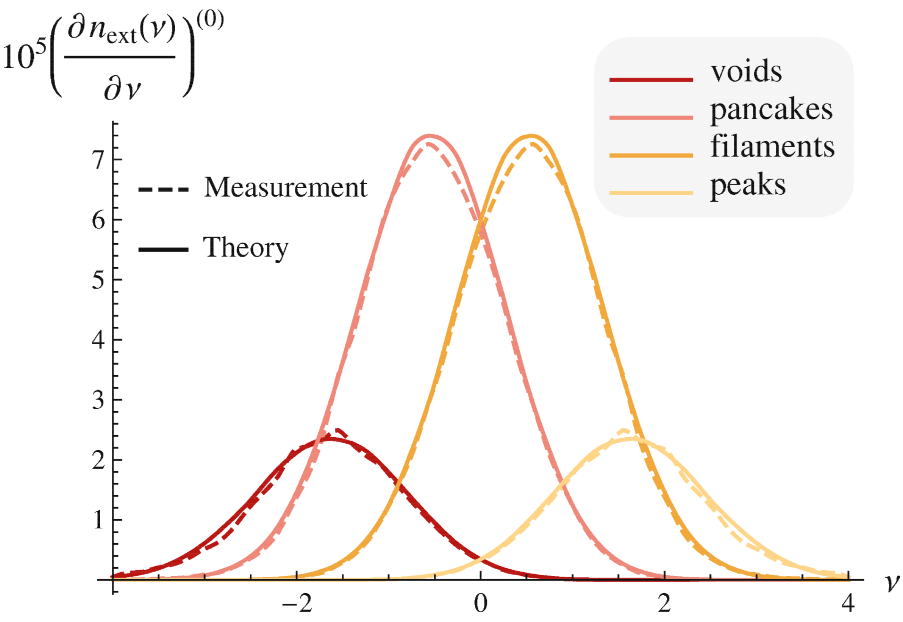}
\includegraphics[width=0.45\textwidth]{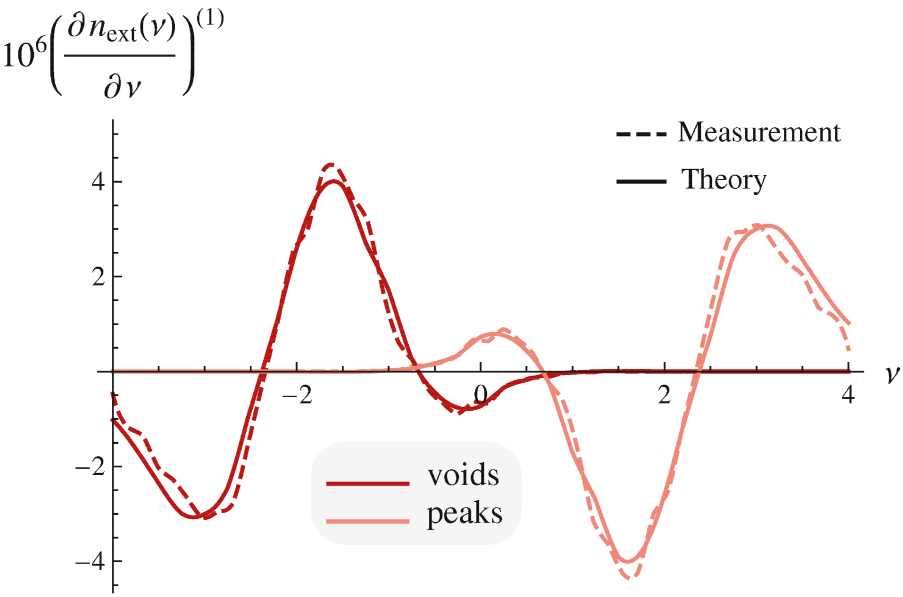}
\caption{
{\sl Top left panel}: The distribution of 2D extrema (voids, saddles and peaks as labeled) as a function of threshold $\nu$
in redshift space for $f=1$ as predicted at Gaussian order (solid) and measured (``even part'' of the data only, see main text for more details).
{\sl Top right panel}:
the corresponding non-Gaussian correction ($f_\mathrm{{nl}}=0.05$) estimated from 20 realisations of the corresponding  scale invariant (n=-1) field of dimension $4096^2$
smoothed over 4 pixels as predicted at first non-Gaussian order (solid) and measured (``odd part'' of the data only).
{\sl Bottom panels}:  Same as top panels for the distribution of 3D extrema (voids, wall saddles, tube saddle and peaks as labeled)  for non-Gaussian scale-invariant ($n=-1$, $256^3$ smoothed over 4 pixels) anisotropic field ($f_\mathrm{{nl}}=0.07$, $f=1$). 
\label{fig:extrema}
}
 \end{center}
\end{figure}

\subsection{  $f_\mathrm{{nl}}$ model for the potential} \label{sec:toyphi}
In Astrophysics, $f_{nl}$ often refers to the potential from which the (density) field derives. Let us thus study in this section the effect of such $f_{nl}$ non-Gaussianities in the potential on the genus of the density field.
If $\Phi$ is a Gaussian (possibly anisotropic) potential field with zero mean and variance $\left\langle \Phi^{2}\right\rangle=\sigma_{0}^{2}$, let us define the following non-Gaussian field
\begin{equation}
\Phi^\mathrm{ NG}=\Phi+\frac{f_\mathrm{{nl}}}{\sigma_{0}}\left( \Phi^{2}-\sigma_{0}^{2}\right)\,.
\end{equation}
Then the resulting non-Gaussian density field reads
\begin{equation}
\delta^\mathrm{ NG}=\Delta\Phi^\mathrm{ NG}=\Delta\Phi+2\frac{f_\mathrm{{nl}}}{\sigma_{0}}\left( \Phi \Delta \Phi+\nabla \Phi\cdot \nabla \Phi\right)\,.
\end{equation}
For the purpose of this paper, let us define again the final density field as a Kaiser transform of this field (c.f. equation~(\ref{eq:Kaiser_mass})) so that in Fourier  space
\begin{equation}\hat
\delta_{k}=(1+f\mu^{2}) \hat \delta^\mathrm{ NG}_{k}, \quad\mathrm{ with}\quad \mu=k_{\parallel}/k.
\end{equation}
Then any Minkowski functional or extrema count of this non-Gaussian field can be written in terms only of 
$f$, $f_\mathrm{{nl}}$ and $\sigma$'s (or $n$ for a scale invariant power spectrum). 
Figure~\ref{fig:realfnl} ({\sl top panels}) shows the prediction for the 2D Euler characteristic 
for different values of $f_\mathrm{{nl}}$ between 0 and 1 and for $f=0$ (isotropic case) and $f=1$ (Kaiser). 
It also compares ({\sl bottom panels}) the prediction with some measurements and shows a great agreement up
to first order (the difference we  observe must come from higher order corrections as the parity in Hermite polynomials
is different). 
\begin{figure}
 \begin{center}
 \includegraphics[width=0.45\textwidth]{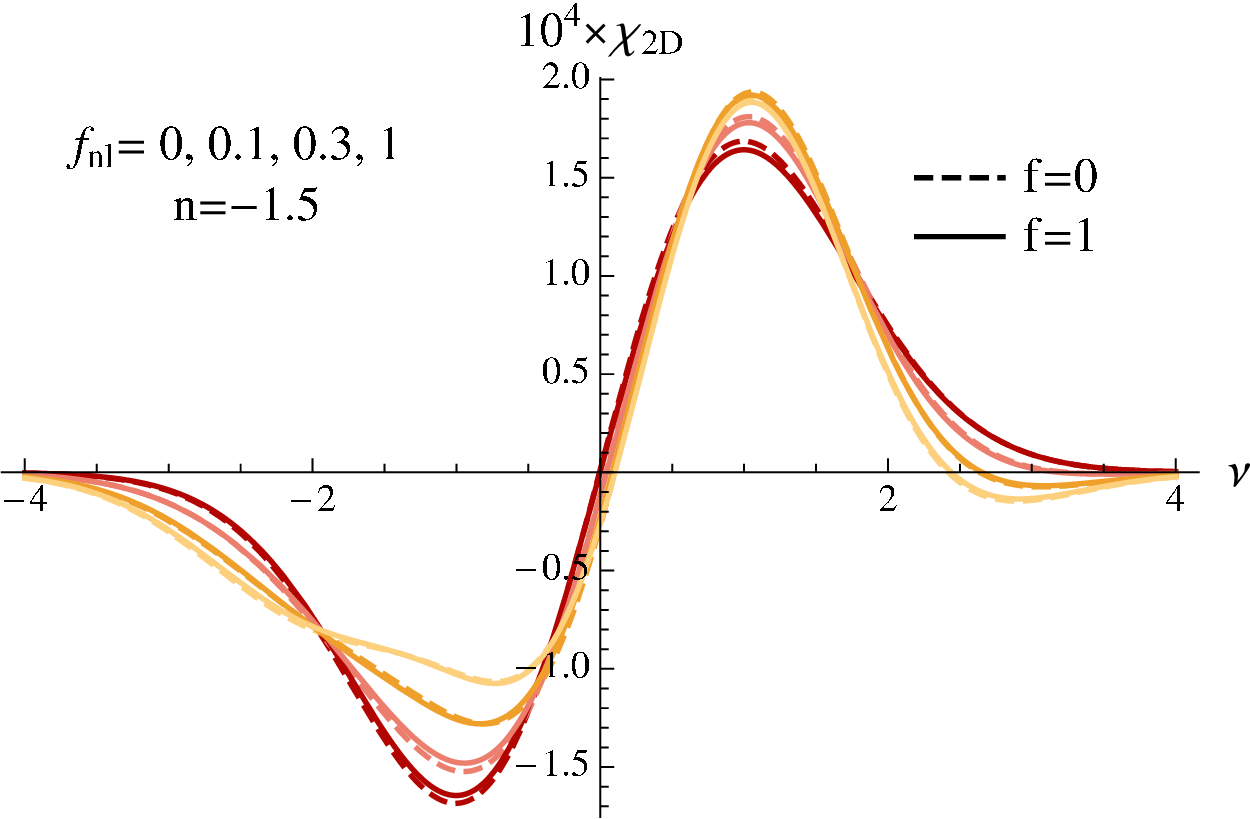}
  \includegraphics[width=0.45\textwidth]{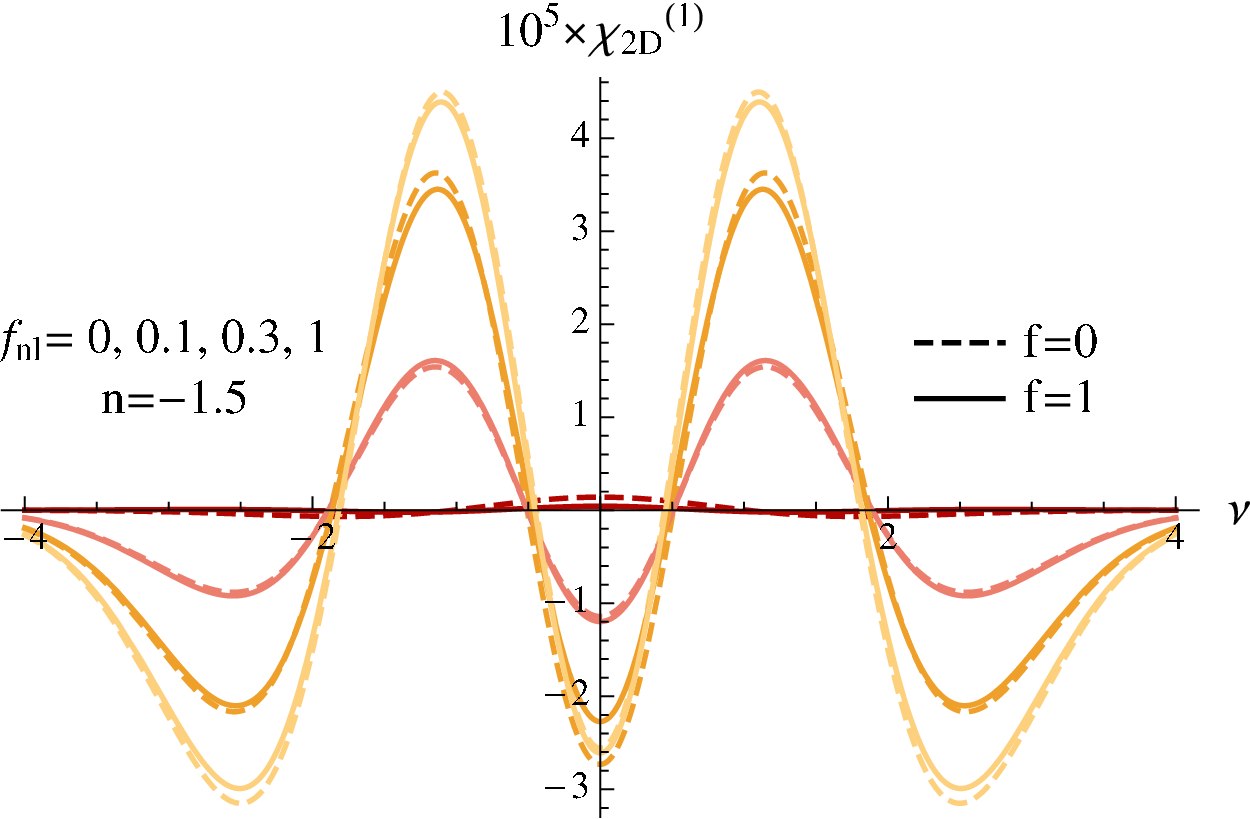}
    \includegraphics[width=0.45\textwidth]{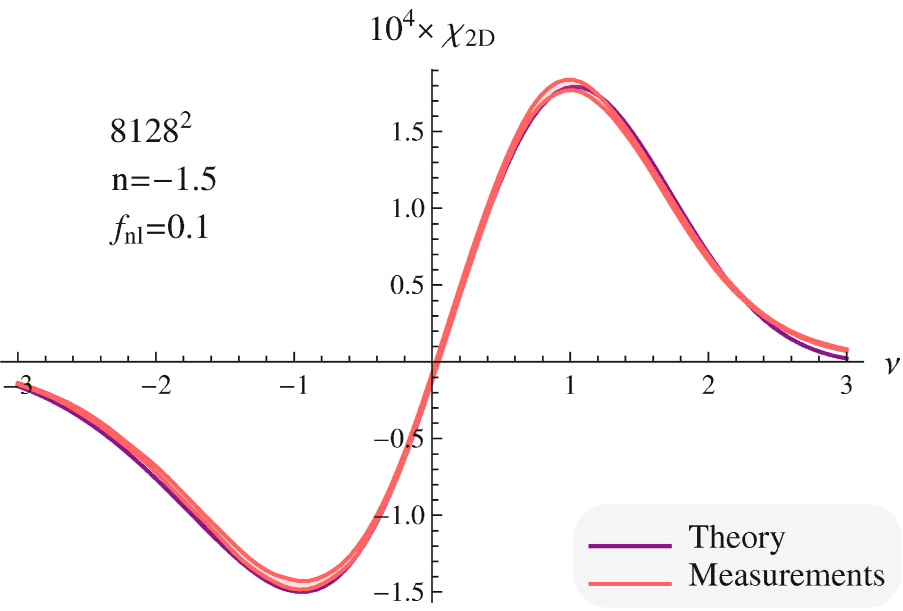}
 \includegraphics[width=0.45\textwidth]{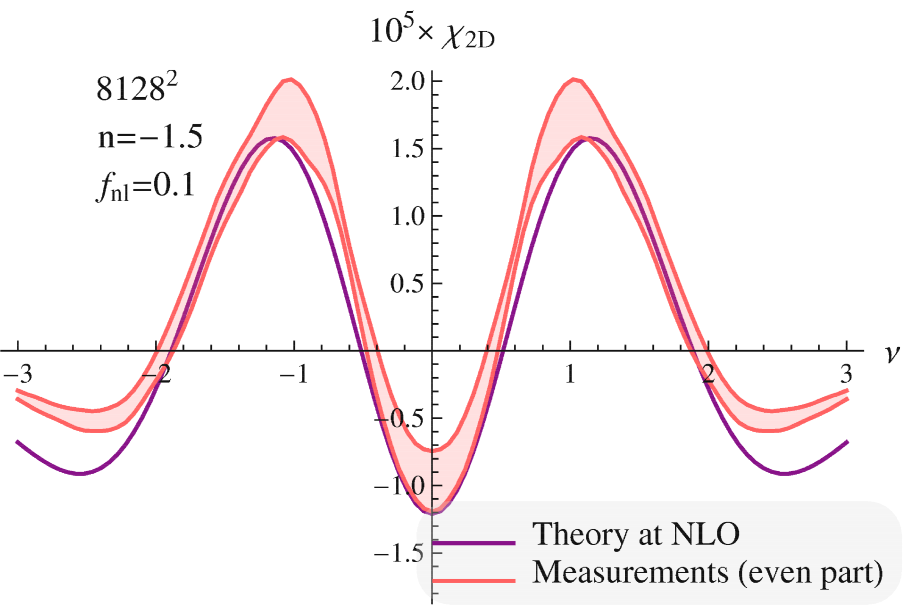}
 \caption{
{\sl Top left:} 2D Euler characteristic of the density field for $f_\mathrm{{nl}}$ between 0 ({\sl red}) and  1 ({\sl yellow}) in the {\sl potential} field and $f =0$ ({\sl dashed line}) or 1 ({\sl solid line}).
{\sl Top right:} same as top left panel for the first non-Gaussian correction.
{\sl Bottom left:} 2D Euler characteristic of the density field for $f_\mathrm{{nl}}=0.1$ in the potential field. Measurements are displayed in {\sl pink} with a shaded area representing the $1-\sigma$ dispersion and prediction in {\sl violet}.
{\sl Bottom right:}  same as left panel for the first non-Gaussian correction.
 \label{fig:realfnl}
}
 \end{center}
\end{figure}
\section{3D Euler characteristic at all orders in non-Gaussianity}
\label{sec:chi3Dappendix}
To compute the 3D Euler characteristics, one needs to take into account the variable $\Upsilon$. Given that
the Gram-Charlier polynomial expansion for the non-Gaussian PDF is obtained by using polynomials orthogonal
with respect to the 
kernel provided by the JPDF in the Gaussian limit, the coupling between $\Upsilon$ and $Q^2, J_{2\perp}$ variables in the Gaussian limit introduces
a set of polynomials $F_{lrs}$ such that
\begin{equation}
 \frac{1}{\pi} \int \dd J_{2\perp} \,
 \dd Q^2  \frac{d\Upsilon}{\sqrt{Q^4 J_{2\perp} - \Upsilon^2}}e^{- J_{2\perp}  - Q^2 }  F_{lrs}\!\left(J_{2\perp},Q^{2},\Upsilon\right)F_{l'r's'}\!\left(J_{2\perp},Q^{2},\Upsilon\right)=\delta_{ll'}\delta_{rr'}\delta_{ss'}
.
\end{equation}
Here $l$ is the power of $J_{2\perp}$, $r$ the power of $Q^{2}$ and $s$ the power of $\Upsilon$.
In particular, $F_{lr0}\!\left(J_{2\perp},Q^{2},\Upsilon\right)=L_{l}\!\left(J_{2\perp}\right)L_{r}\!\left(Q^{2}\right)$
 and $F_{001}\!\left(J_{2\perp},Q^{2},\Upsilon\right)=H_{1}\!\left(\Upsilon\right)$.

The full expansion of 
non-Gaussian JPDF $P=P(x,q^2_{\perp},x_3,\zeta, J_{2\perp}, \xi, Q^2,\Upsilon) $ 
is a series of products of $F_{lrs}$ and familiar Hermite and Laguerre polynomials for the rest of the variables, namely
\begin{equation}
P= 
G\! \left[  1 +\!   \sum_{n=3}^\infty \sum_{\sigma_{n}} 
\!\frac{(-1)^{j+l+r}g_{lrs} }{i!\;j!\; k!\; l!\;m!\;r!\;p!\;s!}
\!\left\langle x^i {q^{2j}_{\perp}} {\zeta}^k x_3^m \xi^p J^{l}_{2\perp}  Q^{2r}\Upsilon^{s}\right\rangle_{\!\mathrm{GC}}\!\!\!\!\!\!
H_i\!\left(x\right)\! L_j\!\left(q_{\perp}^2\right)\!
H_k\!\left(\zeta\right) \!
H_m\!\left(x_3\right)\! H_p\!\left(\xi\right)\!F_{lrs}\!\left(J_{2\perp},Q^{2},\Upsilon\right)
\right], 
\label{eq:defPupsilon}
\end{equation}
where $\sigma_{n}=\{(i,j,k,l,m,p,r,s)\in \mathbb{N}^{7}|i+2 j+k+2 l + m + 2r + p+3s =n\}$ and $G$ is given by
equation~(\ref{eq:3DPDF_Gauss}). The  terms of the
expansion~{\ref{eq:defPupsilon}} are sorted in the order of power in field variables $n$. We immediately see that
at $n=3$ order, the only contribution from $\Upsilon$ comes from the term corresponding to  $i=j=l=m=p=r=0, s=1$.

The Gram-Charlier coefficients are defined by
\begin{equation}
\left\langle x^i {q^{2j}_{\perp}} {\zeta}^k x_3^m \xi^p J^{l}_{2\perp}  Q^{2r}\Upsilon^{s} \right\rangle_{\!\mathrm{GC}}=\frac{(-1)^{j+l+r}}{g_{lrs}}j!\,l!\,r! \,s!\left\langle H_i\!\left(x\right)\! L_j\!\left(q_{\perp}^2\right)\!
H_k\!\left(\zeta\right) \!
H_m\!\left(x_3\right)\! H_p\!\left(\xi\right)\!F_{lrs}\!\left(J_{2\perp},Q^{2},\Upsilon\right)
 \right\rangle\,,
\end{equation}
where $g_{lrs}$ is such that the highest power term in 
$F_{lrs}\!\left(J_{2\perp},Q^{2},\Upsilon\right)$ is $(-1)^{l+r}g_{lrs}J_{2\perp}^{l}Q^{2r}\Upsilon^{s}/(l!r!s!)$. 
Amongst them, we have in particular $g_{lr0}=1$ and $g_{001}=1$.
From equation~(\ref{eq:defchi3D}), it is clear that the 3D Euler characteristic is the sum of a term where Upsilon can be marginalized over (so that the formalism introduced in Section \ref{sec:JPDF} can be used) and a term which integrates $\Upsilon$ under the condition of zero gradient. This latter part is simply computed because $\Upsilon=F_{001}(J_{2\perp},Q^{2},\Upsilon)$ is orthogonal to all other $F_{lrs}$ polynomials in the Gram-Charlier expansion of the JPDF.
After some algebra, it is found that the full 3D Euler characteristic can be eventually written
\begin{equation}
\chi_\mathrm{ 3D}(\nu)
=\frac{ e^{-\nu^{2}/2}}{8\pi^{2}}\frac {\sigma_{1\parallel}\sigma_{1\perp}^{2}} {\sigma^{3}}H_{2}(\nu)+ \sum_{n=3}^\infty\chi_\mathrm{ 3D}^{(n)}\,,
\label{eq:chi3d}
\end{equation}
with $\chi_\mathrm{ 3D}^{(n)}$ given by
\begin{multline}
\chi_\mathrm{ 3D}^{(n)}(\nu) 
= \frac{ e^{-\nu^{2}/2}}{8\pi^{2}}\frac{ \sigma_{2\perp}^{2}\sigma_{2\parallel}}{\sigma_{1\perp}^{2}\sigma_{1\parallel}} 
\Bigg[
\sum_{\sigma_{n}} \frac{(-1)^{j+m}}{2^{m}i!\;j!\;m!} H_{i+2}(\nu) 
\gamma_{\parallel}\gamma_{\perp}^{2}\left\langle  x^{i} q^{2j}_{\perp}  x_3^{2m}   \right\rangle_\mathrm{ GC}
- \sum_{\sigma_{n-1}} \frac{(-1)^{j+m}}{2^{m}i!\;j!\;m!} H_{i+1}(\nu) 
\left\langle  x^{i} q^{2j}_{\perp}  x_3^{2m} \left(\gamma_\perp^2 x_{33} 
+ 2 \gamma_\perp \gamma_\parallel J_{1\perp} \right)  \right\rangle_\mathrm{ GC} \nonumber
\end{multline}
\begin{multline}
+\!\!\sum_{\sigma_{n-2}}\!\! \frac{(-1)^{j+m}}{2^{m}i!\;j!\;m!} H_{i}(\nu)
 \! \left\langle x^{i} q^{2j}_{\perp}  x_3^{2m}  \left(
2 \gamma_\perp (J_{1\perp} x_{33}-\gamma_{2}Q^{2}) + \gamma_\parallel (J_{1\perp}^2
 - {J_{2\perp}})  \right)\right\rangle_{\!\mathrm GC}
-\sum_{\sigma_{n-3}}\!\! \frac{(-1)^{j+m}}{2^{m}i!\;j!\;m!}H_{i-1}
\!\left(\nu\right) \left\langle  x^{i} q^{2j}_{\perp}  x_3^{2m} 
\left( 4 I_3 \right) \right\rangle_\mathrm{\! \!GC}
\Bigg]\,, \label{eq:chi3Dfinalallorder}
\end{multline}
where $I_3=\frac{1}{4} x_{33} ( J^{2}_{1\perp} - J_{2\perp} )-\frac 1 2 \gamma_2 \left( Q^2 J_{1\perp} - \Upsilon \right) $.
In equation~(\ref{eq:chi3Dfinalallorder}), the Gram-Charlier coefficients are written in a concise form but must be interpreted returning from the $(x, x_{33}, J_{1\perp})$ to the $(x,\zeta,\xi)$ variables and using equation~(\ref{eq:GC}).
In the isotropic limit the result is reduced to the following compact form
\begin{multline}
\chi_\mathrm{ 3D}^{(n)}(\nu) 
= \frac{ e^{-\nu^{2}/2}}{8\pi^{2}} \frac{ \sigma_{1}^{3}}{\sigma^3}
\Bigg[ \frac{2}{3\sqrt{3}}
\sum_{\sigma_{n}} \frac{(-1)^{j+m}}{2^{m}i!\;j!\;m!} H_{i+2}(\nu) 
\left\langle  x^{i} q^{2j}_{\perp}  x_3^{2m}   \right\rangle_\mathrm{ GC}
- \frac{2}{\sqrt{3}} \gamma^{-1} \sum_{\sigma_{n-1}} \frac{(-1)^{j+m}}{2^{m}i!\;j!\;m!} H_{i+1}(\nu) 
\left\langle  x^{i} q^{2j}_{\perp}  x_3^{2m} I_1  \right\rangle_\mathrm{ GC}
\\
+2\sqrt{3} \gamma^{-2} \!\!\sum_{\sigma_{n-2}}\!\! \frac{(-1)^{j+m}}{2^{m}i!\;j!\;m!} H_{i}(\nu)
 \! \left\langle x^{i} q^{2j}_{\perp}  x_3^{2m}  I_2 \right\rangle_{\!\mathrm GC}
-\frac{16}{5}\sqrt{\frac{3}{5}} \gamma^{-3} \sum_{\sigma_{n-3}}\!\! \frac{(-1)^{j+m}}{2^{m}i!\;j!\;m!}H_{i-1}
\!\left(\nu\right) \left\langle  x^{i} q^{2j}_{\perp}  x_3^{2m} 
I_3 \right\rangle_\mathrm{\! \!GC}
\Bigg]\,,
\label{eq:chi3Diso}
\end{multline}
where $I_{1}$, $I_{2}$ and $I_{3}$ are the 3 invariants of the Hessian matrix $x_{ij}$, namely $I_{1}=J_{1}=\lambda_{1}+\lambda_{2}+\lambda_{3}=\mathrm{Tr}\, x_{ij}$, $I_{2}=\lambda_{1}\lambda_{2}+\lambda_{2}\lambda_{3}+\lambda_{3}\lambda_{1}$ and $I_{3}=\lambda_{1}\lambda_{2}\lambda_{3}=\det x_{ij}$.
 Note that equations~(\ref{eq:chi3Dfinalallorder}) and (\ref{eq:chi3Diso}) contain some 
$H_{-1}(\nu)=\sqrt{\pi/2}\textrm{ Erfc}(\nu/\sqrt{2})$ which are shown to contribute as a boundary term and
can thus be factorized out of the terms $\chi_\mathrm{ 3D}^{(n)}$ as mentioned in equation~(\ref{eq:Chi3D}).
Moreover, the well known topological relation between the Euler characteristics and the genus allows us to compare the above quantity with the curvature integrated along isocontours (as used in \cite{Matsubara96}). In an anisotropic space, this surface integral is found to be
\begin{multline}
\chi_\mathrm{ 3D}^\mathrm{ s}(\nu)=\frac{ e^{-\nu^{2}/2}}{8\pi^{2}}\frac {\sigma_{1||}\sigma_{1\perp}^{2}} {\sigma^{3}}
\Bigg[ H_{2}(\nu)+\frac{1}{\gamma_{\perp}^{2}}\sum_{n=3}^\infty \sum_{\sigma_{n-2} }
\!\!\!
\frac{(-1)^{j+m}}{i!\;j! m!(2m\!-\!1) 2^{m}} H_{i}(\nu)\left(\left\langle x^i q^{2j}_{\perp}
  {J_{2\perp}} \,x_3^{2m}   \right\rangle_{\!\mathrm{GC}}\!\!\!\!-(1-\gamma_{\perp}^{2}) 
\left\langle x^i q^{2j}_{\perp}  \zeta^{2} x_3^{2m} \right\rangle_{\!\mathrm{GC}}\right) \\
+2\frac{\sqrt{1\!-\!\gamma_{\perp}^{2}}}{\gamma_{\perp}}\sum_{n=3}^\infty \sum_{\sigma_{n-1}} 
\!\!\!
\frac{(-1)^{j+m}}{i!\;j! m!(2m\!-\!1) 2^{m}} \!
\left\langle x^i q^{2j}_{\perp}  \zeta\, x_3^{2m}   \right\rangle_{\!\mathrm{GC}}\!\!\!\!\!\!\!\!H_{i}(\nu) H_{1}(\nu) 
\!-\!\sum_{n=3}^\infty \sum_{\sigma_{n}} 
\!
\frac{(-1)^{j+m}}{i!\;j! m!(2m\!-\!1) 2^{m}} \!
\left\langle x^i q^{2j}_{\perp}   x_3^{2m}   \right\rangle_{\!\mathrm{GC}}\!\!\!\!\!\!\!\!H_{i}(\nu) H_{2}(\nu)\!
\Bigg] \,,
\label{eq:chi3dMatsubara}
 \end{multline}
 where  we use the convention $(-2)!/(-1)!=-1/2$. Equations~(\ref{eq:chi3d}) and (\ref{eq:chi3dMatsubara}) were checked to be equivalent up to at least $n=4$ using some detailed relations between the cumulants (which once again can be established via integrations by parts) assuming that everything vanishes at infinity (in particular $\chi_\mathrm{ 3D}(-\infty)=0$). Note that in this expression $\xi$, $Q^{2}$ and $\Upsilon$ do not appear anymore.

\vfill
\eject

\end{document}